\documentclass[12pt, oneside]{article}
\usepackage{graphicx}
\pdfoutput=1
\usepackage[export]{adjustbox}
\usepackage[paper=a4paper]{geometry}
\usepackage[T1]{fontenc}
\usepackage{setspace} 
\usepackage[hang]{footmisc} 
\usepackage{titling} 
\usepackage{tocloft} 
\usepackage{parskip} 
\usepackage{hyperref}
\usepackage{fourier-orns}
\usepackage{booktabs}
\usepackage{shuffle}
\usepackage{amssymb} 
\usepackage{tabularx}
\usepackage[fixed]{fontawesome5}
\usepackage{twemojis}

\usepackage{tikz}
\usetikzlibrary{arrows.meta,positioning,fit}

\usepackage{amsfonts}
\usepackage{amsmath}
\usepackage{amssymb}
\usepackage{mathrsfs, dsfont, amsthm}
\usepackage{physics}
\usepackage{mathtools}
\numberwithin{equation}{section}

\newtheorem{remark*}{Remark}

\usepackage[dvipsnames]{xcolor}
\definecolor{dark-red}{rgb}{0.50,0.12,0.12}
\definecolor{mblue}{rgb}{0.30, 0.45, 0.70}
\definecolor{mred}{rgb}{0.70, 0.20, 0.20}
\definecolor{mgray}{rgb}{0.63, 0.63, 0.63}
\definecolor{myWhite}{RGB}{255,255,243}

\newcommand*\justify{%
  \fontdimen2\font=0.4em%
  \fontdimen3\font=0.2em%
  \fontdimen4\font=0.1em%
  \fontdimen7\font=0.1em%
  \hyphenchar\font=`\-%
}

\renewcommand{\texttt}[1]{%
  \begingroup
  \ttfamily
\begingroup\lccode`~=`/\lowercase{\endgroup\def~}{/\discretionary{}{}{}}%
\begingroup\lccode`~=`[\lowercase{\endgroup\def~}{[\discretionary{}{}{}}%
\begingroup\lccode`~=`.\lowercase{\endgroup\def~}{.\discretionary{}{}{}}%
\catcode`/=\active\catcode`[=\active\catcode`.=\active
  \justify\scantokens{#1\noexpand}%
  \endgroup
}

\usepackage{hyperref} 
\hypersetup{
colorlinks=true,
allcolors=dark-red,
linktoc=page,
pdftitle={SubTropica},
pdfauthor={Mathieu Giroux, Sebastian Mizera, Giulio Salvatori},
pdfdisplaydoctitle=true,
pdfstartview=FitH
}
\usepackage{cite}

\usepackage{orcidlink}
\definecolor{orcidlogocol}{named}{Maroon}

\usepackage{tikz}
\usepackage{tikzlings}
\usetikzlibrary{calc, patterns, decorations, decorations.markings, shapes, positioning, intersections, quotes, angles}
\usepackage[subrefformat=parens]{subcaption}
\usepackage{pgfplots}
\pgfplotsset{compat=newest}
\newcommand{\mathdefault}[1][]{}

\def\env@sqcases{
  \let\@ifnextchar\new@ifnextchar
  \left\lbrack
  \def\arraystretch{1.2}
  \array{@{}l@{\quad}l@{}}
}
\makeatother

\def \d   {\mathrm{d}}

\newcommand{\D}{\mathrm{D}}

\makeatletter
\newcommand{\subalign}[1]{
  \vcenter{
    \Let@ \restore@math@cr \default@tag
    \baselineskip\fontdimen10 \scriptfont\tw@
    \advance\baselineskip\fontdimen12 \scriptfont\tw@
    \lineskip\thr@@\fontdimen8 \scriptfont\thr@@
    \lineskiplimit\lineskip
    \ialign{\hfil$\m@th\scriptstyle##$&$\m@th\scriptstyle{}##$\hfil\crcr
      #1\crcr
    }
  }
}
\makeatother

\usepackage{mathtools}
\usepackage{comment}
\newcommand{\beq}{\begin{equation}}
\newcommand{\eeq}{\end{equation}}

\renewcommand{\le}{\leqslant}
\renewcommand{\leq}{\leqslant}
\renewcommand{\ge}{\geqslant}
\renewcommand{\geq}{\geqslant}

\usepackage{enumitem}
\setlist[itemize]{leftmargin=1.2em}

\usepackage{tcolorbox}
\usepackage{colortbl} 
\usepackage{listings}
\usepackage{setspace} 
\definecolor{identifiercolor}{rgb}{.4,.6,.56}
\definecolor{stringcolor}{gray}{0.5}
\definecolor{inactivecolor}{rgb}{0.2,0.2,0.2}

\newcommand{\SOFIA}{\adjustbox{valign=t,scale=0.7,raise=0.05em}{\includegraphics[height=1em]{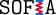}}}
\newcommand{\version}{\texttt{v1.1.0}}

\begin{document}
\raggedbottom
\begin{titlingpage}
    \vspace*{0em}
    \onehalfspacing
    \begin{center}
        \textbf{\Huge 
      {
      \texttt{SubTr}
      {\hspace{-0.3em}{\scalebox{2}{\raisebox{-0.1ex}{\rotatebox{90}{\twemoji{coconut}}}}}\hspace{-0.3em}}
      \texttt{pica}
      }
      }
      \\
      \phantom{ghost}
      \\
      \centering{\large\version}
    \end{center}
    \singlespacing
    \vspace*{2em}
    \begin{center}
        \textbf{
        Mathieu Giroux,$^{1,2,\orcidlink{0000-0002-2672-634X}}$ Sebastian Mizera,$^{1,\orcidlink{0000-0002-8066-5891}}$
        Giulio Salvatori$^{2,3,\orcidlink{0000-0002-5961-3210}}$
        }
    \end{center}
    \vspace*{1em}
    \begin{center}
        \textsl{
        $^1$\ Department of Physics, Columbia University,
        \\538 West 120th Street, New York, NY 10027, USA \\[\baselineskip]
        }
        \textsl{
        $^2$\ Institute for Advanced Study \\
         Einstein Drive, Princeton, NJ 08540, USA \\[\baselineskip]
        }
        \textsl{$^3$\ Max-Planck-Institut f\"ur Physik, Werner-Heisenberg-Institut,\\
        D–85748 Garching bei M\"unchen, Germany
        \\[\baselineskip]
        }
        \href{mailto:giroux@ias.edu}{\small\texttt{giroux@ias.edu}},
        \href{mailto:sebastian.mizera@columbia.edu}{\small\texttt{sebastian.mizera@columbia.edu}},
        \href{mailto:salvatori@ias.edu}{\small\texttt{salvatori@ias.edu}}
    \end{center}
    \vspace*{3em}
    \begin{abstract}
 \noindent 
 We present {\texttt{SubTropica}}, a \textsc{Mathematica} package that performs symbolic integration of multi-polylogarithmic integrals using recent advances in tropical geometry. It focuses on the class of linearly-reducible \emph{Euler integrals}, such as Feynman integrals, and expands them using a tropical subtraction scheme. The engine behind it is \texttt{HyperIntica}, a native  \textsc{Mathematica} package for hyperlogarithm integration that can be used independently.

This paper documents both packages and illustrates their usage on examples from across different physics applications. Additionally, we introduce an AI-driven library of Feynman integrals, which catalogs diagrams discussed in the literature and serves as a database for computed results. Its online version is available at
\begin{center}
\href{https://subtropi.ca}{\texttt{subtropi.ca}}
\end{center}
and features a graphical user interface for diagram input and retrieval of records.
    \end{abstract}
    \vfill
\end{titlingpage}

\newpage
\tableofcontents 
\raggedbottom
\setcounter{page}{2}

\section{Introduction}

A recurring bottleneck in perturbative calculations across high-energy physics is the evaluation of multi-dimensional integrals. Examples include Feynman integrals, phase-space integrals, energy and cosmological correlators, and so on. They belong to a common class known as \emph{Euler integrals}:
\begin{align}
    I(\varepsilon,{\bf s}) = \int_0^\infty \d x_1\cdots \d x_n \prod_i P_i(x,{\bf s})^{a_i \varepsilon + b_i}\,,
    \label{eq:euler}
\end{align}
where $P_i(x,{\bf s})$ are polynomials in the integration variables $x$ with coefficients collectively denoted by~${\bf s}$, the exponents $a_i, b_i$ are fixed constants, and $\varepsilon > 0$ is a small real parameter (typically the dimensional regulator $\varepsilon = (4-\D)/2$ in the Feynman integral context).

\begin{figure}[t]
\centering
\begin{tikzpicture}[scale=1.5]
  \coordinate (v1) at (0.250, 1.750);
  \coordinate (v2) at (2.250, 1.750);
  \coordinate (v3) at (4.500, 1.750);
  \coordinate (v4) at (6.750, 1.750);
  \coordinate (v5) at (3.920, 0.719);
  \coordinate (v6) at (5.500, -0.250);
  \coordinate (v7) at (6.750, -0.250);
  \coordinate (v8) at (3.379, -0.250);
  \coordinate (v9) at (1.250, -0.250);
  \coordinate (v10) at (0.250, -0.250);

  \draw[thick, dashed, line cap=round] (v1) node[left, auto, font=\footnotesize] {$p_{1}$} -- (v2);
  \draw[ultra thick, Maroon] (v2) -- (v3) node[midway, auto, font=\footnotesize] {$-\ell_{2} - \ell_{3} - p_{2} - p_{3}$};
  \draw[thick, dashed, line cap=round] (v4) node[right, auto, font=\footnotesize] {$p_{2}$} -- (v3);
  \draw[thick, dashed, line cap=round] (v5) -- (v3) node[midway, auto, font=\footnotesize] {$-\ell_{1} + \ell_{3}$};
  \draw[thick, dashed] (v5) -- (v6) node[midway, above, font=\footnotesize] {$\ell_{1}$};
  \draw[thick, dashed, line cap=round] (v3) -- (v6) node[midway, auto, font=\footnotesize] {$-\ell_{1} - \ell_{2} - p_{3}$};
  \draw[thick, dashed, line cap=round] (v7) node[right, auto, font=\footnotesize] {$p_{3}$} -- (v6);
  \draw[thick, dashed, line cap=round] (v8) -- (v6) node[midway, below, font=\footnotesize] {$\ell_{2}$};
  \draw[thick, dashed, line cap=round] (v8) -- (v5) node[midway, right, font=\footnotesize] {$\ell_{3}$};
  \draw[thick, dashed, line cap=round] (v2) -- (v8) node[left, yshift=20pt, xshift=-12pt, font=\footnotesize] {$\ell_{2} + \ell_{3} - \ell_{4}$};
  \draw[thick, dashed, line cap=round] (v9) -- (v8) node[midway, below, font=\footnotesize] {$\ell_{4}$};
  \draw[thick, dashed, line cap=round] (v9) -- (v2) node[midway, auto, font=\footnotesize] {$-\ell_{4} + p_{4}$};
  \draw[thick, dashed, line cap=round] (v10) node[left, auto, font=\footnotesize] {$p_{4}$} -- (v9);

  \fill (v2) circle (2pt);
  \fill (v3) circle (2pt);
  \fill (v5) circle (2pt);
  \fill (v6) circle (2pt);
  \fill (v8) circle (2pt);
  \fill (v9) circle (2pt);
\end{tikzpicture}
\caption{A four-loop Feynman diagram with nine propagators, one massive internal (red, solid) edge ($m^2=1$), and four massless (black, dashed) external legs. \texttt{SubTropica} evaluates for the first time the corresponding integral through $\mathcal{O}(\varepsilon^0)$. See Sec.~\ref{sec:examplesCE} for details.}
\label{fig:cuttingEdge}
\end{figure}
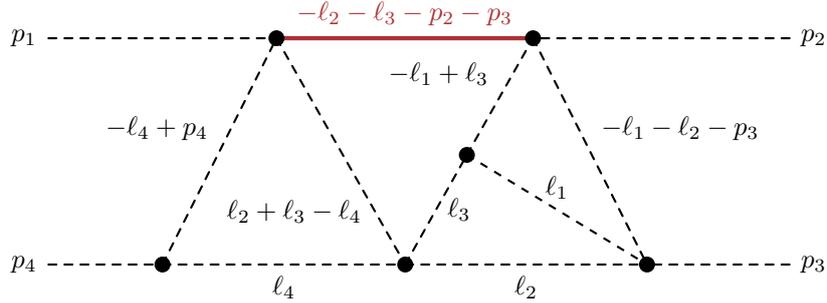

In this paper, we introduce \texttt{SubTropica}, a \textsc{Mathematica} package that evaluates integrals of the form~\eqref{eq:euler} as Laurent expansions in~$\varepsilon$,
\begin{align}
    I(\varepsilon,{\bf s}) = \sum_{i=-N}^\infty \varepsilon^i\, I_i({\bf s})\,,
\end{align}
expressing the coefficients $I_i({\bf s})$ in terms of \emph{hyperlogarithms}. The central difficulty is that, often, the integrand of~\eqref{eq:euler} cannot simply be expanded in $\varepsilon$ and integrated term by term, which would give meaningless expressions. To overcome this, \texttt{SubTropica} implements the tropical subtraction method of~\cite{Salvatori:2024nva} (see also~\cite{Hillman:2023ezp,Brown:2019wna}), which uses the geometry of the Newton polytope associated with the integrand to systematically construct counter-terms that render each integral locally finite. The resulting \emph{locally finite} integrands can then be expanded in $\varepsilon$ at the integrand level and integrated one variable at a time using, e.g., the hyperlogarithm integration algorithms of Brown and Panzer~\cite{Brown:2008um,Panzer_2015}. This last step requires the integral to be \emph{linearly reducible}, meaning that at each stage of the iterated integration, all singularities remain linear in the current integration variable. While not all Euler integrals satisfy this condition, a large and practically important class does, including, e.g., many multi-loop Feynman integrals relevant for phenomenology. 

As a demonstration of its capabilities, \texttt{SubTropica} can tackle cutting-edge examples such as the four-loop, nine-propagator massive diagram shown in Fig.~\ref{fig:cuttingEdge}.

\paragraph{Relation to prior work.}
  While powerful implementations of the hyperlogarithm integration
  algorithm exist in \textsc{Maple} (\texttt{HyperlogProcedures}~\cite{Schnetz:2016fhy}
  and \texttt{HyperInt}~\cite{Panzer:2014caa}) and \textsc{FORM}
  (\texttt{HyperFORM}~\cite{Kardos:2025klp}, \texttt{Forcer}~\cite{Ruijl:2017cxj}),
  a significant part of the high-energy physics community relies on
  \textsc{Mathematica} as its primary computational environment, where tools for
  amplitude generation and symbolic manipulation~\cite{Hahn:2000kx,Hahn:1998yk,Shtabovenko:2023idz,Patel:2015tea,Patel:2016fam,Shtabovenko:2016whf},
  integration-by-parts reduction~\cite{Smirnov:2023yhb,vonManteuffel:2012np,Lee:2013mka,Maierhofer:2017gsa,Klappert:2020nbg,Georgoudis:2016wff,Wu:2023upw,Guan:2024byi,Peraro:2019svx},
  differential equations and canonical forms~\cite{Meyer:2017joq,Lee:2020zfb,Hidding:2020ytt,Gituliar:2017vzm,Prausa:2017ltv,Dlapa:2020cwj},
  numerical evaluation~\cite{Heinrich:2023til,Smirnov:2021rhf,Liu:2022chg,Borinsky:2023jdv},
  evaluation of special functions and iterated integrals~\cite{Maitre:2005uu,Huber:2005yg,Huber:2007dx},
  symbol and singularity analysis~\cite{Fevola:2023kaw,Fevola:2023fzn,Correia:2025wtb,Duhr:2019tlz,Jiang:2024eaj},
  parametric and GKZ methods~\cite{Ananthanarayan:2022ntm},
  spinor-helicity techniques~\cite{Bourjaily:2010wh,Maitre:2007jq,AccettulliHuber:2023ldr,Bourjaily:2023uln},
  and tensor reduction~\cite{Goode:2024cfy}
  are readily available or interoperable. To keep the entire workflow within a
  single platform, \texttt{SubTropica} ships with \texttt{HyperIntica}, a
  self-contained \textsc{Mathematica} reimplementation of the \texttt{HyperInt}
  algorithm that can also be used as a standalone package for evaluating
  iterated integrals of hyperlogarithms.

\paragraph{Repository.} The up-to-date version of \texttt{SubTropica} is hosted open source on \textsc{GitHub}:
\begin{center}
\href{https://github.com/SubTropica/SubTropica}{\texttt{github.com/SubTropica/SubTropica}}
\end{center}
which also contains the associated examples and documentation.
The code is released under the MIT license.

\paragraph{Library.}
Together with \texttt{SubTropica}, we publish an extensive library of diagrams that have been computed in the literature. Each entry contains detailed records about the diagram itself, papers in which it appeared, classes of functions it evaluates to, etc. When available, it also has the result of computation with \texttt{SubTropica} in terms of hyperlogarithms. We invite all readers to submit new entries to the library. The library can be accessed through the graphical user interface (GUI) provided in \texttt{SubTropica}, as well as online at
\begin{center}
\href{https://subtropi.ca}{\texttt{subtropi.ca}}
\end{center}
The online version allows one to draw, browse, and download diagrams, but the integration features require the \textsc{Mathematica} backend. The library is released under the CC BY-NC-SA 4.0 license, including restrictions on machine learning use.

\paragraph{Outline.}
The paper is organized as follows. Sec.~\ref{sec:manual} serves as a quick-start user manual: installation, basic functionality, and an overview of the package architecture. In Sec.~\ref{sec:Preliminary}, we review the theoretical ingredients: Euler integrals and their Newton polytopes, the tropical subtraction algorithm, and the hyperlogarithm integration strategy, including a fully worked example. In Sec.~\ref{sec:examples}, we present detailed examples ranging from Feynman integrals to applications beyond the scattering-amplitude context. We conclude in Sec.~\ref{sec:Conclusion} with a summary and outlook. Technical details on the \texttt{STIntegrate} and \texttt{ConfigureSubTropica} interfaces, as well as \texttt{HyperIntica}'s conventions, are collected in the appendices. To ensure reproducibility, all code snippets quoted in this paper are collected and checked in the accompanying notebook \texttt{PaperChecks.wl} on \textsc{GitHub}.

\paragraph{AI use.} The development of \texttt{SubTropica} was assisted by Claude Opus 4.6--7 at all stages: from algebraic manipulations (cross-checking computations, finding rationalization of square roots, statistical analysis), through coding and algorithms (optimization, benchmarking, debugging, designing the user interface), to the library construction (retrieval and parsing of records, numerical verification, matching system).

\paragraph{Feynman integral conventions.}
Throughout this paper and the accompanying package, we adopt the following normalization convention for Feynman integrals:
\begin{equation}\label{eq:loopMomDef}
I_{\vec{\nu}}(\mathbf{s}) \equiv \frac{\mathrm{e}^{\varepsilon L\gamma_\text{E}}}{(i\pi^{\D/2})^L} \int \frac{\mathcal{N}\, \d^\D \ell_1 \, \d^\D \ell_2 \cdots \d^\D \ell_L}{(q_1^2 - m_1^2 )^{\nu_1} (q_2^2 - m_2^2 )^{\nu_2} \cdots (q_N^2 - m_N^2 )^{\nu_N}}\, ,
\end{equation}
where $\{\ell_a\}_{a=1}^{L}$ represents the set of the $L$ independent loop momenta, and $q_e$ and $m_e$ are the momenta and masses associated with the internal edges. The exponents $\nu_i$ are the integer powers of the respective propagators or irreducible scalar products, while $\mathcal{N}$ is a numerator polynomial in the loop momenta. The vector $\mathbf{s}$ collectively refers to the kinematic variables, such as masses and Mandelstam invariants. The quantity $\gamma_\text{E}$ is a shorthand for the Euler constant $-\Gamma'(1) \approx 0.577$. The Feynman $i\epsilon$ prescription is left implicit. We use mostly-minus $(+-\ldots-)$ conventions.

The resulting Schwinger-parameter representation is
\begin{equation}\label{eq:SPx}
I_{\vec{\nu}}(\mathbf{s}) =   \Gamma(\omega) \left( \prod_{e=1}^{E}\frac{(-1)^{\nu_e} }{\Gamma(\nu_e)} \right) \int \frac{\d^{E} x}{\text{GL(1)}} \prod_{e=1}^{E} x_e^{\nu_e - 1} \frac{\tilde{\mathcal{N}}(\mathbf{s})}{\mathcal{U}^{\D/2 - \omega} [-\mathcal{F}(\mathbf{s})]^{\omega}}\, ,
\end{equation}
where $E$ denotes the number of internal edges (those with $\nu_e > 0$), $x_e$ denotes the Schwinger parameter associated with every internal edge $e$,
and
\begin{equation}
\omega \equiv \sum_{e=1}^{E} \nu_e - \frac{L \D}{2}\,,
\end{equation}
is the superficial degree of divergence, see, e.g., \cite[App.~A]{Hannesdottir_2022}. The factors $\mathcal{U}$ and $\mathcal{F}$ are the first and second Symanzik polynomials respectively, while $\tilde{\mathcal{N}}$ is the numerator polynomial. Modding out by the gauge redundancy $\text{GL}(1)$ can be done by fixing one of the Schwinger parameters, say $x_E=1$.

\paragraph{Measure conventions.}\label{par:measure}
In the text, we write $x_i$ with the standard measure $\prod_i \d x_i$ (as in \eqref{eq:SPx}), and $\alpha_i$ with the logarithmic measure $\prod_i \d\log\alpha_i$ (as in \eqref{eq:defeuler}). The variables are the same ($x_i = \alpha_i$) and the different labels signal which measure is in play. The $\d\log$ form is natural when tropicalizing integrands. The $\d x$ form is standard in the Feynman-integral literature and is used by the integrator \texttt{HyperIntica}. The two measures simply differ by the Jacobian $\prod_i x_i$.

\section{\label{sec:manual}Manual}

This section explains how to install {\texttt{SubTropica}} and its dependencies, outlines its main functions, and provides the complete documentation.

\subsection{Installation}

{\texttt{SubTropica}} can be installed by running
\begin{lstlisting}[extendedchars=true,language=Mathematica]
PacletInstall["https://subtropi.ca/SubTropica.paclet"]
\end{lstlisting}
directly in \textsc{Mathematica}. Alternatively, it is available at the \textsc{GitHub} repository \cite{repo} and can be downloaded or cloned directly. The package then needs to be moved to \textsc{Mathematica}'s \texttt{\$Path} directory manually. {\texttt{SubTropica}} requires \textsc{Mathematica} 12.1 or later.

Once the package is installed,
it can be loaded and configured from anywhere with\footnote{All paths set via \texttt{ConfigureSubTropica} are saved to a persistent configuration file (\texttt{\$UserBaseDirectory/Kernel/SubTropicaConfig.m}) and automatically restored on every subsequent \texttt{Get["SubTropica{\textasciigrave}"]}, so \texttt{ConfigureSubTropica} needs to be called only once, unless the configuration needs updating. The saved configuration can be cleared with \texttt{STResetConfig[]}.}
\begin{lstlisting}[extendedchars=true,language=Mathematica,literate={`}{{\textasciigrave}}1]
Get["SubTropica`"];
\end{lstlisting}
It can be configured by typing
\begin{lstlisting}[extendedchars=true,language=Mathematica,literate={`}{{\textasciigrave}}1]
ConfigureSubTropica[(* required *)
                    PolymakePath   -> "path/to/polymake",
                    (* recommended *)
                    FiniteFlowPath -> "path/to/FiniteFlow",
                    SPQRPath       -> "path/to/SPQR",
                    PythonPath     -> "path/to/python",
                    (* optional *)
                    GinshPath      -> "path/to/ginsh",
                    MaplePath      -> "path/to/maple",
                    FIESTAPath     -> "path/to/FIESTA"
                    ];
\end{lstlisting}

The only required dependency is {\texttt{polymake}} \cite{polymake:2000,polymake:2017}.

All other dependencies are optional. A working installation of \textsc{Python} is highly recommended, as it is needed for the graphical user interface (GUI) and for \texttt{pySecDec} \cite{Heinrich:2023til}. (The latter should be installed via \texttt{pip} into the \textsc{Python} interpreter set as \texttt{PythonPath}.) The package \texttt{ginsh} (the \texttt{GiNaC} interactive shell \cite{Bauer:2000cp}) is needed only to evaluate hyperlogarithms numerically, and \textsc{Maple} can be useful for cross-checks, since {\texttt{SubTropica}} includes {\texttt{HyperIntica}}, a native \textsc{Mathematica} reimplementation of {\texttt{HyperInt}} \cite{Panzer:2014caa}. Finally, \texttt{pySecDec} and \texttt{FIESTA} \cite{Smirnov:2021rhf} can be used to validate the computed results numerically.

For complicated examples, we recommend the user to work with the libraries \texttt{FiniteFlow} \cite{Peraro:2019svx} and \texttt{SPQR} \cite{Chestnov:2025svg} to enable finite-field arithmetic in the partial-fraction step of the integration algorithm within \texttt{HyperIntica}. When the function \texttt{Get["SubTropica`"]}
is called, the package silently checks whether they are already available on \textsc{Mathematica}'s \texttt{\$Path} and loads them automatically if so. In that case, \texttt{FiniteFlowPath} and \texttt{SPQRPath} need not be set. If either library is loaded, \texttt{PartialFractions} automatically switches from \textsc{Mathematica}'s built-in \texttt{PolynomialQuotient} to finite-field reconstruction via \texttt{SPQRPolynomialQuotient}, which avoids intermediate expression swell on complicated integrands.

\paragraph{Tip.} On macOS, the easiest installation route for most packages mentioned above is \textsc{Homebrew} \cite{Homebrew}. For example, the default path for \texttt{polymake} is \texttt{/opt/homebrew/bin/polymake}. After installation, we recommend running
\begin{lstlisting}[extendedchars=true,language=Mathematica,literate={`}{{\textasciigrave}}1]
STBenchmark[]
\end{lstlisting}
which verifies the dependencies are installed correctly and a suite of test examples passes all the checks (\texttt{"Suite" -> "Short"} or \texttt{"Long"} selects the length of the benchmark).

\subsection{Computing Feynman integrals in a hurry}
\label{sec:quickstart}

The entry point of {\texttt{SubTropica}} is the function \texttt{STIntegrate}. While a rich set of calling conventions is available for scripted usage (see Sec.~\ref{sec:functions} and App.~\ref{app:functions}), the simplest way to get started is to call
\begin{lstlisting}[language=Mathematica]
STIntegrate[]
\end{lstlisting}
with no arguments. This command opens a GUI that covers the entire workflow---from drawing the diagram to obtaining the integrated result---without writing any code. 
The GUI features an interactive tutorial after launching it for the first time. After a diagram is drawn and all the necessary options (space-time dimension, requested $\varepsilon$ order) are specified, one simply clicks the ``Integrate'' button to start integrating.

\begin{table}
\begin{center}
\renewcommand{\arraystretch}{1.2}
\begin{tabularx}{\linewidth}{@{} l X l @{}}
\toprule
\textbf{Function} & \textbf{Description} & \textbf{Ref.} \\
\midrule
\multicolumn{3}{@{}l}{\textit{Tropical subtractions}} \\
\texttt{STPreAnalysis}
  & Identify divergent rays and faces of the Newton polytope
  & App.~\ref{app:STPreAnalysis} \\
\texttt{STTropicalContinuation}
  & Perform Nilsson--Passare analytical continuation along specified rays
  & App.~\ref{app:STTropicalContinuation} \\
\texttt{STExpandIntegral}
  & Expand an Euler integrand into locally finite counter-terms via tropical subtraction
  & App.~\ref{app:STExpandIntegral} \\
\texttt{STFactor}
  & Expose $(\cdots)^\varepsilon$ scaling behavior along rays
  & App.~\ref{app:STFactor} \\
  \texttt{STFasterFubini}
  & Find a linearly reducible integration order
  & App.~\ref{app:STFasterFubini} \\
\midrule
\multicolumn{3}{@{}l}{\textit{Integration engine (\texttt{HyperIntica})}} \\[2pt]
\texttt{HyperIntica}
  & Integrate a hyperlogarithm-valued integrand in specified variables
  & App.~\ref{app:hyperintica-fns} \\
\texttt{HyperD}
  & Differentiate a hyperlogarithm expression with respect to a parameter
  & App.~\ref{app:hyperintica-fns} \\
\texttt{HyperSeries}
  & Expand a hyperlogarithm expression as a series around a point
  & App.~\ref{app:hyperintica-fns} \\
\texttt{FibrationBasis}
  & Compute the fibration basis of a hyperlogarithm
  & App.~\ref{app:hyperintica-fns} \\
\texttt{ConvertToSymbol}
  & Convert a hyperlogarithm expression to symbol notation
  & App.~\ref{app:hyperintica-fns} \\
\midrule
\multicolumn{3}{@{}l}{\textit{Numerical evaluation and checks}} \\
\texttt{STNIntegrate}
  & Evaluate an Euler integral numerically via \texttt{pySecDec}, \texttt{FIESTA}, \texttt{AMFlow} or \texttt{feyntrop}
  & Sec.~\ref{sec:numerical} \\
\texttt{STVerify}
  & Cross-check a symbolic result against numerical sector decomposition
  & Sec.~\ref{sec:numerical} \\
\texttt{STToGinsh}
  & Evaluate hyperlogarithms numerically via \texttt{ginsh}
  & App.~\ref{app:hyperintica-fns} \\
\bottomrule
\end{tabularx}
\end{center}
\caption{Main lower-level functions in \texttt{SubTropica} and \texttt{HyperIntica}.}\label{tab:list}
\end{table}

\subsection{\label{sec:functions}Architecture and function overview}

Putting aside the GUI, internally, {\texttt{SubTropica}} consists of three distinct modules. The first implements the tropical subtraction algorithm discussed in Sec.~\ref{sec:tropical}, whose main functions are described further in App.~\ref{app:tropical-utilities}. The second, called \texttt{HyperIntica}, implements the integration of hyperlogarithms \cite{Panzer:2014caa}, reviewed in Sec.~\ref{sec:integration}. A third module provides a Feynman graph interface and combines the previous modules through the single top-level function \texttt{STIntegrate}, which automates $\mathrm{GL}(1)$ gauge fixing for projective integrals, tropical continuation and subtraction, linear reducibility analysis, and (parallel) integration via \texttt{HyperIntica}. 

While the GUI presented above is tailored to Feynman integrals (\hyperref[app:form0-item]{Form~0}), \texttt{STIntegrate} also accepts scripted input in several formats: a Feynman graph specification (\hyperref[app:form1-item]{Form~1}), a propagator list with support for masses, numerators, and linearized propagators (\hyperref[app:form2-item]{Form~2}), or a direct Euler integrand for integrals not tied to a Feynman diagram (\hyperref[app:form3-item]{Form~3}). See the summarizing paragraph below for the full list of calling conventions. 

This variety of allowed inputs for \texttt{STIntegrate} makes the full tropical subtraction and integration pipeline available to generic Euler integrals, as illustrated in the examples of Sec.~\ref{sec:examples}. The main lower-level functions listed in Tab.~\ref{tab:list} can also be called individually for finer control over each step. The complete documentation of \texttt{STIntegrate} (input formats, option reference, checkpointing, and face targeting) is given in App.~\ref{app:STIntegrate-opts}. See Fig.~\ref{fig:flow} for a graphical summary of its internal workflow.

\subsubsection*{\label{app:input-formats}Summary of argument conventions for \texttt{STIntegrate}}

\begin{enumerate}[wide, labelindent=0pt]
\item[\textbf{Form 0.}] \phantomsection\label{app:form0-item}\texttt{STIntegrate[]}. Calling \texttt{STIntegrate[]} with no arguments opens the interactive Feynman diagram interface above, where the user can draw a diagram, assign masses, configure options, and launch the integration from the GUI interactively. In the GUI text fields, the suggested syntax is to use underscore notation for subscripts (e.g., \texttt{m\_1} for $m_1$, \texttt{p\_1} for $p_1$, \texttt{l\_1} for $\ell_1$). Then, masses are automatically squared/homogenized in the output ($m_1 \to \mathtt{mm1}$, $M_1 \to \mathtt{MM1}$). Strings such as \texttt{m1} are treated as literal symbols and will not be automatically homogenized, with the exception of \texttt{m} and \texttt{M} which are recognized as mass parameters. \hfill \decosix

\item[\textbf{Form  1.}] \phantomsection\label{app:form1-item} \texttt{STIntegrate[\{edges, nodes\}, opts]}. The input format follows that of {\SOFIA} \cite{Correia:2025wtb}. A diagram is specified by two lists, \texttt{edges} and \texttt{nodes}. Each entry of \texttt{edges} is a tuple \texttt{\{\{i,~j\},~m\}}, meaning that vertices \texttt{i} and \texttt{j} are connected by a propagator with mass \texttt{m} (order irrelevant). Each entry of \texttt{nodes} is a tuple \texttt{\{i,~M\}}, meaning an external momentum $p_i$ with $p_i^2 = \mathtt{M}^2$ is attached to vertex \texttt{i}. Here, the order \emph{does} matter, as momenta are labeled $p_1, p_2, \ldots$ by their position in the list.

As an example, a two-mass box corresponds to
\begin{lstlisting}[extendedchars=true,mathescape=true,language=Mathematica]
boxEdges = {{{1, 2}, $m_1$}, {{2, 3}, 0}, {{3, 4}, 0}, {{1, 4}, 0}};
boxNodes = {{1, $m_1$}, {2, 0}, {3, 0}, {4, $M_4$}};
STIntegrate[{boxEdges, boxNodes}]
(* Out: $\textcolor{gray}{\frac{\mathtt{Log}[\mathtt{MM4}/\mathtt{s12}] \mathtt{-} \mathtt{Log}[\mathtt{1} \mathtt{-} \mathtt{s23}/\mathtt{mm1}]}{\mathtt{eps}(\mathtt{mm1}(\mathtt{MM4} \mathtt{-} \mathtt{s12}) \mathtt{+} \mathtt{s12}\;\mathtt{s23})}}$ + ... *)
\end{lstlisting}
We observe that the code automatically assigns independent Mandelstam invariants of the form \texttt{sij\ldots}$\,=(p_i+p_j+\cdots)^2$ in a cyclic basis. For example, for six external particles the invariants are \texttt{s12}, \texttt{s23}, \texttt{s34}, \texttt{s45}, \texttt{s56}, \texttt{s16}, \texttt{s123}, \texttt{s234}, and \texttt{s345} (in $\D$ dimensions). We also see that {\texttt{SubTropica}} automatically squares external (and internal) masses $\mathtt{M}_i^2$ ($\mathtt{m}_j^2$) into the symbols $\mathtt{MMi}$ ($\mathtt{mmi}$) and uses $\varepsilon = \mathtt{eps}$ as the default regulator symbol.

Note that the diagram can also be drawn interactively via the {\SOFIA} interface
\begin{lstlisting}[extendedchars=true,language=Mathematica]
{edges, nodes} = FeynmanDraw
\end{lstlisting}
which returns \texttt{\{edges, nodes\}} with generic masses; non-generic mass configurations can then be imposed via usual substitution rules ``\texttt{//.\{...\}}'' on \texttt{\{edges, nodes\}}. The diagram can be visualized with \texttt{FeynmanPlot[\{edges, nodes\}]}. The option \texttt{"LabelInternalEdges" -> True} additionally numbers the internal edges in the order they were drawn. \hfill \decosix

\item[\textbf{Form 2.}]\phantomsection\label{app:form2-item} \texttt{STIntegrate[\{propagators\}, opts]}. The function also accepts a list of propagators and numerators
\begin{lstlisting}[extendedchars=true,mathescape=true,language=Mathematica]
STIntegrate[{propagator1, propagator2, ..., numerator1, ...},
            "Exponents" -> {1, 1, ..., -1, ...}]
\end{lstlisting}
This representation conveniently allows the inclusion of a user-defined numerator depending on the loop momenta. Numerators and (integer powers of) inverse propagators are differentiated by the code using the option \texttt{"Exponents"}, which will be discussed more below.
Similarly, vector numerators can be passed using (arbitrary) reference vectors such as $q^{\mu}$ and $q^{\nu}$. An example illustrating the syntax on the ``squared tadpole'' integral 
\begin{equation}
q^{\mu}q^{\nu}\int \frac{\d^{\D}\ell_1}{i\pi^{\D/2}}\frac{\ell_1^\mu\ell_1^\nu}{[\ell_1^2-m^2]^2} \qquad \text{(with $\D=4-2\varepsilon$)}\,,
\end{equation}
would be 
\begin{lstlisting}[extendedchars=true,mathescape=true,language=Mathematica]
STIntegrate[{$\ell[1]^2$ - mm, $q[1]\cdot\ell[1]$, $q[2]\cdot\ell[1]$}, "Exponents" -> {2, -1, -1}]
(*Out (assuming mm>0): $\textcolor{gray}{\frac{\mathtt{mm}\;\mathtt{q[1]\cdot q[2]}}{\mathtt{2}} \left(\frac{\mathtt{1}}{\mathtt{eps}} \mathtt{-} (\mathtt{Log}[\mathtt{mm}]\mathtt{-}\mathtt{1})\right) \mathtt{+}\; \mathtt{O[eps]}}$ *)
\end{lstlisting}
where the exponent $2$ indicates a squared propagator and exponents $-1$ indicate linear numerators. Note that when tensor numerators are present, Lorentz dot products (see also Sec.~\ref{ex:anomalous}) must be entered using \textsc{Mathematica}'s
\begin{lstlisting}[extendedchars=true,mathescape=true,language=Mathematica]
a \[CenterDot] b      (* or *)       CenterDot[a,b]
\end{lstlisting}
operator or, equivalently, using the dot `` \texttt{a.b} '' symbol on the keyboard. Note that the regular multiplication `` \texttt{a*b} '' is allowed but not recommended since it might lead to contraction ambiguities. \hfill \decosix

\item[\textbf{Form 3.}] \phantomsection\label{app:form3-item}\texttt{STIntegrate[integrand, \{x,0,1\}, \{y,0,$\infty$\}, \ldots, opts]}. In this case \texttt{STIntegrate} takes an integrand in \textsc{Mathematica}-style integration limits. Note that bare symbol inputs (such as \texttt{STIntegrate[integrand, x, y, \ldots]}) default to $[0,\infty)$. 

For generic Euler integrals that are, e.g., not tied to a Feynman diagram, \texttt{STIntegrate} accepts a syntax modeled on \textsc{Mathematica}'s built-in \texttt{Integrate}:
\begin{lstlisting}[extendedchars=true,mathescape=true,language=Mathematica]
STIntegrate[integrand, {x, 0, 1}, {y, 0, $\infty$}, opts]
\end{lstlisting}
Each \texttt{\{var, lo, hi\}} triple specifies an integration variable and its bounds. The allowed bounds are $0$, $1$, and $\infty$. When limits are omitted, bare symbols default to $[0,\infty)$:
\begin{lstlisting}[extendedchars=true,mathescape=true,language=Mathematica]
STIntegrate[integrand, x, y, opts] (* same as {x,0,$\textcolor{gray}{\infty}$}, {y,0,$\textcolor{gray}{\infty}$} *)
\end{lstlisting}
Before moving on, let us note that \texttt{STIntegrate} also accepts a pre-built tuple generated by internal \texttt{SubTropica} functions
\begin{lstlisting}[extendedchars=true,mathescape=true,language=Mathematica]
STIntegrate[{prefactor, integrand, variables, coefficients}]
\end{lstlisting}
where \texttt{prefactor} is a multiplicative factor kept outside the integral, \texttt{variables} lists the variables to integrate from $[0,\infty)$, and \texttt{coefficients} collects all symbolic parameters treated as constants during integration.
 \hfill \decosix

In the next section, we provide the necessary theoretical background to understand the tropical subtraction algorithm and the hyperlogarithm integration procedure that underlie \texttt{STIntegrate}. 

\end{enumerate}

\subsection{Numerical evaluation and verification}\label{sec:numerical}

\texttt{SubTropica} also allows one to numerically evaluate integrals (linearly reducible or not) and verify an analytic output.

In particular, \texttt{STNIntegrate} numerically evaluates integrals by interfacing with \texttt{pySecDec}~\cite{Heinrich:2023til}, \texttt{FIESTA}~\cite{Smirnov:2021rhf}, \texttt{AMFlow}~\cite{Liu:2022chg}, or \texttt{feyntrop}~\cite{feyntrop} as backends. It accepts all the same input formats as \texttt{STIntegrate} (\hyperref[app:form1-item]{Forms~1}--\hyperref[app:form3-item]{3}), including generic parametric representations. This is particularly useful for integrals that are not linearly reducible, where a symbolic evaluation via \texttt{STIntegrate} is not available. For instance, the two-loop generic masses (elliptic) kite
\begin{lstlisting}[extendedchars=true,mathescape=true,language=Mathematica]
edges = {{{1, 2}, $m_1$}, {{1, 3}, $m_2$}, {{1, 4}, $m_3$},
         {{2, 3}, $m_4$}, {{2, 4}, $m_5$}};
nodes = {{3, M}, {4, M}};
STNIntegrate[{edges, nodes},
  "Substitutions" -> {m1 -> 67/23, m2 -> 59/31, m3 -> 159/31, 
                      m4 -> 59/131, m5 -> 117/137, M -> 31 I/5}]
\end{lstlisting}
returns $-0.0708835 + \mathcal{O}(\varepsilon)$ in about 1.2 seconds using \texttt{pySecDec}.

The function \texttt{STVerify} cross-checks a symbolic result from \texttt{STIntegrate} against a numerical evaluation. If it exists, it automatically generates a Euclidean kinematic point, runs \texttt{pySecDec} (or \texttt{FIESTA}/\texttt{AMFlow}/\texttt{feyntrop}), evaluates the symbolic expression numerically via \texttt{ginsh}, and compares the Laurent coefficients order by order. For example,
\begin{lstlisting}[extendedchars=true,mathescape=true,language=Mathematica]
diag = {{{{1,2},0},{{1,2},0},{{2,3},0},{{3,4},0},
         {{3,4},0},{{1,4},0}}, {{1,0},{2,0},{3,0},{4,0}}};
result = STIntegrate[diag];
STVerify[diag, result]
\end{lstlisting}
produces a coefficient-by-coefficient comparison and returns \texttt{"pass" -> True} with a maximum relative error of $\sim 10^{-5}$.

As mentioned above, \texttt{STNIntegrate} also accepts parametric integrands, using the same calling conventions as \texttt{STIntegrate} (\hyperref[app:form3-item]{Form~3}). For the diagram \texttt{diag} in the last example, the Schwinger-parameter representation~\eqref{eq:SPx} (with $x_4 = 1$) can be evaluated as follows:
\begin{lstlisting}[extendedchars=true,mathescape=true,language=Mathematica]
(* Symanzik polynomials of the last "diag" above*)
U = x1 x2 + x1 x3 + x2 x3 + x1 x5 + x2 x5 + x1 x2 x5
    + x1 x3 x5 + x2 x3 x5 + x1 x6 + x2 x6 + x1 x5 x6 + x2 x5 x6;
F = s23 x1 x2 x5 + s12 (x1 x3 x6 + x2 x3 x6 + x1 x3 x5 x6 + x2 x3 x5 x6);
pref = Exp[3 eps EulerGamma] Gamma[3 eps];
integrand = U$^{2(2\;\mathtt{eps} - 1)}$ (-F)$^{-3\;\mathtt{eps}}$;
subs = "Substitutions" -> {s12 -> -7/31, s23 -> -43/89};

(* [0,$\textcolor{gray}{\infty}$): bare symbols *)
STNIntegrate[pref integrand, x1, x2, x3, x5, x6, subs]
(* Out: $\textcolor{gray}{\frac{\mathtt{0.3333}}{\mathtt{eps}^{\mathtt{3}}} \mathtt{+} \frac{\mathtt{4.155}}{\mathtt{eps}^{\mathtt{2}}} \mathtt{+} \frac{\mathtt{32.36}}{\mathtt{eps}} \mathtt{+} \mathtt{183.7} \mathtt{+} \mathtt{O[eps]}}$ *)

(* [0,1]: explicit bounds (for illustration) *)
STNIntegrate[pref integrand, {x1,0,1}, {x2,0,1}, {x3,0,1},
                             {x5,0,1}, {x6,0,1}, subs]
(* Out: $\textcolor{gray}{\frac{\mathtt{0.2195}}{\mathtt{eps}^{\mathtt{2}}} \mathtt{+} \frac{\mathtt{2.987}}{\mathtt{eps}} \mathtt{+} \mathtt{22.59} \mathtt{+} \mathtt{O[eps]}}$ *)

(* Pre-built tuple: {prefactor, integrand, vars, coeffs} *)
STNIntegrate[{pref, integrand, {x1,x2,x3,x5,x6}, {s12,s23}}, subs]
(* same output as bare-symbol call *)
\end{lstlisting}

\subsection{{\texttt{SubTropica}} library}

\texttt{SubTropica} comes with a comprehensive open-source library of diagrams, including references to the most relevant papers that featured them. The library also stores results of computations obtained with \texttt{SubTropica}. Its online version can be accessed at
\begin{center}
\href{https://subtropi.ca}{\texttt{subtropi.ca}}
\end{center}
It is continuously updated with new diagrams and results. We welcome new submissions to the library which can be sent to us simply by clicking the ``Submit to SubTropica library'' button after a successful integration, or alternatively by executing
\begin{lstlisting}[extendedchars=true,mathescape=true,language=Mathematica]
STSubmitResult[]
\end{lstlisting}
after a successful computation. 
\section{Theory and algorithms}
\label{sec:Preliminary}

We begin by defining Euler integrals and their $\varepsilon$-expansion, then introduce the Newton polytope machinery that governs the singularity structure. The Nilsson--Passare continuation and tropical subtraction algorithms convert a divergent Euler integral into a sum of locally finite terms, which are then integrated one variable at a time using the hyperlogarithm calculus implemented in \texttt{HyperIntica} (and \texttt{HyperInt} \cite{Panzer_2014}). We close this section with a discussion of linear reducibility, the key condition that ensures that this sequential integration terminates in closed form.

\subsection{An invitation to Euler integrals}
Mathematically, an Euler integral can be defined as the Mellin transform of a rational function,
\begin{align}
    I({\bf s},c) = \int_{\mathbb{R}^n_{\ge0}}\frac{\d\alpha_1}{\alpha_1}\dots\frac{\d\alpha_n}{\alpha_n} \times \left(\prod_{j=1}^m P_j(\alpha,{\bf s})^{c_j}
    \right)\,,
    \label{eq:defeuler}
\end{align}
where $c_j\in\mathbb{C}$ and $P_j$ are all polynomials in the integration variables $\alpha=(\alpha_1,\dots,\alpha_n)$ with coefficients collectively denoted ${\bf s}$.

An Euler integral is a meromorphic function of the exponent variables $c$~\cite{Salvatori:2024nva}. On the other hand, the dependence on the coefficient variables ${\bf s}$ is much more complicated, involving sophisticated generalizations of the logarithm, which we discuss later in this section (see Sec.~\ref{sec:integration}).

To streamline the discussion in this section, we will restrict ourselves to the case where the exponent variables depend on a common real parameter $\varepsilon$ as\footnote{In the examples section below (see Sec.~\ref{sec:smallx}), we will encounter integrals for which more than one such parameters appear, and show how to deal with them using \texttt{SubTropica}.}
\begin{align}
    c_j=c_j(\varepsilon) = a_j + b_j \varepsilon\, ,
    \label{eq:exponents}
\end{align}
where $a_j, b_j$ are fixed complex numbers.
In quantum field theory, the parameter $\varepsilon$ is often called the \emph{dimensional regulator}. While the subtraction still applies in the case of general $a_j$, the integration discussed below needs integer $a_j$.

With this parametrization, the Euler integral becomes a function $I({\bf s},\varepsilon)$ of finitely many coefficient variables ${\bf s}$ and a single real parameter $\varepsilon$. It turns out that $I({\bf s},\varepsilon)$ extends to a meromorphic function of $\varepsilon$. We have an expansion around $\varepsilon=0$,
\begin{align}
    I({\bf s},\varepsilon)= \sum_{i=-N}^\infty \varepsilon^i I^{(i)}({\bf s})\,.
    \label{eq:epsexpansion}
\end{align}

In this package, we provide an algorithmic and practical way to find an expansion for the coefficient functions $I^{(i)}({\bf s})$ in terms of convergent integrals that can be evaluated with the algorithm implemented in the \textsc{Maple} program \texttt{HyperInt} \cite{Panzer_2014} and the new \textsc{Mathematica} package \texttt{HyperIntica} shipped with \texttt{SubTropica}.

The strategy we follow is based on the \emph{tropical subtraction formula} presented in \cite[Sec.~6]{Salvatori:2024nva} reviewed below (see also \cite{Hillman:2023ezp, Brown:2019wna}). The idea is to add and subtract ``counter-terms'' from the original integrand, according to a combinatorial pattern that reflects the boundary structure of a polytope attached to the integral. Once the dust settles, we are left with an expansion of the Euler integral, 
\begin{align}
    I({\bf s}, \varepsilon) = \sum_i \phi_i(\varepsilon) I_i({\bf s},\varepsilon)\,,
    \label{eq:expansion}
\end{align}
as a combination of \emph{locally finite} integrals $I_i({\bf s},\varepsilon)$. 
By definition, these are integrals whose $\varepsilon$-expansion can be performed directly at the integrand level, providing the desired integral representation for the coefficients $I^{(i)}({\bf s})$.

The tropical subtraction formula is valid only for integrals that satisfy a certain \emph{geometric property}. Luckily, any integral can be written as a combination of such, using the \emph{Nilsson--Passare} analytical continuation \cite{nilsson2013mellin}, or tropical continuation for brevity. In fact, tropical continuation can be used to expand an integral directly in locally finite ones. Therefore  it provides an alternative to tropical subtractions \cite{Panzer_2015}. However this usually results in large complicated expressions, which limits the usefulness of this approach.
Nevertheless, even when expanding a divergent integral in locally finite ones is prohibitive, limiting the tropical continuation to expand in integrals that merely satisfy the geometric property, rather than being locally finite, is often manageable.  

In summary, chaining tropical continuation and tropical subtraction gives an efficient algorithm to expand an arbitrary Euler integral in terms of locally finite ones. This is the core functionality of \verb|SubTropica|.

\paragraph{A comment on positivity.}\label{subsec:positivity}
The treatment of Euler integrals simplifies drastically if each polynomial appearing in the integrand is sign-coherent, that is if its coefficients all have the same sign.\footnote{This requirement can, in fact, be relaxed to allow polynomials that are \emph{completely non-vanishing} \cite{Borinsky:2020rqs}.} In this case the Euler integrand does not vanish in the interior of the domain of integration. 
A crucial consequence is that the singularities of the Euler integral (poles in $\varepsilon$) arise only from the \emph{boundary} of the integration domain. 
This is where tropical geometry enters the game: the \emph{Newton polytope} of the integrand provides the only systematic way to detect these singular corners. 

In the rest of this section, we will discuss two methods to expand an Euler integral in terms of locally finite ones. Both algorithms are rigorously valid only in the case of sign-coherent polynomials. If this property is not satisfied, the Euler integral may have additional poles in $\varepsilon$ that are \emph{not} captured by these methods. This is a famous issue that plagues all direct-integration techniques and their implementation in public software such as \verb|asy| \cite{Jantzen:2012mw}, \verb|FIESTA| \cite{Smirnov:2021rhf} or \verb|pySecDec| \cite{Heinrich:2023til}. 

Currently, the only safe way to deal with this problem (and detect otherwise ``hidden'' singularities) is to perform a careful Landau analysis: one has to enumerate the irreducible components of the Landau variety which have positive solutions. Although this can be done algorithmically with publicly available software, such as \verb|Macaulay2| \cite{M2} and \verb|Bertini| \cite{BHSW06}, it is computationally much more expensive than the tropical analysis discussed here. Therefore, this is to this day no generic practical solution (see, e.g., \cite{Hannesdottir:2022bmo,Gardi:2024axt} for related discussions). 

\begin{figure}
\centering
\begin{tikzpicture}
  \coordinate (v1) at (-2.250, 1.000);
  \coordinate (v2) at (-2.250, -2.000);
  \coordinate (v3) at (0.750, 1.000);
  \coordinate (v4) at (0.750, -2.000);
  \coordinate (v5) at (-0.750, 0.000);
  \coordinate (v6) at (-0.750, -1.000);
  \coordinate (v7) at (-3.000, 1.000);
  \coordinate (v8) at (-3.000, -2.000);
  \coordinate (v9) at (1.500, -2.000);
  \coordinate (v10) at (1.500, 1.000);

  \draw[thick, dashed, line cap=round] (v5) -- (v1) ;
  \draw[thick, dashed, line cap=round] (v6) -- (v2);
  \draw[thick, dashed, line cap=round] (v2) -- (v5) ;
  \draw[thick, dashed, line cap=round] (v1) -- (v6);
  \draw[thick, dashed, line cap=round] (v3) -- (v6) ;
  \draw[thick, dashed, line cap=round] (v3) -- (v5);
  \draw[thick, dashed, line cap=round] (v4) -- (v6) ;
  \draw[thick, dashed, line cap=round] (v4) -- (v5) ;
  \draw[thick, dashed, line cap=round] (v7) -- (v1) ;
  \draw[thick, dashed, line cap=round] (v8) -- (v2);
  \draw[thick, dashed, line cap=round] (v9) -- (v4);
  \draw[thick, dashed, line cap=round] (v10) -- (v3);

  \fill (v1) circle (2pt);
  \fill (v2) circle (2pt);
  \fill (v3) circle (2pt);
  \fill (v4) circle (2pt);
  \fill (v5) circle (2pt);
  \fill (v6) circle (2pt);
\end{tikzpicture}
\caption{\label{fig:crown}The crown diagram features a non-sign-definite Symanzik polynomial.}
\end{figure}
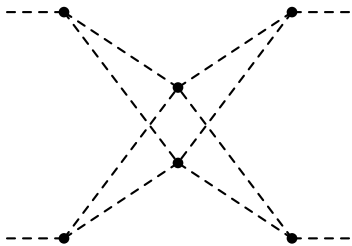

In physical applications, this issue arises only in certain advanced calculations involving highly non-planar massless diagrams, such as the crown diagram in Fig.~\ref{fig:crown}, which is connected to the problem of ``Glauber modes'' (singularities that do not fall into the standard soft-collinear classification) \cite{Gardi:2024axt}. The current version of \texttt{SubTropica} does not handle these divergences and a systematic treatment is left to future work.

\subsection{Tropical algorithms}
\label{sec:tropical}

In this section, we describe how to use rudiments from tropical geometry to detect and calculate divergences of Euler integrals.

\subsubsection{Divergences}

The first step is to understand how divergences originate. 

\paragraph{A warming-up example.}

Let us consider a simple toy example,
\begin{align}
    A &= \int_0^\infty\int_0^\infty  \d\alpha_1\ \d\alpha_2\ \alpha_1^{1+\varepsilon} \alpha_2^{\varepsilon} \left(\alpha_1^2+\alpha_2+\alpha_1\alpha_2\right)^{-(2+\varepsilon)}\,.
    \label{eq:warmup}
\end{align}
The singular behavior of the integrand around the origin suggests that the integral may develop a divergence there. 
In order to expose it, we perform the rescaling 
\begin{align}
    \alpha_2 \to \alpha_1^2 \alpha_2\,.
    \label{eq:exrescaling}
\end{align}
The integral transforms to
\begin{align}
    A &= \int_0^\infty\int_0^\infty  \d\alpha_1\ \d\alpha_2\ \alpha_1^{-1+\varepsilon} \alpha_2^{\varepsilon} \left(1+\alpha_2+\alpha_1 \alpha_2\right)^{-(2+\varepsilon)}\,.
    \label{eq:factorizationsing}
\end{align}
Note how the singular behavior of the integrand is completely captured by the \emph{monomial} $\alpha_1^{-1+\varepsilon}$: the polynomials resulting from the rescaling are non-vanishing at the origin. 
This allows to neatly extract a pole in $\varepsilon$. 
Integrating around $\alpha_1<1$ (which corresponds to a small neighborhood of the origin in the original coordinates) gives
\begin{equation}
\begin{split}
    &\int_0^1\int_0^\infty  \d\alpha_1\ \d\alpha_2\ \alpha_1^{-1+\varepsilon} \alpha_2^{\varepsilon} \left(1+\alpha_2+\alpha_1 \alpha_2\right)^{-(2+\varepsilon)} = \\ 
    &= \int_0^1 \int_0^\infty \d\alpha_1 \d\alpha_2 \left(\alpha_1^{-1+\varepsilon} \alpha_2^\varepsilon (1+\alpha_2)^{-(2+\varepsilon)}\right)\left(1+\mathcal{O}(\alpha_1)\right) \\ 
    &=\frac{1}{\varepsilon} \times \int_0^\infty \d\alpha_2 \alpha_2^\varepsilon (1+\alpha_2)^{-(2+\varepsilon)} + \dots 
\end{split}
\end{equation}

The crucial step was to consider the rescaling \eqref{eq:exrescaling}, whose origin is subtler than it seems. For example, consider the similar rescaling $\alpha_2\to\alpha_1\alpha_2$. As $\alpha_1\to 0$, it does the same job as \eqref{eq:exrescaling} of ``blowing-up'' the origin, but now we have
\begin{align}
        A\to\int_0^\infty \int_0^\infty \d\alpha_1 \d\alpha_2 \ \alpha_1^{\varepsilon}\,\alpha_2^{\varepsilon}\,\left(\alpha_1+\alpha_2+\alpha_1\alpha_2\right)^{-(2+\varepsilon)}\,,
\end{align}
so that integrating around $\alpha_1 \sim 0$ does not produce a pole in $\varepsilon$.

\paragraph{Tropicalization.}
Let us now formulate the problem in more general terms. Consider an Euler integral 
\begin{align}
    \int_{\mathbb{R}^n_{\ge0}} \frac{\d\alpha_1}{\alpha_1}\dots\frac{\d\alpha_n}{\alpha_n}\ \mathcal{I}(\alpha_1,\dots,\alpha_n)\,,
\end{align}
and an arbitrary rescaling
\begin{align}
    \alpha_i \to \alpha_i \lambda^{-\rho_i}, \quad \rho=(\rho_1,\dots,\rho_n) \in \mathbb{Q}^n\,,
    \label{eq:scaling}
\end{align}

Note that we can always eliminate one of the original variables $\alpha_i$ in favor of $\lambda$, and thus regard \eqref{eq:scaling} as a change of coordinates. 
With this understanding the limit $\lambda\to0^{+}$ probes a corner of the boundary of the integration space, where the integration variables scale as \eqref{eq:scaling}.
Clearing denominators and factoring polynomials, we can pull out an overall monomial in $\lambda$,
\begin{align}
    \mathcal{I}(\alpha_i \lambda^{-\rho_i}) = \lambda^{-\mathrm{Trop\ \mathcal{I}(\rho)}} \mathcal{J}(\alpha,\lambda)\,.
    \label{eq:transformation}
\end{align}
We use \eqref{eq:transformation} to define the \emph{tropical integrand}, 
$\mathrm{Trop\ }\mathcal{I}$, a function that associates to the scaling vector $\rho$ minus the exponent of this overall monomial. 
Keeping in mind the overall logarithmic differential form, which is invariant under the transformation, we see that
\begin{align}
    \text{Re}\, \mathrm{Trop\ }\mathcal{I}(\rho)|_{\varepsilon=0} < 0 \quad\text{for all } \rho \in \mathbb{Q}^n \qquad\Rightarrow\qquad A = \mathcal{O}(\varepsilon)\,.
    \label{eq:negtrop}
\end{align}
Due to \eqref{eq:exponents}, the tropical integrand is always at most linear in $\varepsilon$,
\begin{align}
    \mathrm{Trop\ }\mathcal{I}(\rho)=a + b\ \varepsilon\,,
\end{align}
depending on the value of $a$ we say that $\rho$ is either convergent ($\mathrm{Re}\,a<0$), logarithmically divergent ($\mathrm{Re}\, a=0$) or power divergent ($\mathrm{Re}\, a>0$).

As it stands, the criterion \eqref{eq:negtrop} is not particularly helpful, because it requires checking the sign of the tropical integrand over infinitely many vectors $\rho$.
Luckily, it turns out that one only needs to evaluate the tropical integrand on a finite set of ``critical'' vectors \cite{Arkani-Hamed:2022cqe}.
These are the normals to the facets of a polytope canonically attached to the integrand $\mathcal{I}$.

\paragraph{Newton Polytopes.}
The Newton polytope of a polynomial is the convex hull of the exponents of its monomials\cite{ziegler1995lectures}. In other words, if
\begin{align}
    P=\sum_{{\bf m}\in \mathbb{Z}^n}s_{\bf m} \prod_{i=1}^n \alpha_i^{m_i}\,,
    \label{eq:poly}
\end{align}
then the Newton polytope $\mathrm{Newt\ }P$ is the convex hull
\begin{align}
    \mathrm{Newt\ }P\equiv \mathrm{Conv}\{{\bf m}\in \mathbb{Z}^n \mid s_{\bf m} \ne 0\}\,.
    \label{eq:convhull}
\end{align}
For an Euler integrand,
\begin{align}
    \mathcal{I}(\alpha)=\prod_{j} P_j(\alpha)^{a_j + b_j\varepsilon}\,,
\end{align} 
we define its Newton polytope to be $\mathrm{Newt\ }\mathcal{I}\equiv \mathrm{Newt\ }\prod_j P_j$.

The convex hull description \eqref{eq:convhull} is known as the \emph{vertex presentation} of the polytope. Every polytope admits a dual description, or \emph{facet presentation}, as an intersection of half-spaces, 
\begin{align}
    \mathrm{Newt\ } P = \bigcap_j\; \{\alpha=(\alpha_1,\dots,\alpha_n) \;|\; d_j - \rho_j \cdot \alpha\ge0\}.
    \label{eq:hpresentation}
\end{align}
for some constants $d_j \in \mathbb{Q}$ and vectors $\rho_j \in \mathbb{Q}^n$.

Suppose that the locus where  one of inequalities of \eqref{eq:hpresentation} is saturated is a co-dimension one boundary of  $\mathrm{Newt\ } P $, also called a \emph{facet}.
Then the corresponding vector $\rho_j$ can be interpreted geometrically as the outer normal to the facet. 

The collection of all the normals to the facets is a finite set of ``critical'' vectors that control the singularities of the integral. 
If these vectors are convergent (using the terminology introduced below \eqref{eq:negtrop}),  then the Euler integral does not have poles in $\varepsilon$. 

In the case of Newton polytopes, the vertex description \eqref{eq:convhull} is immediately available from the knowledge of the polynomial \eqref{eq:poly}.
The facet description, however, requires some work. The problem involves only linear algebra, but it can be computationally very demanding, depending on the underlying combinatorics. Fortunately, efficient and sophisticated solutions to this problem are implemented in various software. In \verb|SubTropica| we outsource this computational task to \verb|polymake| \cite{polymake:2000,polymake:2017}.

\paragraph{Example.}
Let us illustrate the notions introduced so far by revisiting the example \eqref{eq:warmup}.
The Euler integrand is 
\begin{align}
    \mathcal{I}=\alpha_1^{2+\varepsilon} \alpha_2^{1+\varepsilon} \left(\alpha_1^2+\alpha_2+\alpha_1\alpha_2\right)^{-(2+\varepsilon)}\,,
\end{align}
it tropicalizes to
\begin{align}
    \mathrm{Trop\ }\mathcal{I} = (2+\varepsilon)\alpha_1+(1+\varepsilon)\alpha_2-(2+\varepsilon)\max(2\alpha_1,\alpha_2,\alpha_1+\alpha_2)\,,
    \label{eq:tropexample}
\end{align}
Note the slight abuse of notation, where we keep using the same variables, $\alpha_i$, as arguments of both $\mathcal{I}$ and $\mathrm{Trop\ } \mathcal{I}$.

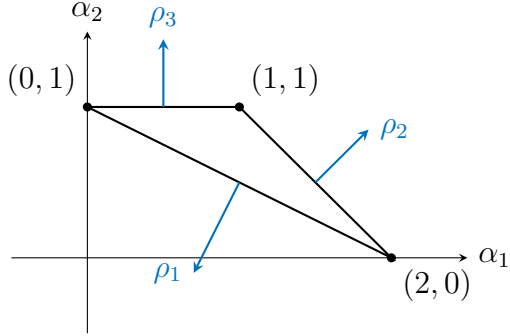
\begin{figure}
    \centering
    \begin{tikzpicture}[scale=2,>=stealth]

  \draw[->] (-0.5,0) -- (2.5,0) node[right] {$\alpha_1$};
  \draw[->] (0,-0.5) -- (0,1.5) node[above] {$\alpha_2$};

  \coordinate (A) at (2,0);
  \coordinate (B) at (1,1);
  \coordinate (C) at (0,1);

  \draw[thick] (A)--(B)--(C)--cycle;

  \fill (A) circle (0.03) node[below right] {$(2,0)$};
  \fill (B) circle (0.03) node[above right] {$(1,1)$};
  \fill (C) circle (0.03) node[above left] {$(0,1)$};

  \coordinate (Mab) at ($(A)!0.5!(B)$);
  \coordinate (Mbc) at ($(B)!0.5!(C)$);
  \coordinate (Mca) at ($(C)!0.5!(A)$);

  \draw[->,RoyalBlue,thick] (Mab) -- ++(0.35,0.35) node[right] {$\rho_2$};
  \draw[->,RoyalBlue,thick] (Mbc) -- ++(0,0.45) node[above] {$\rho_3$};
  \draw[->,RoyalBlue,thick] (Mca) -- ++(-0.3,-0.6) node[left] {$\rho_1$};

\end{tikzpicture}
    \caption{The Newton polytope describing the singularities of the Euler integral \eqref{eq:warmup}}
    \label{fig:trianglepolytope}
\end{figure}
Up to translation, the Newton polytope $\mathrm{Newt\ }\mathcal{I}$ is the triangle of Fig. \ref{fig:trianglepolytope}. At first glance, it is evident that its normals are $\rho_1=(-1,-2)$, $\rho_2=(1,1)$, and $\rho_3=(0,1)$. Plugging these into \eqref{eq:tropexample} it is easy to check that $\rho_1$ is logarithmically divergent, with $\mathrm{Trop\ }\mathcal{I}(\rho_1)=\varepsilon$, while both $\rho_2$ and $\rho_3$ are convergent.
The rescaling \eqref{eq:scaling} associated to $\rho_1$ is
\begin{align}
    \alpha_1\to\alpha_1 \lambda \quad \alpha_2 \to \alpha_2 \lambda^2\,,
\end{align}
which is of course equivalent to the previously mysterious \eqref{eq:exrescaling}.

We have now understood how Newton polytopes encode the singularities of an Euler integral. The poles in the dimensional regulator $\varepsilon$ arise from the corners of the integration space exposed by the rescaling \eqref{eq:scaling}. The exponent vector defining the rescaling is an outer-pointing normal to a facet of the polytope.

\subsubsection{Locally finite expansion}\label{sec:tropical-Cont}

We now turn to the expansion of a divergent Euler integral in terms of locally finite integrals, Eq. \eqref{eq:expansion}. 
As we will see shortly, the Newton polytope of the Euler integral once again plays a central role: it encodes in its boundary structure the data necessary to construct the expansion. 

\paragraph{Tropical continuation.}

Consider an Euler integral $I(\varepsilon)$ and suppose that $\rho$ is a divergent vector. 
Consider the shifted Euler integral
\begin{align}
    I(\lambda;\varepsilon) = \int_{\mathbb{R}^n_{\ge0}} \frac{\d\alpha_1 }{\alpha_1} \dots \frac{\d\alpha_n}{\alpha_n}\,\mathcal{I}(\alpha_i \lambda^{-\rho_i};\varepsilon)\,,
\end{align}
which reduces to the original integral $I(\varepsilon)$ for $\lambda=1$.
In fact, $\frac{\d}{\d\lambda}I(\lambda;\varepsilon)=0$, because the dependence on $\lambda$ can be eliminated at the integrand level by the rescaling $\alpha \to \alpha\lambda^{\rho}$ which leaves invariant both the measure and the integration domain.\footnote{In this context, $\lambda$ should be thought of as a parameter, not an integration variable.}
We have
\begin{align*}
    0 = \int_{\mathbb{R}^n_{\ge 0}}  
    \frac{\d\alpha_1}{\alpha_1}\dots\frac{\d\alpha_n}{\alpha_n}
    \left.\left(\frac{\d}{\d\lambda} \mathcal{I}(\alpha_i\lambda^{-\rho_i},\varepsilon)\right)\right|_{\lambda=1} = -I(\varepsilon)\mathrm{Trop\ \mathcal{I}}(\rho) + \sum_{i} I_{i}(\varepsilon)\,,
\end{align*}
or equivalently,
\begin{align}
   I(\varepsilon) =  \frac{1}{\mathrm{Trop\ \mathcal{I}}(\rho)} \left(\sum_{i} I_{i}(\varepsilon) \right)\,.
\end{align}
Here, $I_i(\varepsilon)$ are new Euler integrals that arise by the action of $\frac{\d}{\d\lambda}$  on $\mathcal{J}_\rho(\alpha,\lambda)$ in \eqref{eq:transformation}.
It turns out that the integrands $\mathcal{I}_i$ for the integrals $I_i$ satisfy $\mathrm{Trop\ }\mathcal{I}_i \le \mathrm{Trop\ } \mathcal{I}$, with the inequality being strict when both sides are evaluated on the ray $\rho$.
This can be seen by considering how the facet $\rho$ of the Newton polytope of $\mathcal{I}(\alpha \lambda^{-\rho};\varepsilon)$ is ``shaved'' by the action of $\frac{\d}{\d\lambda}$.

By iterating this procedure sufficiently many times, we can express $I(\varepsilon)$ as a linear combination of locally finite Euler integrals multiplied by explicit poles in $\varepsilon$. 
This is the tropical continuation due to Nilsson-Passare \cite{nilsson2013mellin,berkesch2014euler,Schultka:2018nrs}.
While it provides a systematic approach to compute the $\varepsilon$-expansion of an Euler integral, tropical continuation is computationally very expensive. This can be traced to two main bottlenecks:
\begin{enumerate}
    \item Iterating derivatives produces an exponential number of integrals.
    \item The pole part $I(\varepsilon)$ is presented in terms of integrals with the same dimension as $I(\varepsilon)$.
\end{enumerate}

We will next describe an alternative strategy which overcomes these limitations. 
It uses more elaborate information about the boundary structure of the Newton polytope: not only which are the facets responsible for divergences, but also how they intersect in lower dimensional faces of the polytope.

\paragraph{Geometric property.}

Let $\mathcal{P}$ be a polytope, and suppose that a subset of its facets, $\{\mathcal{F}_i, \ i=1,\dots,n\}$, share a common vertex. Then we say that these are \emph{compatible} facets. 

If divergent facets are compatible, they can conspire to produce higher order poles. 
In this sense, the Newton polytope leaves an imprint of itself in the pole structure of an Euler integral. 
To capture this pattern we construct the set of all collections of compatible, divergent (normals to the) facets,
\begin{align}
    \hspace{-0.3cm} \Sigma^{\rm div}= \{\sigma | \sigma=(\rho_1,\dots,\rho_m), \ \rho_i \text{ are divergent, compatible normals}\}\cup\{\varnothing\}.
    \label{eq:divergentfan}
\end{align}
Note that we conventionally include the empty set, to streamline formulae down the line.

Consider now an Euler integrand $\mathcal{I}$ with at most logarithmic divergences. 
We say that $\mathcal{I}$ satisfies the \emph{geometric property} if for every divergent vector $\rho$ we are able to find a vector $w_{\rho}$ such that $w_\rho \cdot \rho' = -\delta_{\rho,\rho'}$, for all divergents normals $\rho'$ that are compatible with $\rho$.

The geometric property has an important consequence. 
The set $\Sigma^{\rm div}$ is an abstract simplicial complex. Furthermore, if 
$\sigma = (\rho_1,\dots,\rho_m) \in \Sigma^{\rm div}$, then the normals $\rho_i$ are linearly independent and the corresponding facets meet in a face of codimension $m$, 

In general, Euler integrals do not satisfy the geometric property nor have only logarithmic singularities.  
However, we can always tropically continue along all vectors $\rho$ which are either power-divergent or do not admit an accompanying vector $w_\rho$. By iterating the procedure sufficiently many times, one is eventually left with logarithmically divergent integrals that satisfy the geometric property. 

Our interest in integrals that satisfy the geometric property is due to the existence of an algorithm that allows to expand them in terms of locally finite integrals, and which is substantially more efficient than tropical continuation. This will be discussed next.

\paragraph{Tropical subtraction.\label{par:tropical-subtraction}}

The main idea is to add and subtract suitable ``counter-terms'' $\mathcal{I}^{\rm ct}$ from a divergent Euler integrand. The counter-terms must be chosen so that they cancel the divergences of $\mathcal{I}$, while being simple enough that they can be integrated as many times as the order of the pole they cancel.
In other words, the goal is to write
\begin{align}
    \mathcal{I} = \left[\mathcal{I}-\mathcal{I}^{\rm ct}\right]+\mathcal{I}^{\rm ct}\,,
    \label{eq:subtractionscheme}
\end{align}
in such a way that the ``renormalized'' integrand
\begin{align}
    \mathcal{I}^{\rm ren}\equiv\mathcal{I}-\mathcal{I}^{\rm ct}\,,
\end{align}
is locally finite, and the integral
\begin{align}
    \int_{\mathbb{R}^n_{\ge0}} \frac{\d\alpha_1}{\alpha_1}\dots\frac{\d\alpha_n}{\alpha_n}
\,\mathcal{I}^{\rm ct}\,,
\end{align}
can be partially integrated, dropping one dimension for each pole in $\varepsilon$ exposed. 

If $\mathcal{I}$ is an integrand that satisfies the geometric property, and with at most logarithmic divergences, then we can follow \cite{Salvatori:2024nva} to construct such counter-terms. 
The combinatorics of the subtraction formula \eqref{eq:subtractionscheme} is controlled by the set $\Sigma^{\rm div}$ from \eqref{eq:divergentfan}.
It reflects the intersection pattern of the facets of $\mathrm{Newt\ }\mathcal{I}$ with divergent normal vectors $\rho$.

The counter-terms are the product of two factors. 

The first is the \emph{restriction} of the integrand on a face (also known as \emph{initial form} in the context of tropical and toric geometry \cite{WhatIsToric}).
It is the integrand obtained by keeping from each polynomial $P$ only the monomials that lie on a particular face of $\mathrm{Newt\ }P$.
The fact that the integrand and its restriction have similar Newton polytopes, and thus matching singularities, is what makes them natural candidates in an ansatz for the counter-terms.

Unfortunately, restrictions alone cannot be used as counter-terms. The reason is that they introduce singularities in addition to the ones that they cancel.
This can be easily seen already in the simplest case of one-dimensional integrals. Their Newton polytopes are necessarily always segments. Integrands with  a segment and its vertices for Newton polytopes cannot be combined in a locally finite combination.

 In order to address this issue, we introduce the second factor of the counter-term.
To any vector $\rho$, we associate the function
\begin{align}
    v_{\rho}=\frac{1}{1+\prod_i \alpha_i^{w_i}}\,.
    \label{eq:vfuncts}
\end{align}
To an element $(\rho_1,\dots,\rho_m) \in \Sigma^{\rm div}$ we associate the function
\begin{align}
    v_{(\rho_1,\dots,\rho_m)}=\prod_i v_{\rho_i}^{1-\mathrm{Trop\ }\mathcal{I}(\rho_i)}\,.
\end{align}

We can finally construct the counter-terms. For any $\sigma \in \Sigma^{\rm div}$ we define the counter-term
\begin{align}
    \mathcal{I}_\sigma= v_\sigma\times\mathcal{I}|_{\sigma}\,,
\end{align}
where $|_{\sigma}$ denotes the restriction on the intersection of the compatible facets with normals $\rho_{1,\dots,m}$.
We also set $\mathcal{I}_{\varnothing}=\mathcal{I}$.

For any counter-term integrand $\mathcal{I}_\sigma$ we define a renormalization map 
\begin{align}
    \mathcal{I}_\sigma\to\mathcal{I}_\sigma^{\mathrm{ren}} = \sum_{\sigma' \in \mathrm{\Sigma}^{\rm div}|\sigma \subseteq \sigma'} (-1)^{|\sigma|-|\sigma'|}\mathcal{I}_{\sigma'}.
    \label{eq:renormalizationMap}
\end{align}
A simple application of the inclusion-exclusion principle\footnote{More precisely, this follows from the M\"obius inversion formula applied on the poset $(\Sigma^{\rm div},\supset)$, see \cite{stanley2011enumerative1}.} gives
\begin{align}
    \mathcal{I}=\sum_{\sigma} \mathcal{I}_\sigma^{\rm ren}\,.
    \label{eq:inclexcl}
\end{align}

We now make two important claims. Further details can be found in \cite{Salvatori:2024nva}.
First, the renormalized integrand $\mathcal{I}_\varnothing^{\rm ren}$ is locally finite.
Second, the terms in \eqref{eq:inclexcl} corresponding to $\sigma\neq \varnothing$ can be integrated exactly $|\sigma|$ times, giving a pole of this order. 

This results in the tropical subtraction formula,
\begin{align}
    \int_{\mathbb{R}^n_{\ge 0}} \frac{\d\alpha_1}{\alpha_1}\dots \frac{\d\alpha_n}{\alpha_n} \mathcal{I} = \sum_{\sigma} \mathrm{Vol}(\sigma)\int_{V_\sigma}\left.\frac{\d\alpha_1}{\alpha_1}\dots\frac{\d\alpha_n}{\alpha_n}\right|_{V_\sigma} \prod_{i=1}^{n} \alpha_i^{u_{\sigma,i}} (\mathcal{I}|_\sigma)^{\rm ren} \,, \label{eq:tropicalsubtraction}
\end{align}
where all integrals are locally finite.
The integration of the counter-term $\mathcal{I}_\sigma$ localizes to a subpace $V_\sigma$ where certain variables $\alpha_i$ are set to $1$.
This localization produces poles that are contained in the prefactor, $\mathrm{Vol}(\sigma)$.
An additional monomial, with exponent vector $u_\sigma$, is generated in the process. 

The subspace $V_\sigma$ is not unique. It can be constructed as follows. Let ${\bf e}_i$ be the coordinate vectors of $\mathbb{R}^n$. Choose a subset $\eta=\{e_j\}_{j \in J}$ of $(n-m)$ coordinate vectors that completes the vectors $\sigma=\{\rho_1,\dots,\rho_m\}$ to a basis of $\mathbb{R}^n$. Let $M_\sigma=\sigma|\eta$ be the matrix having as columns the vectors of $\sigma$ followed by those of $\eta$. With this choice the subspace is 
\begin{align}
    V_\sigma =  \{\alpha_j \ge 0\}_{j \in J} \cap \{\alpha_i = 1\}_{i \notin J}\,,
\end{align}
and the prefactor is
\begin{align}
    \mathrm{Vol}(\sigma)=|\det(M_\sigma)|\prod_{\rho \in \sigma} \frac{1}{-\mathrm{Trop\ }\mathcal{I}(\rho)}\,.
    \label{eq:prefactor}
\end{align}
Finally, to compute the exponent vector $u_\sigma$, consider the map $\pi_J : \mathbb{R}^n \to \mathbb{R}^n$ that sets to zero the entries not in $J$. Then 
\begin{align}
    u_\sigma=\sum_{\rho \in \sigma}\mathrm{Trop\ }\mathcal{I}(\rho)\pi_J(w_\rho)\,.
    \label{eq:monomial}
\end{align}

\paragraph{Summary of the main algorithm.}
The algorithms described in this section provide a systematic way to obtain the expansion \eqref{eq:expansion} of an arbitrary Euler integral.
The procedure requires three steps.

\begin{enumerate}
    \item Compute the polytope $\mathrm{Newt\ }\mathcal{I}$, enumerate its divergent facets, and construct the set $\Sigma^{\rm div}$ from \eqref{eq:divergentfan}.
    \item Apply Nilsson--Passare tropical continuation on power divergent vectors, and on sufficiently many vectors that violate the geometric property.
    \item Compute the tropical subtraction formula for each of the resulting integrals.
\end{enumerate}

Local finiteness ensures that we can expand the integrands in $\varepsilon$ and integrate term-by-term the resulting convergent integrals. 
These can then be integrated with the algorithm by Brown and Panzer, which will be reviewed in Sec.~\ref{sec:integration}. 

\subsubsection{A worked out example}

Let us conclude this section with a simple example where we can see all the tropical algorithms introduced so far at work.

Consider the Euler integral
\begin{align}
    A(\varepsilon)=\int_{\mathbb{R}^2_{\ge0}} \frac{\d x_1}{x_1}\frac{\d x_2}{x_2} x_1^\varepsilon x_2^\varepsilon (1+x_1+x_2)^{-3\varepsilon}. 
\end{align}

The first step is to perform the tropical analysis of the singularities of the integrand.

In \verb|SubTropica| this can be done automatically, as follows:

\begin{lstlisting}[extendedchars=true,mathescape=true,language=Mathematica]
integrand = x1$^{\mathtt{eps}}$ x2$^{\mathtt{eps}}$ (1 + x1 + x2)$^{\mathtt{-3\;eps}}$;
xvars = {x1, x2};
coeffs = {};

(* Compute Newton polytope, tropical function, and divergent fan *)
analysis = STPreAnalysis[integrand,xvars,coeffs];

(* Outer normals of Newt[I] *)
vectors = analysis["trData"]["rays"]
(* Out: {{-1,0},{0,-1},{1,1}} *)

(* Values of Trop[I] on the normals *)
tropicalValues = analysis["trops"]
(* Out: {-eps, -eps, -eps} *)

(* Divergent fan *)
sigmaDiv = analysis["faces"]
(* Out: {{}, {1}, {2}, {3}, {1, 2}, {1, 3}, {2, 3}} *)

(* Functions $\textcolor{gray}{v_\rho}$ *)
vFactors = analysis["us"]
(* Out: {"NotFound", "NotFound", "NotFound"} *)
\end{lstlisting}
This prints the outer normals  $\rho_i$ of $\mathrm{Newt\ }\mathcal{I}$, the values that $\mathrm{Trop\ }\mathcal{I}$ takes on them, the set $\Sigma^{\rm div}$ (with vectors $\rho_i$ indicated by the label $i$) and the functions $v_\rho$ from \eqref{eq:vfuncts}, if  it is possible to find them.
From the output we learn that all the normals are logarithmically divergent and pairwise compatible.
From the explicit vectors, it is also easy to see that they all violate the geometric property. Indeed no function $v_\rho$ is found.

We proceed with the second step, for instance by applying the Nilsson--Passare tropical continuation on $\rho=\{1,1\}$:
\begin{lstlisting}[extendedchars=true,mathescape=true,language=Mathematica]
continuation = STTropicalContinuation[{{1,integrand}},xvars,{{1,1}}]
(* Out: {{3, x1$^{\textcolor{gray}{\mathtt{eps}}}$ x2$^{\textcolor{gray}{\mathtt{eps}}}$ (1 + x1 + x2)$^{\textcolor{gray}{\mathtt{-1 - 3\;eps}}}$}} *)
\end{lstlisting}
which represents the identity
\begin{align}
    A(\varepsilon)=3 \int_{\mathbb{R}^2_{\ge0}}
    \frac{\d x_1}{x_1}\frac{\d x_2}{x_2} \mathcal{J}(x_1,x_2), \quad \mathcal{J}(x_1,x_2)=
    x_1^{\varepsilon} x_2^{\varepsilon} (1 + x_1 + x_2)^{-(1+ 3 \varepsilon)}\,.
\end{align}
Repeating the tropical analysis for the new integrand we now find the following functions $v_\rho$
\begin{lstlisting}[extendedchars=true,mathescape=true,language=Mathematica]
STPreAnalysis[continuation[[1,2]], xvars, coeffs]["us"]
(* Out: {$\textcolor{gray}{\frac{\mathtt{x1}}{\mathtt{1} \mathtt{+} \mathtt{x1}}}$, $\textcolor{gray}{\frac{\mathtt{x2}}{\mathtt{1} \mathtt{+} \mathtt{x2}}}$} *)
\end{lstlisting}
This means that we can compute the tropical subtraction without further use of Nilsson--Passare continuation:
\begin{lstlisting}[extendedchars=true,mathescape=true,language=Mathematica]
newintegrand = 3 x1$^{\mathtt{eps}}$ x2$^{\mathtt{eps}}$ (1 + x1 + x2)$^{\mathtt{-1 - 3\;eps}}$;
subtraction = STTropicalSubtraction[newintegrand,xvars,coeffs]
\end{lstlisting}
This returns the right-hand-side of \eqref{eq:tropicalsubtraction},
as a list of pairs $\{\mathrm{Vol}(\sigma)\alpha^{u_\sigma} (\mathcal{I}|_\sigma)^{\rm ren},V_\sigma\}$ indexed consistently with $\Sigma^{\rm div}$.
For example, the term labelled by $\sigma=\{1\}$ is stored in 
\begin{lstlisting}[extendedchars=true,mathescape=true,language=Mathematica]
subtraction[[2]]
(* Out: $\textcolor{gray}{\left\{\left\{\frac{\mathtt{x2}^{\mathtt{-1 + eps}}\; (\mathtt{1} \mathtt{+} \mathtt{x2})^{\mathtt{-1 - 3\;eps}}}{\mathtt{eps}}\mathtt{,}\; \mathtt{-}\frac{\mathtt{x2}^{\mathtt{-1 + eps}}\; \left(\frac{\mathtt{1}}{\mathtt{1} \mathtt{+} \mathtt{x2}}\right)^{\mathtt{1 + eps}}}{\mathtt{eps}}\right\}\mathtt{,}\; \mathtt{\{x2\}}\right\}}$ *)
\end{lstlisting}
It represents a renormalized integrand \eqref{eq:renormalizationMap}, whose counter-terms are stored as a list. 
Note the presence of an explicit pole in $\varepsilon$, which comes from the prefactor $\mathrm{Vol}(\sigma)$.
The expression is to be integrated over $x_2 \in [0,\infty]$.

We end by remarking that the entire procedure is automatized in a single function:
\begin{lstlisting}[extendedchars=true,mathescape=true,language=Mathematica]
STExpandIntegral[integrand,xvars,coeffs]
\end{lstlisting}

\subsection{Integration of locally finite integrands}
\label{sec:integration}
\newcommand{\Hlog}{\mathrm{H}}
\newcommand{\Reg}{\mathrm{Reg}}
\newcommand{\Li}{\mathrm{Li}}

The purpose of this section is to briefly review the by-now well-understood strategy of \cite{brown2010periods,Panzer_2015} (first applied in \cite{Chavez:2012kn,Ablinger:2012qm,Anastasiou:2013srw,Anastasiou:2013mca}) for evaluating locally finite expanded Euler integrals obtained from the procedure discussed in Sec.~\ref{sec:tropical}. We will further restrict ourselves to the cases for which these integrals are of the form
\begin{equation}\label{eq:fnDEF}
    f_n = \int_0^\infty \cdots \int_0^\infty f_{n-1}(x_1, \ldots, x_n) \, \d x_1 \cdots \d x_n\,,
\end{equation}
where $f_{n-1}$ is assumed to be expressible as a product of rational functions and hyperlogarithms (defined below in \eqref{eq:hlogDef} and discussed further in App.~\ref{app:HyperIntica}). Other integration domains $[a,b]$ can be accommodated through the change of variables $x \mapsto t = \frac{a+bx}{1+x}$.

In cases where the linear reducibility condition discussed later in Sec.~\ref{sec:linred} is satisfied, then the integral $f_n$ can be evaluated by integrating one variable at a time. One efficient method to achieve this rests on two key structural properties of hyperlogarithms: their simple behavior under differentiation, which systematizes the construction of primitives/antiderivatives, and the fact that they form a shuffle algebra. The latter property allows one to reorganize products of hyperlogarithms into canonical combinations and, in particular, to ``shuffle regularize'' hyperlogarithms by isolating unambiguously the finite contributions of antiderivatives at (potentially) singular integration boundaries. 

We summarize here how these elements combine to yield a complete integration procedure.

\subsubsection{Hyperlogarithms and their derivatives}

\emph{Hyperlogarithms} are defined as iterated integrals
\begin{equation}\label{eq:hlogDef}
\Hlog(z; \sigma_1, \ldots, \sigma_w) \equiv \int_0^z \frac{\dd t_1}{t_1 - \sigma_1} \int_0^{t_1} \frac{\dd t_2}{t_2 - \sigma_2} \cdots \int_0^{t_{w-1}} \frac{\dd t_w}{t_w - \sigma_w}\,,
\end{equation}
with the convention $\Hlog(z; \varnothing) = 1$. The sequence $\vec{\sigma} = [\sigma_1, \ldots, \sigma_w]$ is called a \emph{word}, and the integer $w$ is its (transcendental) \emph{weight}. 

The entire integration algorithm for the $f_i$'s follows from the action of the total differential. Differentiating with respect to the upper limit $z$ yields
\begin{equation}\label{eq:diff}
\frac{\dd}{\dd z} \Hlog(z; \sigma_1, \ldots, \sigma_w) = \frac{1}{z - \sigma_1} \, \Hlog(z; \sigma_2, \ldots, \sigma_w)\,.
\end{equation}
This relation encodes the fundamental point: differentiation peels off the first letter $\sigma_1$, while integration prepends a new letter. When the letters $\sigma_i$ themselves depend on a parameter, the total differential acquires additional contributions from the variation of each letter (see Eq. (3.3.32) in \cite{Panzer:2015ida} for the complete formula). However, the core principle remains unchanged: the differential structure of hyperlogarithms is entirely determined by the rational one-forms $\d\log(z - \sigma)$.

\subsubsection{The integration strategy}

Suppose now that we wish to integrate the variable $x_n$ from $0$ to $\infty$, given an integrand $f_{n-1}$ that has already been expressed in terms of hyperlogarithms. The strategy proceeds in three steps.

In the first step, the integrand is written as a linear combination of terms
\begin{equation}\label{eq:fibBasis}
f_{n-1} = \sum_{\vec{\sigma}, \tau, k} \frac{\Hlog(x_n; \vec{\sigma})}{(x_n - \tau)^k} \, \lambda_{\vec{\sigma}, \tau, k}\,,
\end{equation}
where the words $\vec{\sigma}$ and poles $\tau$ depend on the remaining variables but not on $x_n$, while $\lambda_{\vec{\sigma}, \tau, k}$ are integration constants (which can themselves involve hyperlogarithms, but in the other integration variables and parameters). This representation is achieved through a transformation that exploits the shuffle algebra and path-composition properties of iterated integrals (see, e.g., \cite{Weinzierl:2022eaz}). These give rise to the \emph{shuffle product}, which, for hyperlogarithms, states that
\begin{equation}\label{eq:shuffle2}
\Hlog(z; u) \times \Hlog(z; v) = \sum_{w \in u \shuffle v} \Hlog(z; w)\,,
\end{equation}
where $u \shuffle v$ denotes the sum over all ``interweavings'' of the original sequence (i.e., all possible ways of merging the two ordered sequences of words $u$ and $v$ into a single sequence while preserving the internal order of $u$ and $v$ respectively; see e.g., \eqref{eq:recSh}).
This identity is used repeatedly to separate the $x_n$-dependent part of each word from the integration constants.

In the second step, we construct an antiderivative $F$ satisfying $\partial_{x_n} F = f_{n-1}$ by ``integrating back'' the differential relation~\eqref{eq:diff}.

\paragraph{Step 2: Antiderivative construction.} We construct an antiderivative $F$ satisfying $\partial_{x_n} F = f_{n-1}$ by processing each term in the sum \eqref{eq:fibBasis}. For a given term, the rational (in $x_n$) coefficient $c(x_n) \equiv \frac{\lambda_{\vec{\sigma}, \tau, k}}{(x_n - \tau)^k}$ is first decomposed via partial fractions:
\begin{equation}\label{eq:PF}
c(x_n) = p(x_n) + \sum_{i,j} \frac{c_{ij}}{(x_n - a_i)^j}\,,
\end{equation}
where $p(x_n)$ is a polynomial in $x_n$, the $a_i$ are the (possibly spurious, meaning it cancels in the sum) pole locations, which may depend on other variables. \emph{Crucially}, the $c_{ij}$ are constants in $x_n$. This last property is guaranteed by linear reducibility: all polynomial factors in the denominator decompose into linear terms $(x_n - a_i)$, and the standard partial fraction theorem then yields constant numerators. We handle the polynomial and pole parts separately.

For the polynomial part, let $P(x_n) \equiv \int_0^{x_n} p(t)\, \dd t$. Integration-by-parts using~\eqref{eq:diff} gives the antiderivative
\begin{equation}\label{eq:polyIBP}
\int p(x_n)\, \Hlog(x_n; \vec{\sigma}) \, \d x_n = P(x_n)\, \Hlog(x_n; \vec{\sigma}) - \int \frac{P(x_n)}{x_n - \sigma_1} \, \Hlog(x_n; \sigma_2, \ldots) \, \d x_n\,.
\end{equation}
The first term contributes directly to the antiderivative; the correction term has a rational prefactor $P(x_n)/(x_n - \sigma_1)$ which is decomposed again via partial fractions, continuing the recursion.
The recursion terminates when the weight reaches zero, and the hyperlogarithm in \eqref{eq:polyIBP} becomes $\Hlog(x_n; \varnothing) = 1$. Indeed, its derivative vanishes
\begin{equation}
\frac{\partial}{\partial x_n} \Hlog(x_n; \varnothing) = \frac{\partial}{\partial x_n} (1) = 0\,.
\end{equation}
and, consequently, no further integration-by-parts corrections are generated and we are left with rational integrands. Assuming linear reducibility, partial fractions allows to decompose these leftover integrands according to \eqref{eq:PF},
which integrates directly since:
\begin{itemize}
\item[$\diamond$] polynomials integrate to polynomials,
\item[$\diamond$] higher-order poles ($j > 1$) integrate to rational functions, via 
\begin{equation}
    \displaystyle\int \frac{c}{(x_n - a)^j} \, \d x_n = \frac{c}{(1-j)(x_n - a)^{j-1}}\,,
\end{equation}
\item[$\diamond$] simple poles ($j = 1$) integrate to logarithms, via
\begin{equation}
    \displaystyle\int \frac{c}{x_n - a} \, \d x_n = c \, \Hlog(x_n; a)\,.
\end{equation}
\end{itemize}

This is all we have to do for the polynomial contribution to \eqref{eq:fibBasis} from \eqref{eq:PF}.

Now we consider the pole part of \eqref{eq:PF}. For simple poles ($j = 1$), inverting~\eqref{eq:diff} gives directly
\begin{equation}\label{eq:simplepole}
\int \frac{c}{x_n - a} \, \Hlog(x_n; \vec{\sigma}) \, \d x_n = c \, \Hlog(x_n; a, \vec{\sigma})\,,
\end{equation}
and it is clear that the pole location $a$ just becomes a new letter prepended to the word. For higher-order poles ($j > 1$), direct integration yields a rational function which generates an integration-by-parts correction on the right-hand side:
\begin{equation}
\begin{split}
    \int \frac{c}{(x_n - \tau)^j}& \, \Hlog(x_n; \vec{\sigma}) \, \d x_n = \frac{c}{(1-j)(x_n - \tau)^{j-1}} \, \Hlog(x_n; \vec{\sigma}) \\&\qquad- \int \frac{c}{(1-j)(x_n - \tau)^{j-1}} \, \frac{\Hlog(x_n; \sigma_2, \ldots)}{x_n - \sigma_1} \, \d x_n\,.
\end{split}
\end{equation}
This correction term, after yet another partial fraction, involves poles of order at most $j-1$ times a hyperlogarithm of lower weight. The recursion thus terminates in the exact same fashion as it did for the polynomial part and the antiderivative $F$ for the original integral is thus obtained in terms of hyperlogarithms.

In the third and final step, we evaluate the antiderivative we found at the integration endpoints. The definite integral is obtained as
\begin{equation}
f_n \equiv \int_0^\infty f_{n-1} \, \d x_n = \lim_{x_n \to \infty} F(x_n) - \lim_{x_n \to 0} F(x_n)\,.
\end{equation}
Both limits require careful treatment, as hyperlogarithms generically diverge when letters coincide with integration boundaries, which we discuss next. 

\subsubsection{Regularization at the endpoints}

The prototypical divergence in hyperlogarithms arises from
\begin{equation}\label{eq:trail0}
\Hlog(z; \underbracket[0.4pt]{0, \ldots, 0}_{k}) = \frac{\log(z)^k}{k!}\,,
\end{equation}
which diverges as $z \to 0$. Shuffle regularization extracts finite values by exploiting \eqref{eq:shuffle} to isolate divergent contributions. For instance, the identity
\begin{equation}
\Hlog(z; a) \cdot \log(z) = \Hlog(z; a, 0) + \Hlog(z; 0, a)\,,
\end{equation}
expresses the divergent integral $\Hlog(z; a, 0)$ as the difference of $\Hlog(z; a) \cdot \log(z)$ and the convergent term $\Hlog(z; 0, a)$. Regularization then discards pure-logarithm divergences while preserving all algebraic relations.

In practice, three regularizations will arise naturally. The regularization $\Reg_0$ handles trailing zeros in \eqref{eq:trail0}, which produce divergences as the argument approaches the lower integration bound.
The regularization $\Reg_\sigma$ handles leading letters equal to the upper limit $\sigma$, which produce divergences as the argument approaches that limit. For instance,
\begin{equation}
\Hlog(z; \sigma) = \int_0^z \frac{\dd t}{t - \sigma} = -\log(\sigma - z) \to +\infty \quad \text{as } z \to \sigma\,.
\end{equation}
Finally, for integrals from $0$ to $\infty$, the regularization $\Reg_\infty$ handles constant trailing of $-\lambda<0$. At large $t$, the kernel $\frac{\d t}{t - (-\lambda)} \sim \frac{\d t}{t}$ has a pole at infinity, so letters $-\lambda$ accumulate logarithmic divergences from the innermost integrations analogously to trailing zeros at the origin:
\begin{equation}
\Hlog(\infty; \underbracket[0.4pt]{-\lambda, \ldots, -\lambda}_{k}) = \int_0^\infty \frac{\dd t_1}{t_1 + \lambda} \cdots \int_0^{t_{k-1}} \frac{\dd t_k}{t_k + \lambda} \sim \frac{\log(z)^k}{k!}\Big|_{z \to \infty} \to +\infty\,.
\end{equation}

We can thus define the \emph{regulated periods} as the fully regularized value that combines both boundary contributions:
\begin{equation}\label{eq:periods}
\mathtt{zeroInfPeriod}(w) \equiv \Reg_\infty \, \Reg_0 \, \Hlog(\infty; w)\,.
\end{equation}
A crucial consistency requirement is that logarithmic divergences from the two boundaries must cancel for the integral to be well-defined. 

\paragraph{A comment on analytic continuation.} When positive letters $\lambda>0$ lie on the integration path $[0, \infty)$, the contour must be deformed around the corresponding singularities. The choice of deformation, i.e., passing above or below each pole, can be tracked by the ``delta-symbol'' $\delta[\sigma] = \pm 1$, representing the sign of $\mathrm{Im}(\sigma + i\varepsilon)$. Each contour deformation contributes terms proportional to $\pm i\pi$, which combine to produce the correct analytic continuation. See \cite{Panzer_2015} for more details. As a practical matter, one can fix the correct sign by matching the analytic formula with numerical evaluation.

\subsubsection{A worked example}\label{sec:workedExample}

To illustrate the full procedure, consider
\begin{equation}\label{eq:workedExample}
f_2 = \int_0^1 \int_0^1 \frac{\d x_1 \, \d x_2}{1 - x_1 + x_1 x_2^2}\,,
\end{equation}
integrated in the order $(x_1, x_2)$ (the opposite order fails; see Sec.~\ref{sec:linred}). The M\"obius map $x_1 = x_3/(1+x_3)$ transforms the $x_1$-integral to $[0,\infty)$, giving $1 - x_1 = 1/(1+x_3)$, $x_1 x_2^2 = x_3 x_2^2/(1+x_3)$, and hence
\begin{equation}
f_0 = \frac{1}{(1+x_3)(1+x_3 x_2^2)}\,.
\end{equation}
Partial fractions in $x_3$ yield two simple poles:
\begin{equation}\label{eq:exPF}
\frac{1}{(1+x_3)(1+x_3 x_2^2)} = \frac{1/(1-x_2^2)}{x_3+1} + \frac{1/(x_2^2-1)}{x_3+1/x_2^2}\,.
\end{equation}
Since the integrand is a pure rational function ($\Hlog(x_3; \varnothing) = 1$, weight zero), the simple-pole rule~\eqref{eq:simplepole} gives the antiderivative directly:
\begin{equation}\label{eq:exPrimitive}
F(x_3) = \frac{1}{1-x_2^2}\,\Hlog(x_3; -1) \;+\; \frac{1}{x_2^2-1}\,\Hlog\!\left(x_3; -\frac{1}{x_2^2}\right)\,.
\end{equation}
We now evaluate at the boundaries. At $x_3 = 0$ both hyperlogarithms vanish (no trailing zeros in the words). At $x_3 \to \infty$, each $\Hlog(x_3; \sigma)$ with $\sigma < 0$ grows as $\log x_3 + \log|\sigma|$, but the leading $\log x_3$ divergences cancel because the two coefficients sum to zero: $\frac{1}{1-x_2^2} + \frac{1}{x_2^2-1} = 0$. The surviving finite parts give
\begin{equation}
f_1 = \frac{\log(x_2^2)}{x_2^2-1}\,.
\end{equation}
For the second variable we integrate $f_1$ over $x_2 \in [0,1]$. Writing $\log(x_2^2) = 2\,\Hlog(x_2; 0)$ and decomposing $\frac{1}{x_2^2-1} = \frac{1}{2(x_2-1)} - \frac{1}{2(x_2+1)}$, the simple-pole rule gives the antiderivative
\begin{equation}\label{eq:exPrimitiveY}
G(x_2) = \Hlog(x_2; 1, 0) - \Hlog(x_2; -1, 0)\,.
\end{equation}
At $x_2=0$, the trailing zeros produce a $\log(x_2)$ divergence. The shuffle identity $\Hlog(x_2; \sigma) \cdot \Hlog(x_2; 0) = \Hlog(x_2; \sigma, 0) + \Hlog(x_2; 0, \sigma)$ gives $\Hlog(x_2; \sigma, 0) = \Hlog(x_2; \sigma) \cdot \log x_2 - \Hlog(x_2; 0, \sigma)$. Then the regularization prescription discussed above discards the $\log x_2$ term, leaving $\Reg_0\,\Hlog(x_2; \sigma, 0) = -\Hlog(x_2; 0, \sigma)$ for $\sigma = \pm 1$. Since $\Hlog(x_2; 0, \sigma) = -\Li_2(x_2/\sigma)$ and $\Li_2(0) = 0$, we get $G(0) = 0$.

At $x_2=1$, we apply $\Reg_0$ (same trailing-zero identity as above) and then $\Reg_1$ (which would strip leading letters equal to the upper limit $1$, but neither $[0, 1]$ nor $[0, -1]$ starts with $1$, so $\Reg_1$ acts trivially). Using $\Hlog(1; 0, \sigma) = -\Li_2(1/\sigma)$, we get
\begin{subequations}
\begin{align}
\Reg_1\,\Reg_0\,\Hlog(1; 1, 0) &= -\Hlog(1; 0, 1) = \Li_2(1) = \zeta(2)\,,\\
\Reg_1\,\Reg_0\,\Hlog(1; -1, 0) &= -\Hlog(1; 0, -1) = \Li_2(-1) = -\frac{\pi^2}{12}\,.
\end{align}
\end{subequations}
Combining these results, we obtain
\begin{equation}
f_2 = G(1) - G(0) = \frac{\pi^2}{4}\,,\label{eq:exResult}
\end{equation}
which is correct. We discuss next what would have happened if the integration order was taken be $(x_2,x_1)$ instead of $(x_1,x_2)$.

\subsection{Integration order and linear reducibility}\label{sec:linred}

The choice of the integration order plays a central role in evaluating hyperlogarithms, as it drastically affects the complexity of intermediate steps, or even whether the procedure explained in Sec.~\ref{sec:integration} terminates at all. For these reasons, the choice of integration order deserves separate attention.

To see this concretely, return to the integral \eqref{eq:workedExample} from Sec.~\ref{sec:workedExample}. In the order $(x_1,x_2)$, the polynomial $1 - x_1 + x_1x_2^2$ is linear in $x_1$. After integrating $x_1$, the resulting letter $x_2^2 - 1 = (x_2-1)(x_2+1)$ factors linearly in $x_2$, and the algorithm runs to completion, yielding $\pi^2/4$.

In the opposite order $(x_2, x_1)$, the polynomial $1 - x_1 + x_1x_2^2$ is quadratic in $x_2$, and one would find the antiderivative
\begin{equation}
\frac{1}{1 - x_1 + x_1 x_2^2} = \frac{\partial}{\partial x_2} \left[ \frac{1}{\sqrt{x_1(1-x_1)}} \arctan \left( \frac{\sqrt{x_1}\,x_2}{\sqrt{1-x_1}} \right) \right]\,.
\end{equation}
The $\arctan$ function cannot be expressed as a hyperlogarithm with rational arguments, since $\arctan(z) = \frac{i}{2} [ \log(1 - i z) - \log(1 + i z)]$ and $z = \sqrt{x_1}\,x_2/\sqrt{1-x_1}$ involves square roots. The integral is not linearly reducible in this order in the $x_1,x_2$ variables.

\subsubsection{Fubini reduction}

The question of finding linearly-reducible (LR) orders amounts to checking if at all steps of the integration, the integrand polynomials are linear in the current integration variable. The key observation is that one does not need to perform full integration to answer this question. This is because for the purposes of finding linearly-reducible orders, we only care about the singularity structure of the integrand, not the form of the integrand itself.

To this effect, we employ the strategy used in \SOFIA, which uses the Fubini reduction algorithm \cite{Brown:2008um,Panzer_2015} to analyze singularities of integrals, see \cite[Sec.~2.2]{Correia:2025wtb}. In the first step, we use dynamical programming to find the set of singularities for every possible integration order. In the second step, we crawl the space of orders discarding those that include at least one non-linear polynomial. At the same time, we score the orders according to the metric described below.

\subsubsection{Ordering orders}
\label{sec:orders}

Once we have determined all the LR orders, we have to order them according to which one is expected to yield the most efficient integration. For this purpose, we assign scores to every LR order, which are meant to be a proxy for how complicated intermediate expressions it is expected to produce.

\begin{figure}
\centering
\includegraphics[width=0.8\textwidth]{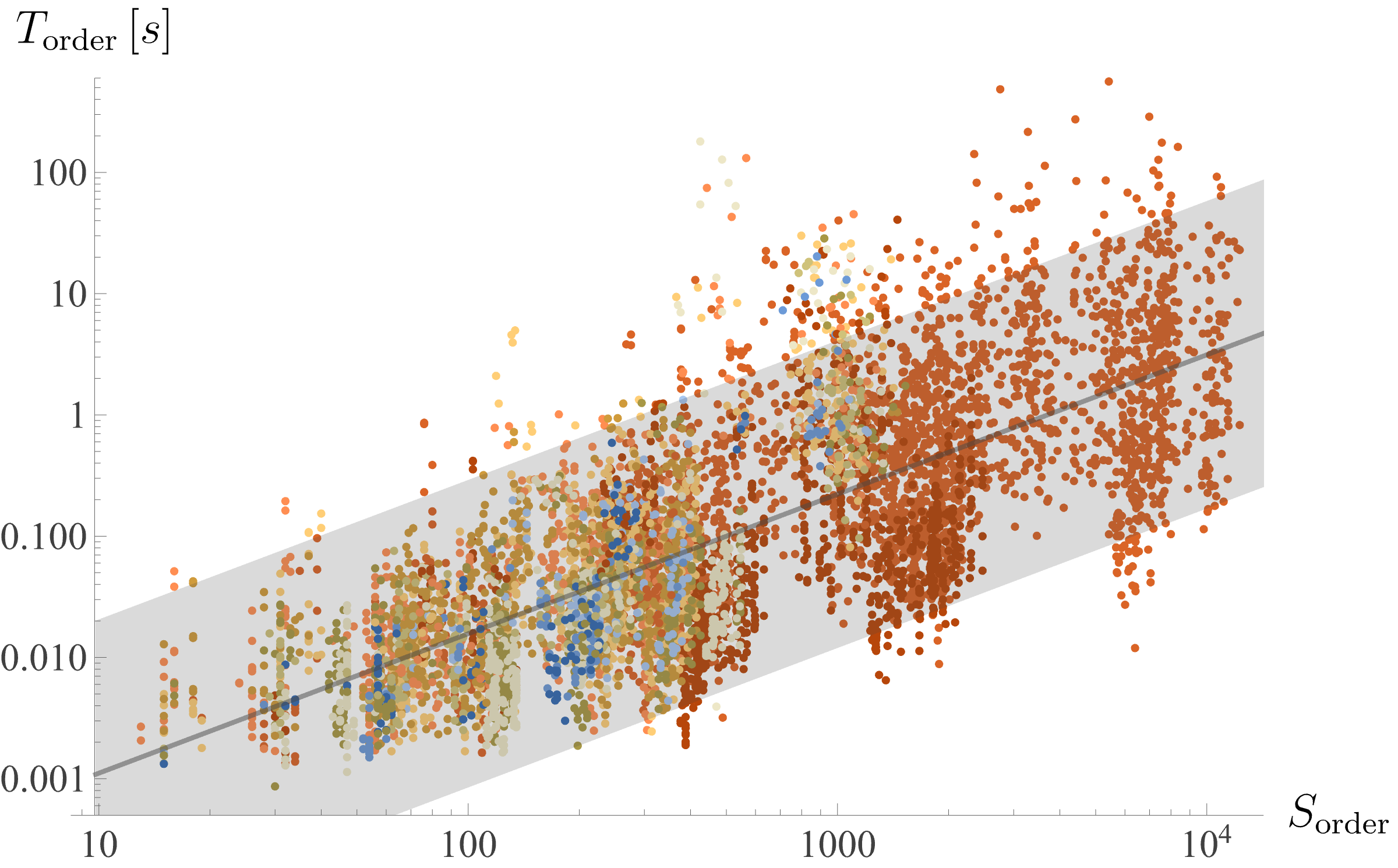}
\caption{\label{fig:correlation}Correlation between the score $S_\text{order}$ given to a given LR order and its resulting timing $T_\text{order}$ in seconds. Different colors represent data points associated with different Feynman integrals. The straight line is the power-law fit and the shaded areas denote the $95\%$ confidence intervals.}
\end{figure}

The intuition is that those LR orders that have large discriminants will also produce large integrands. Hence, we score each LR order by the total size (measured as the number of \verb|Mathematica| expression leaves using \verb|LeafCount|) of its polynomials. 
To corroborate this intuition, we collected a representative list of 34 example Feynman integrals. For each individual integrand term, we integrated it according to a randomly chosen LR order. We recorded the score associated with the LR orders and the eventual integration times using \verb|HyperIntica|. The total number of data points is $10\, 020$. The results are collected in Fig.~\ref{fig:correlation}. Given that the integration times and scores span multiple orders of magnitude, it is more convenient to talk about logarithms of both quantities.

The results in Fig.~\ref{fig:correlation} show a moderately good correlation ($R^2 \approx 0.55$) with a power-law fit between the score $S_{\text{order}}$ and integration time $T_{\text{order}}$ in seconds, for any given LR order. For the mean and variance, we obtain
\begin{subequations}
\begin{align}
\mathbb{E}[\log T_{\text{order}}] &\approx \alpha \log S_{\text{order}} + \beta &&\!\!\!\text{with}\quad \alpha \approx 1.15\, , \quad \beta \approx -9.45\, , \\
\mathbb{V}[\log T_{\text{order}}] &\approx \sigma^2  &&\!\!\!\text{with}\quad \sigma \approx 2.92\, .
\end{align}
\end{subequations}
In particular, we find that the variance is approximately constant, which means that $T_\text{order}$ follows an approximately log-normal distribution.

If we could freely choose an LR order for each integrand term, we would simply pick the one with the smallest score. However, the algorithm described in Sec.~\ref{sec:tropical} forces us to pick the same LR order for every term belonging to the same face $F$ of the polytope. Hence, we are tasked with a problem: how to choose a single LR order that minimizes the total expected integration time for a number of integrands?

Let us denote with $S_\text{order}^{(i)}$ and $T_\text{order}^{(i)}$, respectively, the score and the timing of the integrand $i \in F$ using a fixed order.
We construct the following mean-variance cost function:
\begin{subequations}
\begin{align}
\mathcal{L}_{F,\text{order}} &= \mathbb{E}\left[ \sum_{i \in F} T_{\text{order}}^{(i)} \right] + \lambda \sqrt{ \mathbb{V}\left[ \sum_{i \in F} T_{\text{order}}^{(i)} \right] } \\
&= \mathrm{e}^{\beta + \sigma^2/2} \left[ \sum_{i\in F} (S_{\text{order}}^{(i)} )^\alpha + \lambda \sqrt{\mathrm{e}^{\sigma^2} - 1} \sqrt{\sum_{i \in F} (S_{\text{order}}^{(i)} )^{2\alpha}} \right]\, ,\label{eq:pre-cost-function}
\end{align}
\end{subequations}
where we optimistically assumed that integration times are mutually independent. Here, $\lambda$ is the risk-aversion factor. We set $\lambda = 1.645$ to ensure $95\%$ certainty.

In our setup, we are dominated by the variance since $\lambda \sqrt{\mathrm{e}^{\sigma^2} - 1} \approx 117$. This should be clear from Fig.~\ref{fig:correlation}, since the wide error bars mean we can at best predict $T_\text{order}$ within two orders of magnitude. Hence, in practice, we can essentially ignore the first term in \eqref{eq:pre-cost-function} and instead use
\begin{equation}
\tilde{\mathcal{L}}_{F,\text{order}} = \sum_{i \in F} (S_{\text{order}}^{(i)} )^{2\alpha}\, ,
\end{equation}
where we dropped the irrelevant constants and the square root which is strictly monotonic. This is the cost function we use in the code to determine the optimal LR order.
Empirically, we found that it is good at minimizing the computational effort and penalizing outliers that would otherwise stall the integration.
\section{\label{sec:examples}Detailed
examples}

In this section, we discuss in detail a representative subset of the examples available in the ancillary notebook \texttt{PaperChecks.wl}. The purpose is to illustrate both the computational capabilities and the high level of automation that \texttt{SubTropica} offers for Feynman integrals, as well as its applicability to problems beyond this context.

\subsection{Scalar Feynman integrals}

\begin{figure}[t]
\begin{minipage}[t]{0.45\textwidth}
    \begin{tikzpicture}
  \coordinate (v1) at (2.000, 2.000);
  \coordinate (v2) at (5.750, 2.000);
  \coordinate (v3) at (2.750, 1.250);
  \coordinate (v4) at (2.750, -1.000);
  \coordinate (v5) at (5.000, 1.250);
  \coordinate (v6) at (5.000, -1.000);
  \coordinate (v7) at (5.750, -1.750);
  \coordinate (v8) at (2.000, -1.750);

  \draw[ultra thick, Maroon, line cap=round] (v1) node[left, font=\footnotesize] {$p_{1}$} -- (v3);
  \draw[thick, decorate, decoration={snake, amplitude=1.5pt, segment length=5pt}] (v4) -- (v3) node[midway, auto, font=\footnotesize] {$-\ell - p_{1} - p_{2}$};
  \draw[ultra thick, Maroon, line cap=round] (v2) node[right, font=\footnotesize] {$p_{2}$} -- (v5);
  \draw[ultra thick, Maroon, line cap=round] (v3) -- (v5) node[midway, auto, font=\footnotesize] {$-\ell - p_{2}$};
  \draw[thick, decorate, decoration={snake, amplitude=1.5pt, segment length=5pt}] (v6) -- (v5) node[midway, right, font=\footnotesize] {$\ell$};
  \draw[ultra thick, Maroon, line cap=round] (v7) node[right, font=\footnotesize] {$p_{3}$} -- (v6);
  \draw[ultra thick, Maroon, line cap=round] (v4) -- (v6) node[midway, below, font=\footnotesize] {$\ell - p_{3}$};
  \draw[ultra thick, Maroon ,line cap=round] (v8) node[left, font=\footnotesize] {$p_{4}$} -- (v4);

  \fill (v3) circle (2pt);
  \fill (v4) circle (2pt);
  \fill (v5) circle (2pt);
  \fill (v6) circle (2pt);
\end{tikzpicture}
\end{minipage}
\begin{minipage}[t]{0.45\textwidth}
    \begin{tikzpicture}[scale=1.1]
  \coordinate (v1) at (-4.750, 0.250);
  \coordinate (v2) at (-3.745, 0.257);
  \coordinate (v3) at (-2.250, 1.250);
  \coordinate (v4) at (-2.274, -0.776);
  \coordinate (v5) at (-0.250, -0.750);
  \coordinate (v6) at (-0.250, 1.250);
  \coordinate (v7) at (0.500, 2.000);
  \coordinate (v8) at (0.500, -1.500);

  \draw[ultra thick, gray, line cap=round] (v1) node[left, font=\footnotesize] {$p_{1}$} -- (v2);
  \draw[thick, dashed] (v3) -- (v2) node[midway, above, font=\footnotesize, xshift=-12pt] {$\ell_{1} - p_{1}$};
  \draw[thick, dashed] (v4) -- (v3) node[midway, right, font=\footnotesize] {$\ell_{1} + \ell_{2} + p_{3}$};
  \draw[thick, dashed] (v2) -- (v4) node[midway, below, font=\footnotesize, xshift=-5pt] {$\ell_{1}$};
  \draw[thick, dashed] (v5) -- (v4) node[midway, auto, font=\footnotesize] {$\ell_{2} + p_{3}$};
  \draw[thick, dashed] (v6) -- (v5) node[midway, auto, font=\footnotesize] {$\ell_{2}$};
  \draw[thick, dashed] (v3) -- (v6) node[midway, auto, font=\footnotesize] {$\ell_{2} - p_{2}$};
  \draw[ultra thick, OliveGreen, line cap=round] (v7) node[right, font=\footnotesize] {$p_{2}$} -- (v6);
  \draw[ultra thick, RoyalBlue, line cap=round] (v8) node[right, font=\footnotesize] {$p_{3}$} -- (v5);

  \fill (v2) circle (2pt);
  \fill (v3) circle (2pt);
  \fill (v4) circle (2pt);
  \fill (v5) circle (2pt);
  \fill (v6) circle (2pt);
\end{tikzpicture}
\end{minipage}
\caption{Left: The $e^-e^- \to e^- e^-$ (Møller) scattering one-loop box. Right: a two-loop triangle-box topology with three distinct external masses $\textcolor{gray}{M_1}$, $\textcolor{OliveGreen}{M_2}$, $\textcolor{RoyalBlue}{M_3}$. Wavy and dashed lines are massless.}
\label{fig:triangle}
\end{figure}
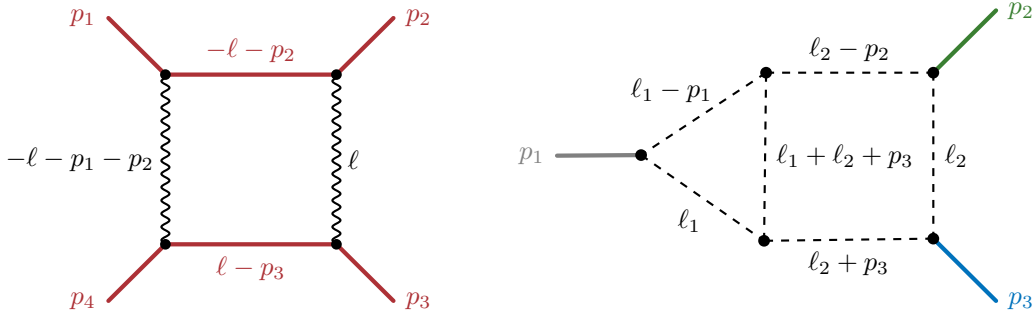

As the first two simple examples, we consider the massless triangle and triangle-box integral in $\D=4-2\varepsilon$ dimensions with three distinct external masses $M_1$, $M_2$, $M_3$ (see Fig.~\ref{fig:triangle}). As the final example of this subsection, a cutting-edge case is also discussed (see Fig.~\ref{fig:cuttingEdge}).

\subsubsection{A one-loop box with algebraic letters (Fig.~\ref{fig:triangle} (left))}
As a first example, consider the one-mass box with equal internal and external masses:
\begin{lstlisting}[extendedchars=true,mathescape=true,language=Mathematica]
diag = {{{{1,2},m},{{2,3},0},{{3,4},m},{{1,4},0}},
        {{1,m},{2,m},{3,m},{4,m}}};
\end{lstlisting}
With the default \texttt{"FindRoots" -> False}, \texttt{STIntegrate} aborts because the integrand contains irreducible quadratic polynomials that obstruct linear reducibility. In this example, this obstruction can be resolved by enabling \texttt{"FindRoots" -> True}, which instructs \texttt{STFasterFubini} to factor these quadratics into linear factors by introducing auxiliary square-root variables, restoring linear reducibility:
\begin{lstlisting}[extendedchars=true,mathescape=true,language=Mathematica]
result = STIntegrate[diag, "FindRoots" -> True]
\end{lstlisting}
The integration completes in about one second and returns
\begin{align}
-\frac{\log(-W_-) - \log(-W_+)}{\mathtt{mm}\,\mathtt{s12}\,(W_- - W_+)}\Big(\frac{1}{\varepsilon}-\log(-s_{12})\Big) + \mathcal{O}(\varepsilon)\,,
\end{align}
where $W_\pm = (-2\,\mathtt{mm} + \mathtt{s23} \pm \sqrt{(\mathtt{s23}-4\,\mathtt{mm})\,\mathtt{s23}})\,/\,(2\,\mathtt{mm})$ are the roots of the quadratic polynomial that obstructed linear reducibility. The result is automatically verified numerically using \texttt{STVerify}:
\begin{lstlisting}[extendedchars=true,mathescape=true,language=Mathematica]
STVerify[diag, result /. GetAlgebraicBackSubRules[]]
(* Out: "pass" -> True *)
\end{lstlisting}
where \texttt{GetAlgebraicBackSubRules} replaces the expressions for $W_{\pm}$ in terms of $m^2$ and $s_{23}$.

\subsubsection{A two-loop triangle-box integral (Fig.~\ref{fig:triangle} (right)).}
As a two-loop example, consider the massless triangle-box topology with three distinct external masses parametrized via the rationalizing variables $z$, $\bar{z}$:
\begin{lstlisting}[extendedchars=true,mathescape=true,language=Mathematica]
answer = STIntegrate[
    {{{{1,4},0},{{1,5},0},{{2,5},0},{{2,3},0},{{3,4},0},{{4,5},0}},
    {{1,M1},{2,M2},{3,M3}}},
    "SetProblemID"  -> "TriangleBox",
    "Substitutions" -> {MM1 -> z zb, MM2 -> (1-z)(1-zb), MM3 -> 1}]
\end{lstlisting}
The raw result at $\mathcal{O}(\varepsilon^0)$ involves weight-4 hyperlogarithms and it can be cross-checked numerically at a specific kinematic point using \texttt{STNIntegrate}:
\begin{lstlisting}[extendedchars=true,mathescape=true,language=Mathematica]
STNIntegrate[diag, "Substitutions" -> {MM1 -> z zb, MM2 -> (1-z)(1-zb),
                                       MM3 -> 1} /. {z -> $\frac{\mathtt{1}}{\mathtt{3}}$, zb -> $\frac{\mathtt{1}}{\mathtt{7}}$}
                 , "Order"         -> 0]
\end{lstlisting}
The numerical output agrees with the analytic result evaluated at the same point. In addition, the known result for $0<z<\overline{z}<1$ (see, e.g., \cite[Sec.~4.3]{Barrera:2025uin}) reads
\begin{align}
\frac{\tfrac{1}{2}\log^2(z\bar z)\big(\mathrm{Li}_2(z){-}\mathrm{Li}_2(\bar z)\big) {-} 3\log(z\bar z)\big(\mathrm{Li}_3(z){-}\mathrm{Li}_3(\bar z)\big) {+} 6\big(\mathrm{Li}_4(z){-}\mathrm{Li}_4(\bar z)\big)}{(1-z)(1-\bar z)(z-\bar z)}\,,
\end{align}
and one can verify that both results agree, e.g., at the symbol-level using \texttt{ConvertToSymbol}.

Having illustrated the use of the package on simple examples, we stress that the package can also be used on certain cutting-edge Feynman integrals.

\subsubsection{A cutting-edge example (Fig.~\ref{fig:cuttingEdge})}\label{sec:examplesCE}
As a demanding benchmark, we evaluate the four-loop, nine-propagator integral
\begin{equation*}
    \int\!\frac{\mathrm{e}^{4\varepsilon \gamma_E} \prod_{i=1}^{4}\frac{\mathrm{d}^{\D}\ell_{i}}{i\pi^{\D/2}}}{[\left(\ell_{2} {+} \ell_{3} {+} p_{2} {+} p_{3}\right)^2 {-} m^2]\left(\ell_{1} {-} \ell_{3}\right)^2\ell_{1}^2\left(\ell_{1} {+} \ell_{2} {+} p_{3}\right)^2\ell_{2}^2\left(\ell_{2} {+} \ell_{3}-\ell_{4}\right)^2\ell_{3}^2\left(\ell_{4} {-} p_{4}\right)^2\ell_{4}^2},
\end{equation*}
shown in Fig.~\ref{fig:cuttingEdge} at the kinematic point $(s_{12},s_{23},m^2)$ where a single internal edge carries squared mass $m^2=1$ and all remaining internal and external masses vanish:
\begin{lstlisting}[extendedchars=true,mathescape=true,language=Mathematica]
diag = {{{{4,5},0},{{1,4},0},{{1,2},1},
         {{2,3},0},{{3,5},0},{{5,6},0},
         {{2,5},0},{{4,6},0},{{1,6},0}},
         {{1,0},{2,0},{3,0},{4,0}}};

STIntegrate[diag, "Dimension"      -> 4 - 2 eps,
                  "Gauge"          -> {x3 -> 1},
                  "SimplifyOutput" -> Identity]
\end{lstlisting}
The tropical subtraction produces
  six locally finite integrands. The timing breakdown is:
\begin{lstlisting}[extendedchars=true,mathescape=true,language=Mathematica]
>> Time to set up integrands and directories:   4.7 s
>> Time to find linearly reducible orderings:   $\approx$ 1.2 h
>> Time to integrate:                           $\approx$ 36.6 h
\end{lstlisting}
The computation completed in approximately 38 hours on a cluster equipped with an Intel Xeon Gold 6326 at 2.9\,GHz (32 cores) and 256\,GB of RAM, with a peak memory usage of approximately 114\,GB. The result starts at $\mathcal{O}(\varepsilon^{-1})$ and is expressed in terms of hyperlogarithms. The result is too large to display here, but it can be found in the \texttt{SubTropica} library.

{\subsection{Linearized propagators and tensor numerators}\label{ex:anomalous}}

\texttt{SubTropica} is not limited to scalar Feynman integrals with quadratic propagators: it also handles \emph{linearized} (eikonal) propagators of the form $v\cdot k$ and loop momenta dependent (tensor) numerators.
As a non-trivial illustration, we consider a family of integrals arising in the calculation of soft anomalous dimensions in \cite[Eq.~(D.11)]{Gardi:2025ule}.

In particular, the family is defined by the integral (in $\D=4-2\varepsilon$):
\begin{align*}
\int \frac{\d^\D k_1}{i\pi^{\D/2}} \frac{\d^\D k_2}{i\pi^{\D/2}}\, \frac{(4\text{e}^{\gamma_\text{E}})^{2\varepsilon}[v_2{\cdot} k_1]^{-r_7}\, [\tilde{\beta}{\cdot} k_2]^{-r_8}\, [v_1{\cdot} k_2]^{-r_9}}{[k_1^2]^{r_1}\,[k_2^2]^{r_2}\,[(k_1{+}k_2)^2]^{r_3}\,[v_1{\cdot} k_1 {-}1]^{r_4}\,[v_2{\cdot}k_2 {-}1]^{r_5}\,[-\tilde{\beta}\cdot(k_1+k_2)]^{r_6}},
\end{align*}
for the single lightcone limit of the cubic ``timelike web'' in \cite[Eq.~(D.8)]{Gardi:2025ule}. Here, $v_1$, $v_2$ are (unit) velocity vectors of timelike Wilson lines with $v_1^2=v_2^2=1$, $\tilde{\beta}$ is the velocity vector for a lightlike Wilson line ($\tilde{\beta}^2=0$), such that the kinematic invariants can be parametrized by $v_1\cdot \tilde{\beta} = -y$, $v_2\cdot\tilde{\beta} = -1$ and $v_1\cdot v_2 = -\tfrac{1}{2}(a_{12} + a_{12}^{-1})$.
The propagators include both standard quadratic denominators $k_i^2$ and eikonalized denominators of the form $v\cdot k - 1$.

In \texttt{SubTropica}, the integral is specified by
\begin{lstlisting}[extendedchars=true,mathescape=true,language=Mathematica]
(* Basis of propagators and numerators *)
basisProp = {k[1]$\cdot$k[1], k[2]$\cdot$k[2], (k[1]+k[2])$\cdot$(k[1]+k[2]),
            v[1]$\cdot$k[1]-1, v[2]$\cdot$k[2]-1, -$\tilde{\beta}\cdot$(k[1]+k[2]),
            v[2]$\cdot$k[1], $\tilde{\beta}\cdot$k[2], v[1]$\cdot$k[2]};
            
(* An optional list of powers for each element of basisProp *)
basisExp = {1,1,1,1,1,1,-1,0,0}; (* default: all 1's *)

(* An optional list of kinematic replacements *)
basisKin = {v[1]$\cdot$v[1] | v[2]$\cdot$v[2] -> 1, 
            v[1]$\cdot\tilde{\beta}$ -> -y, v[2]$\cdot\tilde{\beta}$ -> -1, $\tilde{\beta}\cdot\tilde{\beta}$ -> 0,
            v[1]$\cdot$v[2] -> -$\frac{\mathtt{1}}{\mathtt{2}}$($\frac{\mathtt{1}}{\mathtt{a12}}$ + a12)};
            
STIntegrate[basisProp,
  "Exponents"           -> basisExp,
  "Substitutions"       -> basisKin,
  "LoopMomenta"         -> {k[1],k[2]},
  "Normalization"       -> -(4 Exp[EulerGamma])$^{\mathtt{2\; eps}}$,
  "Gauge"               -> {x5 -> 1},
  "MethodPolysAndPairs" -> "Standard"
  ]
\end{lstlisting}

As for the Feynman diagrammatic inputs considered above,
\texttt{STIntegrate} applied on the propagator and numerator basis \texttt{basisProp} automatically determines the parametric representation, finds a linearly reducible integration order across all counter-term integrands, and evaluates the $\varepsilon$-expansion.
The first two orders read
\begin{align}
-\frac{1}{4\,\varepsilon^4\, y} - \frac{\pi^2(3+7y) + 4(1+y)\log^2 a_{12} + 4y\log^2 y}{8\,\varepsilon^2\, y\,(1+y)} + ...\,,
\end{align}
where the full result through $\mathcal{O}(\varepsilon^{-1})$ is available in the ancillary file. The result is in agreement with the differential equation method of~\cite{Gardi:2025ule}.\footnote{We thank Zehao Zhu and Einan Gardi for providing explicit results for cross-checks.}

\subsection{Advanced examples: Beyond Feynman integrals}
We now illustrate how \texttt{SubTropica} and its lower-level functions from Tab.~\ref{tab:list} can be used to tackle integrals that fall outside the standard Feynman integral class.

\subsubsection{Gravitational energy-energy correlators}

Energy-energy correlators (EECs) are infrared-finite observables that measure correlations between energy deposited at two points on the celestial sphere. In four-dimensional gravity, the reference \cite{Chicherin:2025keq} recently computed EECs at one loop in $\mathcal{N}=8$ supergravity (SG) and in pure Einstein gravity.

In particular, they consider two gravitons scattering in the center-of-mass frame with total energy $2E$, and two calorimeters placed on the celestial sphere along the unit vectors $\vec{n}_1$ and $\vec{n}_2$. The geometry is parameterized by
\begin{equation}\label{eq:eec-kin}
\underbracket[0.4pt]{y_1 = \frac{1-\vec{n}\cdot\vec{n}_1}{2}\,,\quad y_2 = \frac{1-\vec{n}\cdot\vec{n}_2}{2}}_{\text{beam-detector geometry}}\,,\quad \underbracket[0.4pt]{z = \frac{1-\vec{n}_1\cdot\vec{n}_2}{2}}_{\substack{\text{opening angle between}\\\text{the two detectors}}}\,,
\end{equation}
where $\vec{n}$ is the beam axis. The EEC can be generalized to arbitrary (complex) energy weights $J_1$, $J_2$, so that each calorimeter records $\sum_i E_i^{J_k}$ rather than $\sum_i E_i$; the standard EEC corresponds to $J_1=J_2=1$. These weights are related to the spin of twist-2 operators placed at null infinity~\cite{Chicherin:2025keq}.

At NLO in $\mathcal{N}=8$ SG, the real-emission contribution reduces to a univariate integral over the energy fraction $x\in[0,1]$ of the particle hitting detector one, all the other degrees of freedom already integrated out (see \cite[Eq.~(A.14)]{Chicherin:2025keq}):
\begin{equation}\label{eq:EECreal}
\mathrm{EEC}^{\mathrm{real}}_{J_1,J_2} = \frac{E^{-4-4\varepsilon}}{16(2\pi)^{5-4\varepsilon}}\int_0^1 \d x\,\frac{x^{J_1+1-2\varepsilon}(1-x)^{J_2+1-2\varepsilon}}{(1-zx)^{J_2+2-2\varepsilon}}\,\mathcal{M}_{2\to3}\,,
\end{equation}
where $\mathcal{M}_{2\to3}$ is the tree-level $2\to3$ \emph{squared} matrix element summed over the $\mathcal{N}=8$ supermultiplet:
\begin{equation}\label{eq:M2to3}
\mathcal{M}_{2\to3} = \frac{8\,E^4\,\Delta(z,y_1,y_2)}{z(1{-}z)\,y_1\,y_2\,(1{-}y_1)(1{-}y_2)}\,\frac{(1{-}zx)^4}{x^2(1{-}x)^2\,P(x)\,Q(x)}\,.
\end{equation}
Here $\Delta(z,y_1,y_2)$ is the Gram determinant of the beam and detector directions, and $P(x;z,y_1,y_2)$, $Q(x;z,y_1,y_2)$ are quadratic polynomials in $x$ given explicitly in \cite[Eq.~(3.17)]{Chicherin:2025keq} and recorded in a \textsc{Mathematica} syntax below.

\paragraph{Setup.}
As a first step, we encode the integral~\eqref{eq:EECreal} as an Euler integrand for \texttt{STIntegrate}. Since there is a single integration variable, the quadratic polynomials $P$ and $Q$ can be factored into linear factors using the helper function \texttt{STFactorAndTrackRoots}. The symbolic roots are stored in \texttt{explicit[P]} and \texttt{explicit[Q]} and substituted back at the end:
\begin{lstlisting}[extendedchars=true,mathescape=true,language=Mathematica]
(* Factor quadratic polynomials into linear factors: *)
Ppol[x_] := STFactorAndTrackRoots[
  (1-z)(1-y1)+(z+y1-2z y1-y2)(1-x)+z y1(1-x)$^{\mathtt{2}}$, x, P]
Qpol[x_] := STFactorAndTrackRoots[
  (1-z)y1+(y2-z-y1+2z y1)(1-x)+z(1-y1)(1-x)$^{\mathtt{2}}$, x, Q]
\end{lstlisting}
The integral is then cast into the standard Euler form via the M\"obius map $x \to x/(1{+}x)$, which sends $[0,1]\to[0,\infty)$. The result is packaged as a quadruple \texttt{\{prefactor, integrand, variables, coefficients\}}:
\begin{lstlisting}[extendedchars=true,mathescape=true,language=Mathematica]
(* Prefactor: *)
pref = $\frac{\mathtt{ee}^{-\mathtt{4}-\mathtt{4}\;\mathtt{eps}}}{\mathtt{16}(\mathtt{2}\textcolor{black}{\mathtt{Pi}})^{\mathtt{5}-\mathtt{4}\;\mathtt{eps}}}$$\frac{\mathtt{8}\;\mathtt{ee}^{\mathtt{4}}\; \Delta}{\mathtt{z}(\mathtt{1}\mathtt{-}\mathtt{z})\; \mathtt{y1}\; \mathtt{y2}\; (\mathtt{1}\mathtt{-}\mathtt{y1})(\mathtt{1}\mathtt{-}\mathtt{y2})}$;

(* Integrand on [0,$\textcolor{gray}{\infty}$) after x -> x/(1+x): *)
integrand = d[x] $\frac{\mathtt{x}^{\mathtt{J1}+\mathtt{1}-\mathtt{2}\;\mathtt{eps}}(\mathtt{1}\mathtt{-}\mathtt{x})^{\mathtt{J2}+\mathtt{1}-\mathtt{2}\;\mathtt{eps}}}{(\mathtt{1}\mathtt{-}\mathtt{z\; x})^{\mathtt{J2}+\mathtt{2}-\mathtt{2}\;\mathtt{eps}}}$$\frac{(\mathtt{1}\mathtt{-}\mathtt{z\; x})^{\mathtt{4}}}{\mathtt{x}^{\mathtt{2}}(\mathtt{1}\mathtt{-}\mathtt{x})^{\mathtt{2}}\; \mathtt{Ppol[x]\; Qpol[x]}}$
                /. x -> x/(1+x) // Applyd[#, {x}] //. d[x_] :> 1 &;
(* Applyd computes the Jacobian of the substitution on d[x] = $\textcolor{gray}{\d x}$. *)

(* Variables and coefficients: *)
xvars  = {x};
coeffs = {ee, $\Delta$, z, y1, y2};
\end{lstlisting}

\paragraph{Tropical analysis.}
For positive integer energy weights, we expect no divergences in the region $0<z<1$, so that the dimensional regulator $\varepsilon$ is unnecessary. Within \texttt{SubTropica}, we can verify this explicitly with \texttt{STPreAnalysis}. For example, working near the integer values of energy weights $J_1=1$ and $J_2=2$ this command (note the Jacobian \texttt{Times @@ xvars} converting to the $\d\log$ convention required by \texttt{STPreAnalysis}, see Sec.~\ref{par:measure} and App.~\ref{app:STPreAnalysis})
\begin{lstlisting}[extendedchars=true,mathescape=true,language=Mathematica]
(* Jacobian between the d[Log[$\textcolor{gray}{x}$]] and  d[x] measures *)
jac = (Times @@ xvars)
(* Tropical analysis *)
STPreAnalysis[jac integrand /.{J1->1+eps, J2->2+eps} 
                            /. FactorCompletely2[a_, b_, c_] :> a,
              xvars, coeffs]
\end{lstlisting}
returns \texttt{trops~->~\{\}}, confirming that no divergences are present. In contrast, setting, e.g., $J_1=0$ produces a non-empty \texttt{trops} list with entries proportional to $\varepsilon$, signaling a divergence regulated by the dimensional regulator alone. (A more severe situation, where an additional regulator beyond $\varepsilon$ is required, is encountered in the small-$x$ example discussed below.)

\paragraph{Expansion and integration.}
Since no tropical subtraction is needed around  $J_1=1$ and $J_2=2$ (and, in fact, around any strictly positive integers), we can expand the integrand directly in $\varepsilon$ and in the small deformation parameters $\delta J_1 = J_1 - 1$, $\delta J_2 = J_2 - 2$ before integrating:
\newpage
\begin{lstlisting}[extendedchars=true,mathescape=true,language=Mathematica]
(* Expand around J1 = 1, J2 = 2 *)
seriesJ1J2 = Series[integrand /. {J1 $\to$ 1 - $\delta$J1, J2 $\to$ 2 - $\delta$J2},
                    {eps, 0, 0}, {$\delta$J1, 0, 1}, {$\delta$J2, 0, 1}]
\end{lstlisting}
The triple expansion is then passed to \texttt{STIntegrate} order by order in $(\varepsilon, \delta J_1, \delta J_2)$.

\paragraph{Result.} We find that the dependence on $y_1$ and $y_2$ cancels entirely, and the result depends only on the detector-detector angle $z$:
\begin{align}\label{eq:eec-result}
&\mathrm{EEC}^{\mathrm{real}}_{J_1,J_2}\big|_{\substack{J_1\to 1 \\ J_2\to 2}} = \frac{E^{-4-4\varepsilon}}{16(2\pi)^{5-4\varepsilon}}\frac{8\,E^4\,\Delta(z,y_1,y_2)}{z(1{-}z)\,y_1\,y_2\,(1{-}y_1)(1{-}y_2)}\Big[\frac{1}{2}+\frac{3}{4}(1{-}J_1) 
\\&+ \frac{(2{-}J_2)(1{-}z)\big(z+(1{-}z)\log(1-z)\big)}{2z^2}+\frac{(1{-}J_1)(2{-}J_2)}{4z^2}\Big[\big(1{-}4z{+}3z^2\big)\log(1{-}z)\notag
\\&+ z\big(3{-}3z{-}2z\,\zeta_2\big) + (4z{-}2)\,\mathrm{Li}_2(z)\Big] + \mathcal{O}\big(\delta J_i \delta J_j\big)\Big]\,,\notag
\end{align}
up to the overall prefactor in~\eqref{eq:EECreal}.

\subsubsection{Fourier transforms in small-\texorpdfstring{$x$}{x} physics}\label{sec:smallx}
We first turn our attention to dimensionally regulated Fourier transforms arising in small-$x$ physics. In the scattering of a high-energy quark-antiquark dipole off a hadronic target (e.g., a proton or a nucleus), there exists a frame in which the target appears Lorentz-contracted to a thin null sheet, so that the non-trivial QCD dynamics between the dipole and the target is confined to the $\D=2-2\varepsilon$ dimensional Euclidean transverse plane. A particularly interesting integral arises in the calculation of real-virtual corrections \cite[Eq.~(B.9)]{Brunello:2025rhh} (see also \cite{Balitsky:2007feb}):
\begin{align}
I = \frac{\text{e}^{-2\varepsilon\gamma_E}}{\pi^{\D}} \int \frac{\d^\D \ell_1\, \d^\D \ell_2}{\ell_1^2(\ell_1-\ell_2)^2}\, \log\!\left(\frac{\ell_1^2}{\ell_2^2}\right)\, \text{e}^{i(\ell_1\cdot X_1 + \ell_2\cdot X_2)}\,,
\label{eq:smallx}
\end{align}
where $X_i$ and $\ell_i$ denote transverse positions (relative to the parent dipole) and momenta of the real and virtual emissions, respectively.

Physics aside, the presence of the logarithm in \eqref{eq:smallx} makes it naively difficult to achieve a standard parametric form that \texttt{SubTropica} could try to evaluate directly. A natural solution is to introduce a new regulator $q$, so that the $q$-series of
\begin{align}
F_q = \frac{\text{e}^{-2\varepsilon\gamma_E}}{\pi^{\D}}\int \frac{\d^\D \ell_1\, \d^\D \ell_2}{(\ell_2^2)^{1-q}(\ell_1-\ell_2)^2(\ell_2^2)^q}\, \text{e}^{i(\ell_1\cdot X_1 + \ell_2\cdot X_2)}\,,
\end{align}
reproduces the original integral at $\mathcal{O}(q)$. After using the usual Schwinger trick to put all the momentum dependence of the integrand in a single exponential, we can perform the loop integrals with standard Gaussian integration tricks and obtain the parametric form (given $X_{ij}\equiv X_i\cdot X_j$)
\begin{align}\label{eq:Fq}
F_q = \frac{\text{e}^{-2\varepsilon\gamma_E}\,\Gamma(\D-2)}{\Gamma(q)\,\Gamma(1-q)} \int &\frac{\d^3x}{\mathrm{GL}(1)}\, x_1^{-q}\,x_3^{q-1}\, (x_1x_2+x_1x_3+x_2x_3)^{\D/2-2} 
\\&
\times \, \big[(x_2+x_3)\, X_{11} + 2x_2\, X_{12} + (x_1+x_2)\, X_{22}\big]^{2-\D}\,.\notag
\end{align}

\paragraph{Tropical analysis.}
If one naively expands the integrand \eqref{eq:Fq} in $q$ and passes it to \texttt{STIntegrate} (e.g., as we did earlier for $J_1$ and $J_2$ in the EEC example), the computation is aborted internally because a $q$-regulated divergence at the integration contour endpoint is not regulated by the (default) regulator $\varepsilon$.
Once again, to diagnose this \emph{a priori}, we use the lower-level function \texttt{STPreAnalysis} (see App.~\ref{app:STPreAnalysis}). A first call,
\begin{lstlisting}[extendedchars=true,mathescape=true,language=Mathematica]
(* Jacobian between the d[Log[$\textcolor{gray}{x}$]] and  d[x] measures *)
jac = (Times @@ xvars)

(* Tropical analysis *)
analysis = STPreAnalysis[jac integrand, xvars, coeffs]
\end{lstlisting}
returns \texttt{trops -> \{eps\}} with a single divergent ray: the tropical analysis only sees the $\varepsilon$-regulated divergence and is blind to the singularities involving $q$. To reveal the full singularity structure, we promote $q$ to an $\varepsilon$-dependent regulator via, e.g., \texttt{q -> q eps}:
\begin{lstlisting}[extendedchars=true,mathescape=true,language=Mathematica]
analysis = STPreAnalysis[jac integrand /. q -> q eps, xvars, coeffs]
\end{lstlisting}
Now \texttt{trops -> \{eps, -eps(-1+q), -eps q\}} with three divergent rays. Recall from \eqref{eq:scaling}, that each ray $\rho=(\rho_1,\rho_2)$ of the Newton polytope probes a boundary of the integration domain via the rescaling $x_i\to x_i\,\lambda^{-\rho_i}$ as $\lambda\to 0$. The associated \texttt{trops} entry gives the leading power of $\lambda$ in the integrand under this rescaling: if it vanishes, the divergence along that ray is unregulated.

To see this concretely, we can use \texttt{STFactor} (see App.~\ref{app:STFactor}) to inspect the $\lambda$-dependence of the integrand along each ray:\footnote{In contrast, the ray $\{-1,-1\}$ ($x_1\to x_1\lambda,\, x_2\to x_2\lambda$) makes the integrand scale as $\lambda^{1-\varepsilon-q}$, which vanishes as $\lambda\to 0$ and thus sources no divergence.}
\begin{lstlisting}[extendedchars=true,mathescape=true,language=Mathematica]
(* Ray {1,1}: x1 -> x1/$\textcolor{gray}{\lambda}$, x2 -> x2/$\textcolor{gray}{\lambda}$ *)
STFactor[jac integrand/. {x1 -> x1/$\lambda$, x2 -> x2/$\lambda$}]
  (* $\textcolor{gray}{\to}$ $\textcolor{gray}{\lambda}^{\textcolor{gray}{q}}$: regulated by q only *)
  
(* Ray {0,1}: x1 fixed, x2 -> x2/$\textcolor{gray}{\lambda}$ *)
STFactor[jac integrand/. {x1 -> x1, x2 -> x2/$\lambda$}]
  (* $\textcolor{gray}{\to}$ $\textcolor{gray}{\lambda}^{\textcolor{gray}{-\varepsilon}}$: regulated by $\textcolor{gray}{\varepsilon}$ *)
  
(* Ray {1,0}: x1 -> x1/$\textcolor{gray}{\lambda}$, x2 fixed *)
STFactor[jac integrand/. {x1 -> x1/$\lambda$, x2 -> x2}]
  (* $\textcolor{gray}{\to}$ $\textcolor{gray}{\lambda}^{\textcolor{gray}{-\varepsilon+q}}$: (already) regulated by $\textcolor{gray}{\varepsilon}$ *)

\end{lstlisting}
Clearly, setting $q=0$, only the entry of \texttt{trops} along ray $\{1,1\}$ vanishes, confirming that this is the only divergence genuinely regulated by $q$. The other rays remain regulated by $\varepsilon$ alone. This tells us that the ray $\{1,1\}$ must be treated separately before expanding in $q$, which is what we discuss next.

\paragraph{Nilsson--Passare continuation.}

The $q$-regulated divergence along ray $\{1,1\}$ is handled by the tropical continuation (see Sec.~\ref{sec:tropical-Cont}), applied via \texttt{STTropicalContinuation} (see App.~\ref{app:STTropicalContinuation}):
\begin{lstlisting}[extendedchars=true,mathescape=true,language=Mathematica]
{{pref1, int1}, {pref2, int2}} = 
    STTropicalContinuation[{{pref,jac integrand}},xvars,{{1,1}}]
    
(* pref1 $\textcolor{gray}{=}$ $\textcolor{gray}{\mathtt{Exp}}$[$\textcolor{gray}{\mathtt{-2}}$ $\textcolor{gray}{\mathtt{eps}}$ $\textcolor{gray}{\mathtt{EulerGamma}}$]($\textcolor{gray}{\mathtt{1}}$ $\textcolor{gray}{\mathtt{+}}$ $\textcolor{gray}{\mathtt{eps}}$) $\textcolor{gray}{\mathtt{Gamma}}$[$\textcolor{gray}{\mathtt{-2}}$ $\textcolor{gray}{\mathtt{eps}}$] $\textcolor{gray}{\frac{\mathtt{Sin}[\pi\, \mathtt{q}]}{\pi\, \mathtt{q}}}$        *)

(* int1  $\textcolor{gray}{=}$ $\textcolor{gray}{\mathtt{x1}^{\mathtt{1}-\mathtt{q}}}$ $\textcolor{gray}{\mathtt{x2}}$ ($\textcolor{gray}{\mathtt{x1}}$ $\textcolor{gray}{\mathtt{+}}$ $\textcolor{gray}{\mathtt{x2}}$)$\textcolor{gray}{(\mathtt{x1}\mathtt{+}\mathtt{x2}\mathtt{+}\mathtt{x1\; x2})^{-\mathtt{2}-\mathtt{eps}}}$
            $\textcolor{gray}{(\mathtt{x1}\mathtt{+}\mathtt{x2}\mathtt{-}\mathtt{x2\; z}\mathtt{-}\mathtt{x2\; zb}\mathtt{+}\mathtt{z\; zb}\mathtt{+}\mathtt{x2\; z\; zb})^{\mathtt{2\; eps}}}$ *)

(* pref2 $\textcolor{gray}{=}$ $\textcolor{gray}{\mathtt{Exp}}$[$\textcolor{gray}{\mathtt{-2}}$ $\textcolor{gray}{\mathtt{eps}}$ $\textcolor{gray}{\mathtt{EulerGamma}}$]$\textcolor{gray}{\mathtt{z}}$ $\textcolor{gray}{\mathtt{zb}}$ $\textcolor{gray}{\mathtt{Gamma}}$[$\textcolor{gray}{\mathtt{1}}$ $\textcolor{gray}{\mathtt{-}}$ $\textcolor{gray}{\mathtt{2}}$ $\textcolor{gray}{\mathtt{eps}}$] $\textcolor{gray}{\frac{\mathtt{Sin}[\pi\, \mathtt{q}]}{\pi\, \mathtt{q}}}$          *)
(* int2  $\textcolor{gray}{=}$ $\textcolor{gray}{\mathtt{x1}^{\mathtt{1}-\mathtt{q}}}$ $\textcolor{gray}{\mathtt{x2}}$$\textcolor{gray}{(\mathtt{x1}\mathtt{+}\mathtt{x2}\mathtt{+}\mathtt{x1\; x2})^{-\mathtt{1}-\mathtt{eps}}}$
            $\textcolor{gray}{(\mathtt{x1}\mathtt{+}\mathtt{x2}\mathtt{-}\mathtt{x2\; z}\mathtt{-}\mathtt{x2\; zb}\mathtt{+}\mathtt{z\; zb}\mathtt{+}\mathtt{x2\; z\; zb})^{-\mathtt{1}+\mathtt{2\; eps}}}$ *)
\end{lstlisting}
The output is a list of two \texttt{\{prefactor, integrand\}} pairs.

The continuation extracts the $q$-pole along ray $\{1,1\}$ and decomposes the original integral into a sum of terms where the singular $q$-dependence (including the $\Gamma(q)\Gamma(1-q)$ poles from~\eqref{eq:Fq}) is isolated in the prefactors. In both resulting integrands, the only remaining $q$-dependence is a factor $x_1^{1-q}$ in the $\d\log$ measure, i.e., $x_1^{-q}$ in the standard $\d x$ measure (see Sec.~\ref{par:measure}). One can verify that this does not introduce any new unregulated divergence by running \texttt{STPreAnalysis} on the two integrands: all \texttt{trops} entries are nonzero at $q=0$ (or absent altogether for the second term) with no new divergent ray detected, confirming that every divergence is regulated by $\varepsilon$ alone.

\paragraph{Expansion in $q$ and tropical subtraction in $\varepsilon$.}

With the $q$-pole now isolated in the prefactors, we extract the $\mathcal{O}(q)$ coefficient of each term via \texttt{SeriesCoefficient}, recovering the logarithmic integrand of the original problem~\eqref{eq:smallx} and obtaining $q$-independent integrands that still carry $\varepsilon$-divergences. These are then passed to \texttt{STExpandIntegral}, which performs the standard tropical subtraction to render them locally finite in $\varepsilon$:
\begin{lstlisting}[extendedchars=true,mathescape=true,language=Mathematica]
(* Extract O(q) coefficient of each term *)
int1 = SeriesCoefficient[pref1 int1, {q, 0, 1}];
int2 = SeriesCoefficient[pref2 int2, {q, 0, 1}];

(* Tropical subtraction in eps *)
serI1 = STExpandIntegral[int1, xvars, coeffs]
serI2 = STExpandIntegral[int2, xvars, coeffs]
\end{lstlisting}
The locally finite integrands in \texttt{serI1} and \texttt{serI2} are then integrated order-by-order in $\varepsilon$ using \texttt{HyperIntica} (see \texttt{PaperChecks.wl} for details).

The raw result is a lengthy expression of hyperlogarithms. After simplification via symbol and constant matching (see \texttt{PaperChecks.wl} for the detailed procedure), the result takes the compact form (in the region where $(z-\overline{z})^2<0$ enforcing that $\overline{z}=z^*$ or, equivalently the Cauchy–Schwarz inequality at the level of the transverse distances squared $X_{12}^4\le X_{11}^2 X_{22}^2$ in \eqref{eq:smallx})
\begin{equation}
\begin{split}
    I &= \frac{1}{2\varepsilon^3}+\frac{\zeta_2}{2\varepsilon} + 4\,\mathrm{Re}\!\left[\mathrm{Li}_3(1{-}z)\right] + 4\,\mathrm{Re}\!\left[\log(1{-}z)\right]\mathrm{Re}\!\left[\mathrm{Li}_2(z)\right] \\& + 4\,\mathrm{Re}\!\left[\log(1{-}z)\right]\!\left(\mathrm{Re}\!\left[\log(1{-}z)\log z\right] - \zeta_2\right) + \frac{\zeta_3}{3} + \mathcal{O}(\varepsilon)\,.
    \end{split}
\end{equation}
We checked that this result agrees perfectly with the one computed using differential equations in \cite{Brunello:2025rhh}.

{\subsection{A standalone {\texttt{HyperIntica}} example}\label{sec:hyperInticaEx}}

Finally, we demonstrate \texttt{HyperIntica} as a standalone integration engine, independent of \texttt{SubTropica}'s Feynman integral pipeline. Consider the five-variable integral with a cubic logarithmic numerator:\footnote{We would like to thank Vladimir A. Smirnov for suggesting this example to us.}
\begin{lstlisting}[extendedchars=true,mathescape=true,language=Mathematica]
num = t2$^{\mathtt{2}}$ t3$^{\mathtt{4}}$ t4$^{\mathtt{2}}$ t5
  (3 Log[t1] + 5 Log[t2] - 87 Log[t3] - 17 Log[t1+t2+t3] - 12 Log[t4]
   + 19 Log[t5] + 9 Log[1+t5] + 24 Log[(1+t1+t2) t3 + t2 t4 t5]
   + 50 Log[(1+t1+t2) t3+t2 t4 (1+t5)] - 3 Log[t1 t3+(t2+t3)(t3+t4 t5)])$^{\mathtt{3}}$;

den = 6 (t1+t2+t3)$^{\mathtt{2}}$ (1+t5)((1+t1+t2) t3 + t2 t4 t5)$^{\mathtt{2}}$
        ((1+t1+t2) t3 + t2 t4 (1+t5))$^{\mathtt{2}}$(t1 t3 + (t2+t3)(t3 + t4 t5))$^{\mathtt{2}}$;

integrand = $\frac{\mathtt{num}}{\mathtt{den}}$;
\end{lstlisting}
This integrand is a hyperlogarithm-valued rational function in the five variables $t_1,\ldots,t_5$, and its integral over $[0,\infty)^5$ is a period (i.e., a number). Linear reducibility requires that the set of irreducible polynomial factors remains linear in the next integration variable at every step. Here the relevant polynomials are the non-monomial arguments of the logarithms and the irreducible factors of the denominator. Since these coincide, the polynomial set is
\begin{lstlisting}[extendedchars=true,mathescape=true,language=Mathematica]
polys = {t1+t2+t3, 1+t5, (1+t1+t2) t3 + t2 t4 t5,
         (1+t1+t2) t3 + t2 t4 (1+t5), t1 t3 + (t2+t3)(t3 + t4 t5)};
\end{lstlisting}
Passing these (together with the variables) to \texttt{STFasterFubini} (App.~\ref{app:STFasterFubini}) finds a linearly reducible order:
\begin{lstlisting}[extendedchars=true,mathescape=true,language=Mathematica]
STFasterFubini[{Join[polys, {t1,t2,t3,t4,t5}]},
                 {t1,t2,t3,t4,t5}]
(* Out: {t4, t5, t1, t3, t2} *)
\end{lstlisting}
The integral is then evaluated by passing it directly to \texttt{HyperIntica}:
\begin{lstlisting}[extendedchars=true,mathescape=true,language=Mathematica]
HyperIntica[integrand, {t4, t5, t1, t3, t2}, "Monitor" -> True]
\end{lstlisting}
It is worth mentioning that for this integral, the built-in heuristic in \texttt{STFasterFubini} (see Sec.~\ref{sec:linred}) does \emph{not} return the fastest integration order, which turns out to be \texttt{\{t4, t5, t1, t2, t3\}} (note the swapped last two variables). This illustrates the difficulty of predicting the optimal order from compatibility graph reduction and the heuristic implemented in \texttt{SubTropica} alone; we return to this point in Sec.~\ref{sec:Conclusion}.

For both orders, the computation processes each variable sequentially
(approximately 6300 seconds total on a laptop for the optimal order)
and returns the same linear combination of multiple zeta values:
\begin{align}
\int_{[0,\infty)^5}\!\text{\texttt{integrand}}\;\d^5 t \;=\;& \tfrac{744301}{8} + \tfrac{67803}{2}\,\zeta_2 - \tfrac{1375959}{20}\,\zeta_2^2 + \tfrac{1234511}{15}\,\zeta_2^3 - \tfrac{3889163}{24}\,\zeta_3 \notag\\
&- \tfrac{43941}{2}\,\zeta_2\,\zeta_3 + \tfrac{584512}{5}\,\zeta_2^2\,\zeta_3 + 3344\,\zeta_3^2 - \tfrac{109324}{3}\,\zeta_5 \notag\\
&- 27354\,\zeta_2\,\zeta_5 - \tfrac{1142764}{3}\,\zeta_7\,,
\end{align}

which was cross-verified with \texttt{HyperInt}.

\section{Conclusion}
\label{sec:Conclusion}

We have presented \texttt{SubTropica}, a \textsc{Mathematica} package for the automatic evaluation of linearly reducible Euler integrals via tropical subtractions. The package combines the tropical subtraction algorithm of~\cite{Salvatori:2024nva} with \texttt{HyperIntica}, a self-contained reimplementation of~\cite{Panzer:2014caa}, providing an end-to-end workflow entirely within \textsc{Mathematica}. As demonstrated through examples ranging from multi-loop Feynman integrals to
phase-space integrals, Fourier transforms in high-energy QCD, and energy correlators, the package achieves a high level of automation while remaining accessible through a graphical interface. We have also released the \texttt{SubTropica} library, a repository of Feynman integrals computed in the literature together with their results.

Several directions for improvement and extension are worth mentioning.

\paragraph{Extending the domain of physical applications.}
While a large number of physical computations are naturally formulated in terms of Euler integrals already, we expect that the class of applications can be enlarged further. For instance, phase-space integrals have, in some cases, been rewritten in this form, see, e.g.,~\cite{Anastasiou:2013srw,Herzog:2018ily,Smirnov:2024pbj}. 
Other examples include correlation functions for large-scale structure in cosmology, which can often be brought into Euler integral form~\cite{Simonovic:2017mhp} (see also \cite{Herderschee:2025znl}).
With or without \texttt{SubTropica}, being able to achieve this systematically would unlock a whole new class of computational tools. We leave it as a
challenge to the community.

\paragraph{Improving tropical subtraction schemes.} Recall from Sec.~\ref{sec:tropical-Cont} that one of the current bottlenecks is the need for Nilsson--Passare continuation whenever the integrand does not satisfy the geometric property. Circumventing this limitation would yield enormous speedups for certain classes of integrals, and hence should be explored more thoroughly. As an encouraging example, Hillman's tropical subtraction formula~\cite{Hillman:2023ezp} achieves this
objective for integrals with UV divergences. A concrete goal would be to extend it to IR divergences as well.

\paragraph{{\texttt{C++}} backend.} Another bottleneck, specific to \textsc{Mathematica}, is its slow implementation of polynomial operations such as \texttt{Series}, \texttt{Apart}, \texttt{Together}, and \texttt{Factor}. The clear path forward is to introduce a \texttt{C++} backend implementing these functions. We plan to pursue this in the immediate future.

\paragraph{Parallelization.} For every face of the Newton polytope, \texttt{HyperIntica} is launched in parallel on multiple integrals. In practice, we find that a small subset of integrals consumes most of the compute. A better strategy would weight integrals by a predicted-complexity estimate to balance the load. Additionally, it would be desirable to explore possible parallelization of the standalone package \verb|HyperIntica|. 

\paragraph{Extending the class of integrals.} The requirement of linear reducibility, while satisfied by a large class of physically relevant integrals, is a genuine limitation. For example, transverse $\d z\wedge \d \overline{z}$ integrals in, e.g., high-energy QCD \cite{Caron-Huot:2016tzz} would require \texttt{HyperIntica} to account systematically for holomorphic anomalies. Moreover, although \texttt{SubTropica} deals with some integrals involving algebraic (e.g., square-root) letters, automatic detection of such features is not yet systematic. On the other hand, algorithmic methods of rationalization do exist, see, e.g.,~\cite{Bogner:2007cr}. We have implemented prototype AI-assisted
strategies to address this problem, which will be featured in a future release of \texttt{SubTropica}.

\paragraph{Finding linear orders.} Even though we have found simple-to-compute heuristic metrics that correlate reasonably well with the eventual integration time (see Sec.~\ref{sec:orders}), the predictor still has an enormous variance, spanning three orders of magnitude. In practice, this means a single integral can occasionally dominate the wall-clock budget, clogging the pipeline. More reliable ways of finding linearly-reducible orders, perhaps AI-assisted, therefore remain a
  priority.

\paragraph{Extensions beyond hyperlogarithms?}
The tropical subtraction scheme is completely general and it can be applied to any Euler integral, regardless of whether it is multi-polylogarithmic or not. From this point of view, it can be used in conjunction with any future direct-integration software, e.g., for elliptic iterated integrals. We look forward to seeing progress in this direction.

\paragraph{Expansion by regions.} A natural extension of \texttt{SubTropica} would be to implement the expansion by regions (see, e.g.,~\cite{Jantzen:2012mw}), which is already formulated in terms of polyhedral geometry. We intend to add this feature in a future release.

\paragraph{Crown problems.} The issue of non-positive integrands mentioned in Sec.~\ref{subsec:positivity} is not only of practical concern, but also tied to a deep problem of defining perturbative quantum field theory in the first place. It is not entirely clear that in such cases Feynman integrals are analytic, or, more precisely, boundary values of a single analytic function~\cite{Hannesdottir:2022bmo}. It is plausible that one is forced to split the integral into multiple analytic functions, each with a different kinematic $i\epsilon$ prescription. Curiously, one of the simplest examples of this problem, the crown diagram (Fig.~\ref{fig:crown}), belongs to a master-integral family computed in~\cite{Henn:2020lye,Bargiela:2021wuy}, suggesting that differential equations might serve as a guide for understanding the issue of non-positive Feynman integrals.

\paragraph{\texttt{SubTropica} as an end-to-end tool for scattering amplitudes.} With a simple click-to-integrate approach, we demonstrated that one does not need to be an expert on Feynman integrals to be able to compute them. There is no reason to stop here. In the future, we plan to extend this approach to scattering amplitudes wholesale, streamlining the full pipeline from the Lagrangian to the observable. We imagine a future in which every computation in quantum field theory will be only a single \texttt{Shift+Enter} away.

\paragraph{Acknowledgments.} We thank Luca Abu El-Haj, Wayne Ariston, Miguel Correia, Giulio Crisanti, Einan Gardi, Johannes Henn, Won Lim, Ian Moult, Audrey Pepin, Vladimir Smirnov, John Staunton, Zehao Zhu for discussions, suggesting example applications, feature requests, and beta-testing. M.G. is supported by the U.S. Department of Energy (DOE) grant No. DE-SC0011941. The work of G.S. is also supported in part by the DOE (Grant No. DE-SC0009988), further support was made possible by the Carl B. Feinberg cross-disciplinary program in innovation at the IAS. The work of G.S. is part of the PositiveWorld project funded by the European Union’s Horizon 2023
research and innovation programme under the Marie Sk\l{}odowska-Curie grant agreement 101151760. Funded by the European Union (ERC, UNIVERSE PLUS, 101118787).

{\footnotesize{
Views and opinions of the authors expressed are those of the author(s) only and do not necessarily reflect those of the European Union or the European Research Council Executive Agency. Neither the European Union nor the granting authority can be held responsible for them.
}}

\appendix
\section{\label{app:functions}Function reference}

\subsection{\label{app:Config-opts}Detailed package configuration: \texttt{ConfigureSubTropica}}

The full list of possible options supplied to  \texttt{ConfigureSubTropica} is:
\begin{lstlisting}[extendedchars=true,language=Mathematica,literate={`}{{\textasciigrave}}1]
  ConfigureSubTropica[
      (* required *)
      PolymakePath                -> "path/to/polymake",
      
      (* polynomial-arithmetic backend (recommended) *)
      FiniteFlowPath              -> "path/to/finiteflow/mathlink",
      SPQRPath                    -> "path/to/SPQR",

      (* numerical verification backends (optional) *)
      GinshPath                   -> "path/to/ginsh",
      FIESTAPath                  -> "path/to/FIESTA5",
      AMFlowPath                  -> "path/to/amflow",
      FeyntropPath                -> "path/to/feyntrop",
      PythonPath                  -> "path/to/python3",

      (* symbolic integration (optional) *)
      MaplePath                   -> "path/to/maple",
      HyperIntPath                -> "path/to/HyperInt.mpl",

      (* IBP reduction (optional, used by AMFlow) *)
      FIREPath                    -> "path/to/FIRE6",
      LiteRedPath                 -> "path/to/LiteRed",
      LiteIBPPath                 -> "path/to/finiteflow-mathtools/packages",

      (* tuning *)
      PolymakeConcurrencyFraction -> 0.75,
      BenchmarkNudge              -> True
  ];
  \end{lstlisting}
  
The options of \texttt{ConfigureSubTropica} listed above fall into three
  groups. The first group consists of paths to tools already introduced in
  the main text: \texttt{PolymakePath}, \texttt{GinshPath},
  \texttt{MaplePath}, \texttt{PythonPath}, and the pair
  \texttt{FiniteFlowPath} and \texttt{SPQRPath}. Two of these options deserve a
  brief comment. \texttt{HyperIntPath} points at \texttt{HyperInt.mpl}
  file itself rather than a containing directory; leaving this option empty
  triggers auto-discovery under \texttt{\textasciitilde/HyperInt},
  \texttt{\textasciitilde/hyperint}, and \texttt{/opt/HyperInt}.
  \texttt{PythonPath} selects a specific Python \emph{interpreter}, not a
  package root: the canonical use case is a virtual environment whose
  \texttt{site-packages} directory contains \texttt{pySecDec}.

  The second group configures the numerical verification backends invoked by
  \texttt{STVerify} and by the GUI validator. In addition to \texttt{FIESTAPath},
  which points at the directory containing \texttt{FIESTA5.m}, one may set
  \texttt{AMFlowPath} and \texttt{FeyntropPath}, which enable two further
  backends: \texttt{AMFlow} \cite{Liu:2022chg}, based on auxiliary mass flow,
  and \texttt{feyntrop} \cite{Borinsky:2023jdv}, based on tropical Monte--Carlo
  sampling. \texttt{AMFlowPath} points at the root directory of
  \texttt{AMFlow} (containing \texttt{AMFlow.m}), while \texttt{FeyntropPath}
  points at the directory holding the compiled \texttt{feyntrop} binary.
  Because \texttt{AMFlow} delegates its IBP-reduction step to an external
  reducer, three further path options are provided: \texttt{FIREPath},
  \texttt{LiteRedPath}, and \texttt{LiteIBPPath}, which locate
  \texttt{FIRE6} \cite{Smirnov:2023yhb}, \texttt{LiteRed}
  \cite{Lee:2013mka}, and \texttt{finiteflow-mathtools}
  \cite{Peraro:2019svx} package, respectively. When more than one reducer is
  available, \texttt{SubTropica} dispatches automatically.

  The third group consists of two runtime knobs.
  \texttt{PolymakeConcurrencyFraction} (default \texttt{0.75}) caps the
  fraction of the available CPU cores consumed by concurrent \texttt{polymake}
  jobs, and the boolean \texttt{BenchmarkNudge} (default \texttt{True})
  toggles a one-line reminder in the welcome banner suggesting a call to
  \texttt{STBenchmark[]} on first use of a fresh installation. All options
  discussed above are auto-persisted to the file \texttt{\$STConfigFile} and
  replayed on subsequent loads, so \texttt{ConfigureSubTropica} need only be
  called once per machine, or whenever a dependency is moved or upgraded.

\subsection{\label{app:STIntegrate-opts}Complete option reference: \texttt{STIntegrate}}

\paragraph{Option reference.}

The available options, grouped by purpose, are:
\begin{lstlisting}[extendedchars=true,mathescape=true,language=Mathematica]
	STIntegrate[input,
	(* A: Core settings *)
	"Order"                     -> Automatic,
	"Dimension"                 -> 4 - 2 eps,
	"Normalization"             -> Automatic,
	"Substitutions"             -> {},
	"Representation"            -> "Schwinger",
	"Integrator"                -> "HyperIntica",
	(* A': Propagator input only *)
	"Exponents"                 -> Automatic,
	"LoopMomenta"               -> Automatic,
	(* B: Gauge selection and scoring *)
	"Gauge"                     -> Automatic,
	"IncludeGauges"             -> All,
	"Heuristic"                 -> "LeafCountLinear",
	"ScanScoreInterval"         -> {1, 3},
	"ScoreInParallel"           -> All,
	"TimeUpperBound"            -> $10^{17}$,
	"ScoringMemoryFraction"     -> 0.5,
	"MethodLR"                  -> "Lungo",
	"MethodPolysAndPairs"       -> "Fast",
	"FindRoots"                 -> True,
	(* C: Parallelization and memory *)
	"KernelsAvailable"          -> $\$$ProcessorCount - 1,
	"ClearCachesPerIntegrand"   -> False,
	(* D: Output and diagnostics *)
	"Verbose"                   -> False,
	"ShowTimings"               -> True,
	"ShowIntegrands"            -> False,
	"SaveSlowestIntegrand"      -> False,
	"SetProblemID"              -> Automatic,
	"SimplifyOutput"            -> Simplify,
	"CleanOutput"               -> False,
	"ContourHandling"           -> "Abort",
	(* E: Checkpointing and workflow *)
	"StopAt"                    -> Automatic,
	"StartAt"                   -> None,
	"SelectFaces"               -> All,
	"ReuseExistingResults"      -> True]
\end{lstlisting}
This list, as well as all the values each option can take can be called in a session using \texttt{STOptionValues[]}.

\paragraph{A: Core settings.}

\texttt{"Order" -> Automatic} (default) returns the result through the finite part, i.e.,\ through $\mathcal{O}(\varepsilon^0)$. Setting \texttt{"Order" -> n} returns the result through $\mathcal{O}(\varepsilon^n)$. The required integration depth is determined automatically from the pole structure of the overall prefactor (e.g.,\ from $\Gamma(\varepsilon)$ factors), so the user never needs to worry about how many orders \texttt{HyperIntica} must compute internally not to miss $\varepsilon/\varepsilon$ contributions. \hfill \decosix

\texttt{"Dimension"} sets the spacetime dimension, e.g.\ \texttt{"Dimension" -> 6 - 2*eps}. \texttt{"Normalization" -> Automatic} includes the standard $\mathrm{e}^{L\gamma_E\varepsilon}$ prefactor for an $L$-loop integral. \texttt{"Substitutions"} accepts replacement rules applied to kinematic invariants before integration, e.g.\ \texttt{\{s12 -> 1, mm -> 0\}}. \texttt{"Representation"} selects \texttt{"Schwinger"} or \texttt{"LeePomeransky"} \cite{Lee:2013hzt}, where the Schwinger representation typically yields simpler polynomial structures and thus simpler integration. \texttt{"Integrator"} selects the symbolic integration backend: \texttt{"HyperIntica"} (default) uses the built-in \textsc{Mathematica} implementation, while \texttt{"HyperInt"} delegates to the \texttt{Maple} implementation of~\cite{Panzer:2014caa} (requires \texttt{MaplePath} and \texttt{HyperIntPath} to be set via \texttt{ConfigureSubTropica}). \hfill \decosix

\paragraph{A$'$: Propagator input only.}

\texttt{"Exponents" -> Automatic} assigns unit exponent to every propagator by default. Setting \texttt{"Exponents" -> \{$n_1$, $n_2$, \ldots\}} assigns the exponent $n_i$ to the $i^{\text{th}}$ entry. Negative values insert the corresponding polynomial as a (possibly tensor) numerator factor. 
\hfill \decosix

\begin{figure}
	\centering
	\begin{tikzpicture}[
    >=Latex,
    node distance=0.7cm,
    every node/.style={font=\small},
    stepbox/.style={
        draw=Maroon!55!black,
        fill=Maroon!7,
        rounded corners=5pt,
        minimum width=6.2cm,
        minimum height=1.05cm,
        align=center,
        line width=0.55pt
    },
    intbox/.style={
        draw=Maroon!55!black,
        fill=Maroon!7,
        rounded corners=5pt,
        minimum width=6.2cm,
        minimum height=1.4cm,
        align=center,
        line width=0.55pt
    },
    port/.style={
        draw=RoyalBlue!80!black,
        fill=RoyalBlue!22,
        circle,
        minimum size=7mm,
        inner sep=0pt,
        font=\footnotesize\bfseries
    },
    inoutbox/.style={
        draw=black!35,
        fill=black!4,
        rounded corners=3pt,
        minimum width=5cm,
        minimum height=0.8cm,
        align=center
    },
    sidelab/.style={
        draw=RoyalBlue!65!black,
        fill=RoyalBlue!10,
        rounded corners=3pt,
        minimum width=4.4cm,
        minimum height=0.75cm,
        align=center
    }
]

\node[stepbox] (s1) {\texttt{STExpandIntegral} $+$ \texttt{STsetupDirectoryExpansion}:\\[-1pt] tropical subtraction and directory setup};
\node[port, below=of s1] (check1) {1};
\node[stepbox, below=of check1] (s2) {\texttt{STfindLinearlyReducibleOrders}: find linearly reducible orders};
\node[port, below=of s2] (check2) {2};
\node[intbox, below=of check2] (s3) {
    \texttt{STLaunchHyperIntica}: integrate faces\\[3pt]
    {\footnotesize\color{RoyalBlue!70!black}\texttt{SelectFaces},\quad\texttt{ReuseExistingResults}}
};

\node[inoutbox, above=1.2cm of s1] (input) {Input to \texttt{STIntegrate[...]}: \\[2pt] Feynman diagram, propagator list or Euler integrand \eqref{eq:defeuler}};
\node[inoutbox, below=1.2cm of s3] (output) {Output: Laurent expansion \eqref{eq:epsexpansion}};

\draw[->, line width=0.8pt] (input) -- (s1);
\draw[->, line width=0.8pt] (s1) -- (check1);
\draw[->, line width=0.8pt] (check1) -- (s2);
\draw[->, line width=0.8pt] (s2) -- (check2);
\draw[->, line width=0.8pt] (check2) -- (s3);
\draw[->, line width=0.8pt] (s3) -- (output);

\node[sidelab, right=1.2cm of check1] (p1) {\texttt{"AfterExpansion"}\\[2pt]\footnotesize locally finite expansion \eqref{eq:expansion}};
\node[sidelab, right=1.2cm of check2] (p2) {\texttt{"AfterLinearOrder"}\\[2pt]\footnotesize integration orders found};

\draw[<->, dashed, draw=gray!60] (check1) -- (p1);
\draw[<->, dashed, draw=gray!60] (check2) -- (p2);

\draw[->, dashed, draw=RoyalBlue!70!black, line width=0.6pt]
    (s3.west)
    to[out=180, in=180, looseness=1.8]
    node[left, align=right, font=\scriptsize, text=RoyalBlue!70!black,xshift=-10] {\large$\substack{\text{partial}\\\text{run}}$}
    (check2.west);

\end{tikzpicture}
	\caption{The algorithmic steps of \texttt{STIntegrate} described in Sec.~\ref{sec:Preliminary}. Numbered circles mark the two checkpoints where the pipeline can be paused via \texttt{"StopAt"} and resumed later via \texttt{"StartAt"}. The single-headed dashed feedback arrow indicates that the integration step can be run multiple times with different \texttt{"SelectFaces"} selections, with completed faces reused via \texttt{"ReuseExistingResults"}.}
	\label{fig:flow}
\end{figure}
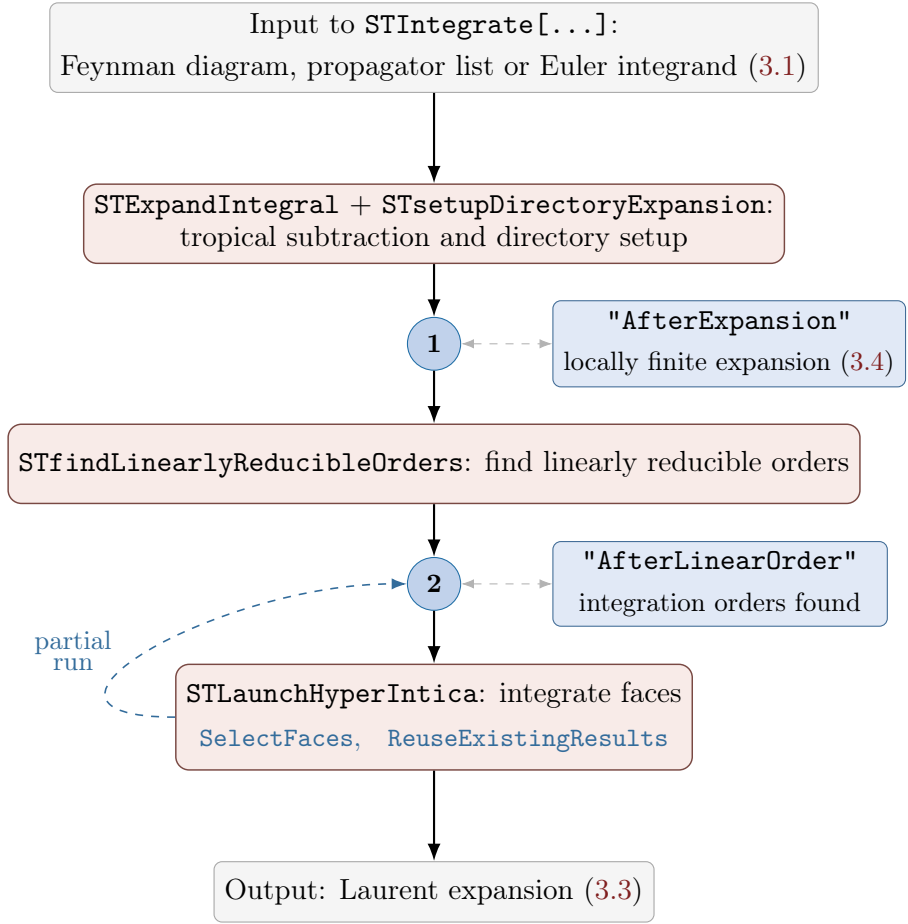

The default option \texttt{"LoopMomenta" -> Automatic} assumes that the loop momenta are labeled by \texttt{\{l[1], l[2], ... \}} in the propagator list. The user can set the option to e.g., \texttt{\{k[1], k[2], ... \}}, to specify them manually. \hfill \decosix

\paragraph{\label{par:B}B: Gauge selection and scoring.}
\texttt{SubTropica} automatically detects whether the input integrand is homogeneous in the integration  variables. When it is (e.g., for integrands arising from the Schwinger representation of a Feynman graph),  the code assumes an overall $\mathrm{GL}(1)$ redundancy and searches for the gauge choice $x_i = 1$ that  yields the simplest integration (according to the scoring metric set by \texttt{"Heuristic"}, which is discussed below).  

\texttt{"Gauge" -> Automatic} (default) scans all candidate gauges, scores them, and selects the best one, while \texttt{"Gauge" -> \{x3 -> 1\}} fixes the gauge directly. \texttt{"IncludeGauges" -> \{1,2,5\}} restricts the automatic search to $x_1=1$, $x_2=1$, $x_5=1$, whereas \texttt{All} (default) considers every Feynman  parameter. 

Note that if the integrand is \emph{not} homogeneous (e.g., the Lee--Pomeransky representation of a Feynman graph),  \texttt{SubTropica} detects this automatically and integrates over all parameters without gauge fixing. When necessary, this behavior can also be forced manually even on homogeneous integrands via \texttt{"Gauge" -> \{\}}.  \hfill \decosix

\texttt{"Heuristic"} controls how candidate linearly reducible integration orders are ranked. The default \texttt{"LeafCountLinear"} minimizes the sum of leaf counts of intermediate polynomials, while alternatives include \texttt{"ByteCountLinear"}, \texttt{"TermCountLinear"}, \texttt{"TotalDegreeLinear"}, and root-mean-square variants. \hfill \decosix

On complicated examples, gauge scoring can be memory-intensive. \texttt{SubTropica} implements options to spread the memory pressure when needed. The option \texttt{"ScanScoreInterval" -> \{m,n\}} restricts scoring to faces $m$ through $n$, reducing cost. \texttt{"ScoreInParallel" -> k} splits candidates into $k$ parallel chunks, while \texttt{All} (default) processes all gauges in one batch. \texttt{"ScoringMemoryFraction" -> f} ($0 < f \leq 1$, default $0.5$) sets the fraction of available RAM budgeted for parallel scoring, controlling the number of gauges scored concurrently.

The option \texttt{"TimeUpperBound" -> T} sets a per-gauge time limit in seconds and candidates exceeding it are deprioritized.
\hfill \decosix

\texttt{"MethodLR"} selects the algorithm used to find linearly reducible integration orders. The default \texttt{"Lungo"} uses an optimized Fubini reduction algorithm. The alternative \texttt{"Espresso"} combines it with deduplication via proportional-polynomial intersection, which is typically faster but might lead to sub-optimal integration orders. \hfill \decosix

\texttt{"MethodPolysAndPairs"} controls how the polynomial sets used for the linear reducibility check are extracted from the integrands and counter-terms.
 The default \texttt{"Fast"} reads the irreducible polynomial factors directly from the internal monomial-polynomial decomposition (\texttt{STtoCoeffMonPols}) of
  each counter-term integrand \emph{before} expanding in $\varepsilon$, which avoids the cost of assembling and re-factoring the summed integrand at each $\varepsilon$-order.
   The alternative \texttt{"Standard"} first sums all integrands and counter-terms at each $\varepsilon$-order and then extracts the polynomial set via \texttt{STpreparePolysAndPairs},
   which can detect cancellations between counter-terms that reduce the polynomial set. When \texttt{"Fast"} is selected (default) and no linearly reducible order is found, 
   \texttt{STIntegrate} automatically falls back to \texttt{"Standard"} and retries, since the reduced polynomial set relevant up to the specified \texttt{Order} may admit a linearly reducible order that the all-orders 
   \texttt{"Fast"} set does not allow. \hfill \decosix

\texttt{"FindRoots" -> True} (default) attempts to factor univariate polynomials of degree greater than one during the linear-order search by introducing auxiliary root variables (e.g., \texttt{Wm[i]}/\texttt{Wp[i]}), extending the class of linearly reducible integrals at the cost of increased complexity. Setting it to \texttt{False} restricts the search to orders that are linearly reducible without algebraic letters. \hfill \decosix

\paragraph{C: Parallelization and memory.}
\texttt{"KernelsAvailable" -> n} sets the number of sub-kernels (default $\$\mathtt{ProcessorCount}-1$). Setting \texttt{"KernelsAvailable" -> 0} or \texttt{1} runs the entire pipeline in serial on the main kernel: no sub-kernels are launched, and all \texttt{ParallelTable}/\texttt{ParallelMap} calls fall back to their serial equivalents automatically.
Sub-kernels are launched on the first \texttt{STIntegrate} call and reused across subsequent calls in the same session; it is advisable to restart kernels after large runs to release accumulated memory. \hfill \decosix

\texttt{"ClearCachesPerIntegrand" -> False} (default) clears all sub-kernel memoization caches only once, after all counter-term integrands for a given diagram have been integrated, maximizing cache reuse. Setting it to \texttt{True} clears caches after each individual integrand, which reduces peak memory at the cost of losing intermediate integrand cache reuse. \hfill \decosix

When \texttt{FiniteFlow} \cite{Peraro:2019svx} and \texttt{SPQR} \cite{Chestnov:2025svg} are loaded, the potentially costly partial-fraction step inside \texttt{HyperIntica} uses finite-field reconstruction via \texttt{SPQRPolynomialQuotient}, avoiding intermediate expression swell. This introduces an independent layer of parallelism (via \texttt{FiniteFlow}'s internal \texttt{C++} threads) that is non-interfering with \textsc{Mathematica}'s sub-kernel scheduler. The weight range used by the reconstruction is controlled by \texttt{\$PartialFractionsMaxWeight} (default \texttt{\{1,100\}}) and can be overridden per call via \texttt{PartialFractions[f, x, "MaxWeight" -> \{1, 200\}]}. \hfill \decosix

\paragraph{D: Output and diagnostics.}

\texttt{"Verbose" -> True} enables detailed print output throughout the computation. 
\texttt{"ShowTimings" -> True} (default) prints timing for each stage, and \texttt{"ShowIntegrands" -> True} shows a live monitor of the integrand running on each kernel, which is useful for spotting bottlenecks in real time. \hfill \decosix

\texttt{"SaveSlowestIntegrand" -> "file.m"} records the slowest integrand after the run (default \texttt{False}). \hfill \decosix

\texttt{"SetProblemID" -> Automatic} generates a unique run identifier used to name intermediate directories, and can be set to any string. \texttt{"SimplifyOutput" -> f} applies \texttt{f} to the final result. Passing \texttt{Identity} skips simplification entirely. \hfill \decosix

\texttt{"CleanOutput" -> True} applies \texttt{CleanZeroInf[]} to resolve any reducible remaining \texttt{zeroInfPeriod} expressions into explicit multi-zeta values before \texttt{"SimplifyOutput"} is applied. \hfill \decosix

\texttt{"ContourHandling" -> "Abort"} (default) controls how the code responds when the contour direction around a singularity on the real axis cannot be determined unambiguously during the evaluation of $\Hlog(\infty; \ldots)$ limits. The default \texttt{"Abort"} halts the computation with a diagnostic message identifying the problematic letter. Setting \texttt{"ContourHandling" -> "Continue"} instead leaves the offending $\Hlog(\infty; \ldots)$ unevaluated and continues, which can be useful for debugging or when the user intends to resolve the contour prescription externally. \hfill \decosix

\paragraph{E: Checkpointing and workflow.}
For difficult or long-running computations, the pipeline can be paused at intermediate stages and resumed later, and individual tropical faces can be targeted selectively.

\texttt{"StopAt"} interrupts the pipeline and saves an intermediate result to disk as a checkpoint under \texttt{checkpoints/<ID>.m} (where \texttt{<ID>} is the problem identifier set by \texttt{"SetProblemID"}) that can be reloaded later via \texttt{"StartAt"}. The accepted values are \texttt{"AfterBuildingIntegrand"} (returns the raw parametric integrand \texttt{\{prefactor, integrand, variables, coefficients\}}), \texttt{"AfterExpansion"} (returns the list of locally finite integrands from the tropical subtraction without launching \texttt{HyperIntica}), and \texttt{"AfterLinearOrder"} (additionally records the linearly reducible integration orders). The default \texttt{Automatic} runs the full pipeline without saving any checkpoint. The lower-level functions invoked at each stage, \texttt{STExpandIntegral} (at \texttt{"AfterExpansion"}) and \texttt{STFasterFubini} (at \texttt{"AfterLinearOrder"}), are documented in App.~\ref{app:STExpandIntegral} and~\ref{app:STFasterFubini} respectively (see also Fig.~\ref{fig:flow}). The valid values for \texttt{"StopAt"} and all other options with enumerable values can be queried interactively via the helper function \texttt{STOptionValues[]}. For instance, \texttt{STOptionValues["StopAt"]} returns the list of accepted strings. \hfill \decosix

\texttt{"StartAt"} resumes the pipeline from a checkpoint previously saved by \texttt{"StopAt"}. The argument is the checkpoint identifier returned by the earlier call. \hfill \decosix

\texttt{"SelectFaces" -> All} (default) passes all tropical face integrands to \texttt{HyperIntica}. Setting it to a single integer $n$ or a list $\{n_1, n_2, \ldots\}$ restricts integration to the faces with those indices. The syntax \texttt{o -> i} selects face $i$ at $\varepsilon$-order $o$ only, and lists of such pairs are accepted as well. The pattern \texttt{Except[f]} integrates every face not matched by \texttt{f}, where \texttt{f} can be any of the above forms. Face contributions that have not yet been integrated appear as \texttt{STwrapError[\ldots]} placeholders in the assembled output until the missing faces are completed. \hfill \decosix

\texttt{"ReuseExistingResults" -> True} (default) checks, before calling \texttt{HyperIntica} on each face, whether a completed \texttt{result.m} and \texttt{successQ.m} already exist on disk. If so, the face is skipped and its stored result is read back at assembly time. Setting this to \texttt{False} forces every selected face to be (re)integrated regardless of what is on disk. This option is most useful in combination with \texttt{"SelectFaces"} and \texttt{"StartAt"}, as illustrated below. \hfill \decosix

The following example illustrates a typical multi-stage workflow for a diagram ``\texttt{diag}'':
\newpage
\begin{lstlisting}[language=Mathematica]
	(* Stage 1: run tropical subtraction and stop before integration *)
	stage1 = STIntegrate[diag, "StopAt" -> "AfterExpansion"];
	
	(*Stage 2: additionally find linearly reducible integration orders*)
	stage2 = STIntegrate[diag, "StopAt" -> "AfterLinearOrder"];
	
	(*Stage 3: resume from any checkpoint and complete the integration*)
	res = STIntegrate[diag, "StartAt" -> stage1["CheckpointID"]];
	res = STIntegrate[diag, "StartAt" -> stage2["CheckpointID"]];
	
	(* Stages can be combined: stop and resume in a single call *)
	stage2 = STIntegrate[diag, "StartAt" -> stage1["CheckpointID"],
	                            "StopAt" -> "AfterLinearOrder"];
	
	(* To perform the integration from any of the stages, simply run *)
	res = STIntegrate[diag, "StartAt" -> stage2["CheckpointID"]];
\end{lstlisting}
Note that checkpoints persist between different examples and across kernel restarts, since they are written to \texttt{checkpoints/<ID>.m}. \hfill \decosix

\paragraph{Targeting specific faces.}
Similarly, for difficult computations it can also be useful to integrate one subset of faces at a time, for instance when a single face is slow or failing while the others complete quickly. This requires first anchoring all calls to the same problem directory via \texttt{"StopAt"}/\texttt{"StartAt"}, and then using \texttt{"SelectFaces"} to route each call to the desired subset. A typical workflow for a diagram with four faces is:
\begin{lstlisting}[language=Mathematica]
	(* First obtain a checkpoint to anchor all subsequent calls *)
	stage = STIntegrate[diag, "StopAt" -> "AfterLinearOrder"]
	
	(* Integrate only faces 1 and 4 (all epsilon orders): *)
	STIntegrate[diag, "StartAt" -> stage["CheckpointID"],
	              "SelectFaces" -> {1, 4}]
	(*Contributions from skipped faces are flagged by STwrapError[...]*)
	
	(* Integrate the remaining faces reusing faces 1 and 4 on disk *)
	STIntegrate[diag, "StartAt" -> stage["CheckpointID"],
				  "SelectFaces" -> Except[{1, 4}],
	     "ReuseExistingResults" -> True]
	
	(* Collect the full result; all faces are now on disk *)
	res = STIntegrate[diag, "StartAt" -> stage["CheckpointID"],
	           "ReuseExistingResults" -> True]
	
	(* Faces can also be selected by {epsOrder -> faceIndex} pairs *)
	STIntegrate[diag, "StartAt" -> stage["CheckpointID"],
	              "SelectFaces" -> {0 -> 1, 0 -> 2}]
\end{lstlisting}
The accepted \texttt{"SelectFaces"} syntax is summarized below:
\begin{itemize}
	\item[$\diamond$] \texttt{All}: every face at every $\varepsilon$-order (default).
	\item[$\diamond$] \texttt{n} or \texttt{\{n\}}: face index $n$ at all $\varepsilon$-orders.
	\item[$\diamond$] \texttt{\{n1, n2, \ldots\}}: face indices $n_1, n_2, \ldots$ at all $\varepsilon$-orders.
	\item[$\diamond$] \texttt{o -> i}: face index $i$ at $\varepsilon$-order $o$ only.
	\item[$\diamond$] \texttt{\{o1->i1, o2->i2, \ldots\}}: specific $(\varepsilon\text{-order},\,\text{face})$ pairs.
	\item[$\diamond$] \texttt{Except[f]}: every face not matched by \texttt{f}.
\end{itemize}

\paragraph{\label{sec:directories} Directories.}
Each \texttt{STIntegrate} call creates a working directory at
\begin{lstlisting}[language=Mathematica]
	integrands/<ID>/
\end{lstlisting}
where \texttt{<ID>} is the problem identifier set by \texttt{"SetProblemID"}. Inside, the package creates one subdirectory per face integrand of the selected gauge, e.g.:
\begin{lstlisting}[language=Mathematica]
	integrands/SubTropicaID$42_x3_1/ord_0_face_1/
\end{lstlisting}
Each face subdirectory contains:
\begin{itemize}
	\item[$\diamond$] \texttt{counter\_terms\_integrands.m}: the list of counter-term integrands for that face
	\item[$\diamond$] \texttt{bestOrder.m}: the linearly reducible integration order found by \texttt{STFasterFubini} (see App.~\ref{app:STFasterFubini})
	\item[$\diamond$] \texttt{partial\_results/result\_ct\_<n>.m}: the result of each counter-term integration, written as it completes
	\item[$\diamond$] \texttt{result.m} and \texttt{successQ.m}: the combined face result and a boolean success flag, written once all counter-terms are done. The flag in \texttt{successQ.m} is read by \texttt{"ReuseExistingResults"} to decide whether a face can be skipped in a subsequent call
\end{itemize}
These files persist after the computation and can be read back at any time via \texttt{STReadResults[id]}. This is also how \texttt{STIntegrate} assembles the final answer: it reads and sums all \texttt{result.m} files at the end of the run.

By default, the directory is wiped at the start of each call, so re-running the same diagram always starts from scratch. Setting
\begin{lstlisting}[language=Mathematica]
	$STOverwritePreviousDirectories = False (* Default: True *);
\end{lstlisting}
before a computation preserves any existing directory from the time it is ran. Note that when working with \texttt{"StartAt"}, the preferred way to skip already-completed faces is \texttt{"ReuseExistingResults" -> True}, which operates face by face and does not require setting this global flag. The flag \texttt{\$STOverwritePreviousDirectories = False} is more suited to manual workflows and debugging, where one wants to inspect or partially reuse a directory outside of the \texttt{"StartAt"} mechanism. \hfill \decosix

\subsection{\label{app:STNIntegrate}Numerical evaluation: \texttt{STNIntegrate}}

\texttt{STNIntegrate} numerically evaluates integrals. It accepts all the same input formats as \texttt{STIntegrate} (\hyperref[app:form1-item]{Forms~1}--\hyperref[app:form3-item]{3}), including generic parametric representations.
The syntax is:
\begin{lstlisting}[extendedchars=true,mathescape=true,language=Mathematica]
    STNIntegrate[input,
    "Method"             -> "pySecDec",
    "Substitutions"      -> {},
    "Order"              -> 0,
    "MaxEval"            -> $10^7$,
    "Accuracy"           -> $10^{-4}$,
    "Integrator"         -> "Qmc",
    "ContourDeformation" -> True,
    "Verbose"            -> False,
    "Dimension"          -> 4 - 2 eps,
    "Normalization"      -> Automatic,
    "Exponents"          -> Automatic,
    "LoopMomenta"        -> Automatic,
    "NumSamples"         -> 1,
    "Seed"               -> Automatic,
    "CacheDirectory"     -> Automatic]
\end{lstlisting}

\paragraph{Options.}
\texttt{"Method"} selects the backend: \texttt{"pySecDec"} (default), \texttt{"FIESTA"}, \texttt{"AMFlow"} or  \texttt{"feyntrop"}. \texttt{"Substitutions"} specifies the numerical values for kinematic variables as a list of rules (e.g., \texttt{\{s12 -> -7, mm -> 3\}}). \texttt{"Order"} sets the $\varepsilon$-expansion depth. \texttt{"MaxEval"} caps the number of integrand evaluations. \texttt{"Integrator"} selects the integration algorithm: \texttt{"Qmc"} (default, quasi-Monte Carlo), \texttt{"Vegas"}, or \texttt{"Disteval"}. \texttt{"Qmc"} is the default as it typically provides the fastest convergence. \texttt{"LoopMomenta"} follows the same convention as in \texttt{STIntegrate}. \texttt{"NumSamples" -> n} ($n>1$) re-evaluates the integral at $n$ independent kinematic points with prime-ratio components; the results are returned under the key \texttt{"samples"}. \texttt{"Seed"} overrides the default graph-hash-based random seed. \hfill \decosix

\paragraph{Return value.}
An \texttt{Association} with keys \texttt{"Value"} (the numerical result as a \texttt{SeriesData}), \texttt{"Error"} (integration uncertainties), \texttt{"Method"}, \texttt{"Integrator"}, \texttt{"Substitutions"}, and \texttt{"Timing"} (wall-clock time in seconds). \hfill \decosix

\paragraph{Numerical evaluation of hyperlogarithms: \texttt{STToGinsh}.}
\texttt{STToGinsh[expr]} numerically evaluates hyperlogarithm expressions via \texttt{ginsh}, the \texttt{GiNaC} interactive shell (requires \texttt{ginsh} installed and on the system path). It accepts \texttt{Hlog}, \texttt{Mpl}, \texttt{mzv}/\texttt{zeta}, and \texttt{Log} objects, converts them to \texttt{GiNaC} format internally, and returns a floating-point approximation. Expressions containing none of these are returned unchanged. A simple example of usage is:
\begin{lstlisting}[extendedchars=true,mathescape=true,language=Mathematica]
	(* Evaluate an expression with Hlogs and MZVs *)
	result = STToGinsh[3 Hlog[2, {0, 1}] + mzv[3]]
    (* Out: $\textcolor{gray}{\mathtt{-6.20015 + 6.53276\; I}}$ *)
\end{lstlisting}

\subsection{\label{app:STVerify}Verification: \texttt{STVerify}}
\texttt{STVerify} cross-checks a symbolic result from \texttt{STIntegrate} against a numerical evaluation. It automatically generates a Euclidean kinematic point, runs \texttt{STNIntegrate} for the numerical side and \texttt{STToGinsh} for the symbolic side, and compares Laurent coefficients order by order. The syntax is:
\begin{lstlisting}[extendedchars=true,mathescape=true,language=Mathematica]
    STVerify[input, result,
    "Order"      -> 0,
    "Dimension"  -> 4 - 2 eps,
    "Tolerance"  -> $10^{-3}$,
    "MaxEval"    -> Automatic,
    "Accuracy"   -> $10^{-4}$,
    "Integrator"     -> "Qmc",
    "Method"         -> "pySecDec",
    "Verbose"        -> False,
    "NumSamples"     -> 1,
    "Seed"           -> Automatic,
    "Substitutions"  -> Automatic,
    "Exponents"      -> Automatic]
\end{lstlisting}

\paragraph{Options.}
\texttt{"Tolerance"} (default $10^{-3}$) sets the threshold for the relative error between numerical and symbolic coefficients. \texttt{"MaxEval" -> Automatic} defaults to $10^6$ for one-loop integrals and $10^7$ otherwise. \texttt{"NumSamples" -> n} ($n>1$) re-verifies at $n$ independent kinematic points and aggregates the results. \texttt{"Seed"} overrides the default hash-based random seed. \texttt{"Substitutions"} and \texttt{"Exponents"} are forwarded to \texttt{STNIntegrate} when present; \texttt{"Substitutions" -> Automatic} (default) auto-generates a Euclidean kinematic point. The remaining options are inherited from \texttt{STNIntegrate}. \hfill \decosix

\paragraph{Return value.}
An \texttt{Association} with keys \texttt{"pass"} (\texttt{True}/\texttt{False}), \texttt{"reason"}, \texttt{"maxRelErr"}, \texttt{"relErr"} (per-coefficient relative errors), \texttt{"coefficients"} (detailed comparison at each $\varepsilon$-order), \texttt{"kinPoint"} (the kinematic point used), and \texttt{"symbolic"} (the evaluated symbolic result). \hfill \decosix

\subsection{\label{app:hyperintica-fns}Integration functions: \texttt{HyperIntica}}

While \texttt{HyperIntica} is the main (analytic) integration routine working under the hood of the \texttt{STIntegrate} function, it can also be used as a standalone function. It evaluates iterated integrals $\int_0^\infty \cdots \int_0^\infty f\, \mathrm{d}x_1 \cdots \mathrm{d}x_n$ in \eqref{eq:fnDEF} for expressions built from hyperlogarithms, polylogarithms, logarithms, and rational functions, returning results in terms of multi-zeta values and hyperlogarithms in the original integrand $f$ parameters (if any). The syntax is:
\begin{lstlisting}[extendedchars=true,mathescape=true,language=Mathematica]
	HyperIntica[integrand, variables, "Monitor" -> False]
\end{lstlisting}

\paragraph{Required arguments.}
The argument
\texttt{integrand} accepts either a symbolic expression involving \texttt{Hlog}, \texttt{Mpl}, \texttt{PolyLog}, logarithms and/or rational functions. The argument \texttt{variables}$\,= \{x_1, ..., x_n\}$ is the ordered list of integration variables, each $x_i$ integrated, left to right, from $0$ to $\infty$ by default.  \hfill \decosix

\paragraph{Optional arguments.}
Integration bounds other than $[0,\infty)$ can be specified in \textsc{Mathematica} \texttt{Integrate}-style syntax: \texttt{HyperIntica[f, \{x, 0, 1\}, \{y, 0, $\infty$\}]}, where the allowed bounds are $0$, $1$, and $\infty$. The equivalent rule-based syntax \texttt{\{x -> \{0, 1\}, y -> \{0, $\infty$\}\}} is also accepted. The option \texttt{"Monitor" -> False} (default) prints progress to standard output as each variable is processed when set to \texttt{True}, useful for tracking, e.g., long multi-variable integrations. \hfill \decosix

\paragraph{Return value.}
The function returns a combination of multi-zeta values, transcendental constants (such as $\log(2)$), and hyperlogarithms in any remaining parameters. Some simple examples of usage are:
\begin{lstlisting}[extendedchars=true,mathescape=true,language=Mathematica]
	(* Univariate integral *)
	HyperIntica[$\frac{\textcolor{Maroon}{\mathtt{Log}}[\mathtt{z}]}{\mathtt{1} \mathtt{-} \mathtt{z}^{\mathtt{2}}}$, {z}] (* Out: $\textcolor{gray}{\mathtt{-}\frac{\mathtt{3}\;\mathtt{mzv[2]}}{\mathtt{2}}}$ *)

	(* Two-variable integral with finite bounds *)
	HyperIntica[$\frac{\mathtt{1}}{(\mathtt{1}\mathtt{+}\mathtt{x})(\mathtt{1}\mathtt{+}\mathtt{y})}$, {x, 0, 1}, {y, 0, 1}]  (* Out: $\textcolor{gray}{\mathtt{Log[2]}^{\mathtt{2}}}$ *)

	(* Monitor progress for a less trivial example *)
	HyperIntica[$\frac{\textcolor{Maroon}{\mathtt{Log}}[\mathtt{1} \mathtt{+} \mathtt{x}/\mathtt{y}]^{\mathtt{2}}\; \textcolor{Maroon}{\mathtt{Log}}[\mathtt{1} \mathtt{+} \mathtt{1}/\mathtt{x}]\; \textcolor{Maroon}{\mathtt{Log}}[\mathtt{y}]}{
		\mathtt{x}\; (\mathtt{1} \mathtt{+} \mathtt{y})\; (\mathtt{1} \mathtt{+} \mathtt{x} \mathtt{+} \mathtt{y})}$, {x, y}, "Monitor" -> True]
	(* Out: $\textcolor{gray}{\frac{\mathtt{24}\;\mathtt{mzv[2]}^{\mathtt{2}}}{\mathtt{5}} \mathtt{-} \mathtt{9}\;\mathtt{Log[2]}\;\mathtt{mzv[2]}^{\mathtt{2}} \mathtt{-} \frac{\mathtt{27}}{\mathtt{2}}\;\mathtt{mzv[2]}\;\mathtt{mzv[3]} \mathtt{+} \frac{\mathtt{127}\;\mathtt{mzv[5]}}{\mathtt{8}}}$ *)
\end{lstlisting}
If the output of \texttt{HyperIntica} depends on \texttt{zeroInfPeriod}, one can try to simplify it in terms of multi-zeta values by applying \texttt{CleanZeroInf[]} to the output.
\hfill \decosix

\paragraph{Linear reducibility.}
The order of \texttt{variables} (i.e., the order in which we integrate) matters a lot: the algorithm requires each polynomial encountered during integration to factor linearly the current variable. When calling \texttt{HyperIntica} directly, a wrong order can silently return an incomplete (and thus incorrect) result. The two global flags \texttt{\$NoAlgebraicRootsContributions} and \texttt{\$HyperWarnZeroed} control the response to a non-linearly-factoring polynomial:
\begin{itemize}
	\item[$\diamond$] \texttt{\$NoAlgebraicRootsContributions = True} (default): the non-linear factor is set to zero. If \texttt{\$HyperWarnZeroed = True} (default), two messages fire: \texttt{LinearFactors::nonlinear} identifies the offending polynomial and \texttt{LinearFactors::zeroed} reminds the user to verify the ordering. Setting \texttt{\$HyperWarnZeroed = False} suppresses both reminders. In either case the result is $0$, which is incorrect if the ordering is not linearly reducible.
	\item[$\diamond$] \texttt{\$NoAlgebraicRootsContributions = False}: \texttt{HyperIntica} returns \texttt{\$Failed} immediately upon encountering the non-linear factor, with message \texttt{LinearFactorsE::nonlinear}. This makes an incorrect ordering immediately visible.
\end{itemize}
 Note that using \texttt{\$NoAlgebraicRootsContributions = True} is safe when \texttt{HyperIntica} is called from \texttt{STIntegrate}, which verifies a valid order exists first, but it is in general advisable to verify a chosen order is linearly reducible by setting \texttt{\$NoAlgebraicRootsContributions = False} before calling \texttt{HyperIntica} directly as a standalone function.

Both settings are illustrated by the following example:
\begin{lstlisting}[extendedchars=true,mathescape=true,language=Mathematica]
    (*Consider the following function to be integrated over x and y: *)
	f = 1/(1 - x + x y$^{\mathtt{2}}$);

	(* Good order: 1-x+x*y$\textcolor{gray}{{}^2}$ is linear in x; after integrating x the resulting letter y$\textcolor{gray}{{}^2}$-1 = (y-1)(y+1) factors linearly in y *)
	HyperIntica[f, {x, 0, 1}, {y, 0, 1}]   (* Out: $\textcolor{gray}{\frac{\mathtt{3}\;\mathtt{mzv[2]}}{\mathtt{2}}}$ *)

	(* Bad order: x*y^2+(1-x) does not factor linearly in y *)

	(* With $\textcolor{gray}{\$}$NoAlgebraicRootsContributions = True,
	        $\textcolor{gray}{\$}$HyperWarnZeroed = True (defaults): *)
	HyperIntica[f, {y, 0, 1}, {x, 0, 1}]
	(* LinearFactors::nonlinear and LinearFactors::zeroed warnings fire
	Out: 0 (incorrect!) *)

	(* With $\textcolor{gray}{\$}$NoAlgebraicRootsContributions = False: *)
	HyperIntica[f, {y, 0, 1}, {x, 0, 1}]
	(* LinearFactorsE::nonlinear fires
	Out: $\textcolor{gray}{\$}$Failed *)

	(* Cross-check *)
	Integrate[f, {x, 0, 1}, {y, 0, 1}]
    (* Out: $\textcolor{gray}{\frac{\mathtt{Pi}^{\mathtt{2}}}{\mathtt{4}} = \frac{\mathtt{3}\;\mathtt{mzv[2]}}{\mathtt{2}}}$ *)
\end{lstlisting}

\paragraph{Divergence checking.}
The global flag \texttt{\$HyperInticaCheckDivergences} controls whether \texttt{HyperIntica} computes and verifies boundary expansions at each integration step. In \texttt{HyperIntica.wl} the default is \texttt{True}: after finding the primitive in each variable, the code expands the result at $z=0$ and $z=\infty$, collects all pole terms, and checks that they cancel. If a divergence is detected and \texttt{\$HyperInticaAbortOnDivergence = True} (default), the computation aborts with \texttt{\$Failed}. Setting \texttt{\$HyperInticaAbortOnDivergence = False} instead stores the divergence (if any) in the association \texttt{\$HyperInticaDivergences} and continues.

When the flag is set to \texttt{False}, the boundary expansions themselves are skipped entirely, not just the final cancellation check, so a speedup generally comes from avoiding the computation of the expansions altogether.

When \texttt{HyperIntica} is called from within \texttt{STIntegrate}, the flag is automatically set to \texttt{False}, since the tropical subtraction algorithm guarantees that every counter-term integrand passed to \texttt{HyperIntica} is locally finite. When using \texttt{HyperIntica} as a standalone function on an integral whose finiteness is not known \emph{a priori}, the user should keep \texttt{\$HyperInticaCheckDivergences = True} and \texttt{\$HyperInticaAbortOnDivergence = True} (the default in standalone mode). \hfill \decosix

\paragraph{Memoization and cache management.}
During integration, {\texttt{HyperIntica}} accumulates several internal memoization caches. These persist for the duration of the session, which is beneficial when evaluating many related integrals, and can be cleared at any time by calling \texttt{ForgetAllMemo} in the notebook where calculations are done. \hfill \decosix

\paragraph{Derivatives.} \texttt{HyperD[expr, var]} computes the derivative of \texttt{expr} with respect to \texttt{var}, correctly implementing differentiation rules for \texttt{Hlog} and \texttt{Mpl} that the built-in \textsc{Mathematica} \texttt{D[]} does not handle. \texttt{expr} may contain \texttt{Hlog}, \texttt{Mpl}, \texttt{PolyLog}, logarithms, and algebraic functions, while \texttt{var} can appear in both the upper limit and the letters of hyperlogarithms. Some examples are:
\begin{lstlisting}[extendedchars=true,mathescape=true,language=Mathematica]
	(* Differentiate Li$\textcolor{gray}{{}_2}$(1-x/u) with respect to x *)
	HyperD[Mpl[{2}, {1 - x/u}], x]
	(* Out: $\textcolor{gray}{\mathtt{-}\frac{\mathtt{Mpl[\{1\}, \{1 - x/u\}]}}{\mathtt{u}(\mathtt{1} \mathtt{-} \mathtt{x}/\mathtt{u})}}$ *)

	(* Differentiate Hlog with var-dependent letters *)
	HyperD[Hlog[z, {a, b}], a]
	(* Out: $\textcolor{gray}{\frac{\mathtt{Hlog[z, \{a\}]} \mathtt{-} \mathtt{Hlog[z, \{b\}]}}{\mathtt{a} \mathtt{-} \mathtt{b}} \mathtt{-} \frac{\mathtt{Hlog[z, \{b\}]}}{\mathtt{z}\mathtt{-}\mathtt{a}}}$ *)

	(* Product involving Mpl *)
	HyperD[Log[x] Mpl[{1,2}, {x, y}], x]
	(* Out: $\textcolor{gray}{\frac{\mathtt{Log[x]}}{\mathtt{x}\mathtt{-}\mathtt{1}} \left(\frac{\mathtt{Mpl[\{2\}, \{x\;y\}]}}{\mathtt{x}}\mathtt{-}\mathtt{Mpl[\{2\},\{y\}]}\right) \mathtt{+} \frac{\mathtt{Mpl[\{1,2\},\{x,y\}]}}{\mathtt{x}}}$ *)
\end{lstlisting}

\paragraph{Series.} \texttt{HyperSeries[expr, \{var, point, order\}]} expands \texttt{expr} around \texttt{var = point} through $\mathcal{O}((\mathtt{var}-\mathtt{point})^{\mathtt{order}})$, handling \texttt{Hlog} and \texttt{Mpl} when their arguments depend on the expansion variable (unlike the built-in \texttt{Series}). Some examples are:
\begin{lstlisting}[extendedchars=true,mathescape=true,language=Mathematica]
	(* Expand Li$\textcolor{gray}{{}_2}$(z) around z = 0 *)
	HyperSeries[Mpl[{2}, {z}], {z, 0, 5}]
	(* Out: $\textcolor{gray}{\mathtt{z} \mathtt{+} \frac{\mathtt{z}^{\mathtt{2}}}{\mathtt{4}} \mathtt{+} \frac{\mathtt{z}^{\mathtt{3}}}{\mathtt{9}} \mathtt{+} \frac{\mathtt{z}^{\mathtt{4}}}{\mathtt{16}} \mathtt{+} \frac{\mathtt{z}^{\mathtt{5}}}{\mathtt{25}}}$ *)

	(* Expand with parametric dependence *)
	HyperSeries[Mpl[{2}, {1 - x/u}], {x, 0, 3}]
	(* Out: $\textcolor{gray}{\frac{\mathtt{Pi}^{\mathtt{2}}}{\mathtt{6}} \mathtt{+} \frac{\mathtt{x}(\mathtt{-1} \mathtt{+} \mathtt{Log[x/u]})}{\mathtt{u}} \mathtt{+} \frac{\mathtt{x}^{\mathtt{2}}(\mathtt{-1} \mathtt{+} \mathtt{2}\;\mathtt{Log[x/u]})}{\mathtt{4}\;\mathtt{u}^{\mathtt{2}}} \mathtt{+} \frac{\mathtt{x}^{\mathtt{3}}(\mathtt{-1} \mathtt{+} \mathtt{3}\;\mathtt{Log[x/u]})}{\mathtt{9}\;\mathtt{u}^{\mathtt{3}}}}$ *)

	(* Expand Hlog with argument approaching 0 *)
	HyperSeries[Hlog[z, {0, 1}]/z, {z, 0, 4}]
	(* Out: $\textcolor{gray}{\mathtt{-1} \mathtt{-} \frac{\mathtt{z}}{\mathtt{4}} \mathtt{-} \frac{\mathtt{z}^{\mathtt{2}}}{\mathtt{9}} \mathtt{-} \frac{\mathtt{z}^{\mathtt{3}}}{\mathtt{16}}}$ *)
\end{lstlisting}

\paragraph{Fibration basis.} \texttt{FibrationBasis[expr, variables]} rewrites an expression in a fibration basis, often making non-trivial functional identities between hyperlogarithms manifest. The argument \texttt{expr} accepts a symbolic expression (e.g., \texttt{PolyLog} or \texttt{Hlog}). The \texttt{variables} argument defines the fibration order $x_1 \prec x_2 \prec \cdots \prec x_n$, where $w_j = w_j(x_1,\ldots,x_n)$. The output can vary significantly with this ordering. The decomposition is based on the shuffle identity introduced earlier in \eqref{eq:shuffle2},
which allows one to iteratively fiber off letters depending on $x_k$ by evaluating them at the leading letter $w_{j_1}$,
\begin{equation*}
	\Hlog\!\left(x_k;\, w_{j_1}, w_{j_2}, \ldots, w_{j_w}\right)
	=
	\sum_{\vec{u}\,\shuffle\,\vec{v}\,=\,(w_{j_2},\ldots,w_{j_w})}
	\Hlog\!\left(x_k;\, w_{j_1},\, \vec{u}\,\right)
	\cdot
	\Hlog\!\left(w_{j_1};\, \vec{v}\,\right),
\end{equation*}
where $\vec{u}$ contains only letters at levels ${\prec}\, x_k$ and $\Hlog(w_{j_1};\vec{v})$ provides the coefficient in the lower-level variables. Applying this recursively expresses any $\Hlog$ as a polynomial in a basis of $\Hlog$'s, where each word is lexicographically smaller than all of its proper suffixes.
For example:
\begin{lstlisting}[extendedchars=true,mathescape=true,language=Mathematica]
	(* Single-variable: Li$\textcolor{gray}{{}_5}$(-1/z) expressed as sum of  $\textcolor{gray}{c_w}$ * Hlog[z, w] *)
	FibrationBasis[PolyLog[5, -1/z], {z}]
	(* Out: $\textcolor{gray}{\mathtt{-Hlog[z, \{0, 0, 0, 0, -1\}]} \mathtt{+} \mathtt{Hlog[z, \{0, 0, 0, 0, 0\}]}}$
	        $\textcolor{gray}{\mathtt{+}\; \mathtt{Hlog[z, \{0, 0, 0\}]}\;\mathtt{mzv[2]} \mathtt{+} \frac{\mathtt{7}\;\mathtt{Hlog[z, \{0\}]}\;\mathtt{mzv[2]}^{\mathtt{2}}}{\mathtt{10}}}$ *)
	
	(* Two-variable: *)
	FibrationBasis[Hlog[1, {1 + x, 1 + y}], {x, y}]
	(* Out: $\textcolor{gray}{\mathtt{-Hlog[x, \{y, -1\}]} \mathtt{+} \mathtt{Hlog[x, \{y, 0\}]} \mathtt{-} \mathtt{Hlog[x, \{0\}]}\;\mathtt{Hlog[y, \{-1\}]}}$
	$\textcolor{gray}{\mathtt{+}\; \mathtt{Hlog[x, \{y\}]}\;\mathtt{Hlog[y, \{-1\}]} \mathtt{+} \mathtt{Hlog[x, \{0\}]}\;\mathtt{Hlog[y, \{0\}]}}$
	$\textcolor{gray}{\mathtt{-}\; \mathtt{Hlog[x, \{y\}]}\;\mathtt{Hlog[y, \{0\}]} \mathtt{+} \mathtt{Hlog[y, \{0, -1\}]} \mathtt{-} \mathtt{Hlog[y, \{0, 0\}]} \mathtt{-} \mathtt{mzv[2]}}$ *)
\end{lstlisting}

\paragraph{Symbol.} \texttt{ConvertToSymbol[expr]} converts an expression containing hyperlogarithms and multiple polylogarithms to its ``symbol'' \cite{Goncharov:2010jf,Duhr:2011zq}. The available options are:
\begin{lstlisting}[extendedchars=true,mathescape=true,language=Mathematica]
	ConvertToSymbol[expr, "Expand" -> True,
	               "DropConstants" -> False]
\end{lstlisting}
\texttt{"Expand" -> True} (default) factors each symbol letter into irreducible polynomials via \texttt{FactorList}, revealing the so-called ``symbol alphabet.'' 
\texttt{"DropConstants" -> True} removes entries with only constant letters, following the same convention as \texttt{PolyLogTools} \cite{Duhr:2019tlz}. Some examples are:
\begin{lstlisting}[extendedchars=true,mathescape=true,language=Mathematica]
	(* Symbol of Li$\textcolor{gray}{{}_2}$(z) *)
	ConvertToSymbol[PolyLog[2, z]]  (* Out: $\textcolor{gray}{\mathtt{-Sym[\{z, 1-z\}]}}$ *)

	(* Symbol *)
	ConvertToSymbol[Log[x] Log[1-x]]
	(* Out: $\textcolor{gray}{\mathtt{Sym[\{1 - x, x\}]} \mathtt{+} \mathtt{Sym[\{x, 1 - x\}]}}$ *)

	(* Without factoring letters *)
	ConvertToSymbol[Hlog[1, {0, 1-x}], "Expand" -> False]
	(* Out: $\textcolor{gray}{\mathtt{Sym[\{}\frac{\mathtt{1}}{\mathtt{x} \mathtt{-} \mathtt{1}}\mathtt{,}\; \frac{\mathtt{x}}{\mathtt{1} \mathtt{-} \mathtt{x}}\mathtt{\}]}}$ *)
\end{lstlisting}

\paragraph{Notation conversions.}
\texttt{HyperIntica} uses two internal representations for iterated integrals: \texttt{Hlog[x, \{letters\}]} (hyperlogarithms with explicit upper limit) and \texttt{Mpl[\{weights\}, \{args\}]} (multiple polylogarithms in the $\mathrm{Li}_{n_1,\ldots}(z_1,\ldots)$ convention). The following replacement rules convert between them:
\begin{lstlisting}[extendedchars=true,mathescape=true,language=Mathematica]
expr /. Hlog -> HlogAsMpl   (* Hlog $\textcolor{gray}{\to}$ Mpl notation *)
expr /. Mpl -> MplAsHlog    (* Mpl $\textcolor{gray}{\to}$ Hlog notation *)
\end{lstlisting}
These are used, for instance, when comparing results expressed in different bases or when interfacing with external tools that expect a specific notation.

When integrating over $[0,\infty)$, boundary contributions at $0$ and $\infty$ may arise as formal \texttt{ZeroInfPeriod} objects. The replacement rule
\begin{lstlisting}[extendedchars=true,mathescape=true,language=Mathematica]
expr /. ZeroInfPeriod -> ZeroInfPeriodAsMpl
\end{lstlisting}
expresses these boundary periods in terms of \texttt{Mpl} values (typically multiple zeta values), allowing the full result to be written in a uniform notation. \hfill \decosix

\subsection{\label{app:tropical-utilities}Tropical subtraction utilities}

This section documents the lower-level functions that implement the tropical subtraction pipeline. When using \texttt{STIntegrate}, these are called internally. They can also be invoked directly for non-standard integrals or debugging purposes (see Sec.~\ref{sec:examples} for a worked example).

\emph{Measure convention.} The tropical utility functions (\texttt{STPreAnalysis}, \texttt{STTropicalContinuation}, \texttt{STExpandIntegral}) expect the integrand in the $\d\log\alpha_i$ convention (Sec.~\ref{par:measure}). When calling them directly on an integrand written in the $\d x_i$ convention, multiply by $\prod_i x_i$. The output of \texttt{STExpandIntegral} is returned in the $\d x_i$ convention, ready to be passed directly to \texttt{HyperIntica}. \hfill \decosix

\subsubsection{\label{app:STPreAnalysis}Tropical pre-analysis: \texttt{STPreAnalysis}}

\texttt{STPreAnalysis} performs an extended analysis of the singularity structure of an Euler integrand. It is useful for diagnosing which divergences are and are not regulated by a given set of regulators.
\begin{lstlisting}[extendedchars=true,mathescape=true,language=Mathematica]
	STPreAnalysis[integrand, variables, coefficients
	, True       (* facesQ, optional *)
	, {eps}  ]   (* regulators, optional *)
\end{lstlisting}

\paragraph{Arguments.}
\texttt{integrand}, \texttt{variables}, and \texttt{coefficients} follow the same conventions as in \texttt{STExpandIntegral} below. The optional argument \texttt{facesQ} (default \texttt{True}) controls whether to compute the divergent faces. The optional \texttt{regulators} (default \texttt{\{eps\}}) lists the parameters used to determine which rays correspond to divergences. \hfill \decosix

\paragraph{Return value.}
An \texttt{Association} with keys:
\texttt{"trops"} (the value of the tropical integrand $\mathrm{Trop}\,\mathcal{I}(\rho)$ evaluated on each divergent ray~$\rho$), \texttt{"rays"} (indices of the divergent rays), \texttt{"faces"} (the set $\Sigma^{\rm div}$ from \eqref{eq:divergentfan}), \texttt{"us"} (the $u$-variables), and \texttt{"trData"} (the full tropical data structure including the rays, facets, vertices, and equations). 
A zero entry in \texttt{"trops"} indicates that the integrand has a logarithmic singularity along that ray which is \emph{not} regulated by the specified regulators: the $\varepsilon$-dependent part of the tropical integrand vanishes on this ray, so $\varepsilon$ alone cannot render the integral finite. Such rays typically signal the need of a secondary regulator (e.g., the parameter $q$ in the small-$x$ example of Sec.~\ref{sec:examples}) and must be handled via \texttt{STTropicalContinuation} (App.~\ref{app:STTropicalContinuation}) before proceeding with the tropical subtraction. \hfill \decosix

\subsubsection{\label{app:STTropicalContinuation}Nilsson--Passare continuation: \texttt{STTropicalContinuation}}

\texttt{STTropicalContinuation} performs the Nilsson--Passare analytical continuation along a specified set of rays of the Newton polytope, extracting pole contributions before the subtraction formula is applied. It is called internally by \texttt{STExpandIntegral}, but can also be invoked directly when finer control over the continuation is needed (see the small-$x$ example in Sec.~\ref{sec:examples} for a worked illustration). 
\begin{lstlisting}[extendedchars=true,mathescape=true,language=Mathematica]
	STTropicalContinuation[prefsInts, variables, rays]
\end{lstlisting}

\paragraph{Arguments.}
\texttt{prefsInts} is a list of pairs \texttt{\{prefactor, integrand\}}, where each pair represents a term in the Euler integral. \texttt{variables} is the list of integration variables. \texttt{rays} is a list of rays (integer vectors) along which the continuation is to be performed. Inputting an empty list \texttt{\{\}} returns the input unchanged. The function processes the rays recursively, one at a time. \hfill \decosix

\paragraph{Return value.}
A list of pairs \texttt{\{prefactor, integrand\}} after the continuation has been performed. Each resulting term can then be passed to \texttt{STExpandIntegral} for tropical subtraction. \hfill \decosix

\subsubsection{\label{app:STExpandIntegral}Tropical subtraction: \texttt{STExpandIntegral}}

\texttt{STExpandIntegral} is the core routine that reduces a (potentially divergent) Euler integral to a sum of locally finite integrals, applying the Nilsson--Passare procedure followed by tropical subtractions as described in Sec.~\ref{sec:tropical-Cont}. It takes three required and three optional positional arguments:
\begin{lstlisting}[extendedchars=true,mathescape=true,language=Mathematica]
	STExpandIntegral[integrand, variables, coeffs
	, {eps}  (* regulators, optional *)
	, {}     (* trDataGiven, optional *)
	, {}   ] (* forceUs, optional *)
\end{lstlisting}

\paragraph{Required arguments.}
\texttt{integrand} is an Euler-type integrand, expressed as a product of polynomials raised to powers that depend on the dimensional regulator (see \eqref{eq:euler}). \texttt{variables} is the list of integration variables (Feynman or Schwinger parameters, denoted $x_j=\mathtt{xj}$), and \texttt{coeffs} lists the kinematic coefficients appearing in those polynomials. \hfill \decosix

\paragraph{Return value.}
The function returns a list \texttt{\{\{prefactor, subtraction\}, ...\}}, one entry per Nilsson--Passare continuation term. Each \texttt{subtraction} is itself a list \texttt{\{\{faceIntegrand, faceVariables\}, ...\}} indexed by the divergent faces of the Newton polytope. Here \texttt{faceIntegrand} is a list of additive terms whose sum is locally finite and ready to be passed to \texttt{HyperIntica}. \hfill \decosix

\paragraph{Optional arguments.}
\texttt{regulators} (default \texttt{\{eps\}}) lists the dimensional regulators used to identify which rays of the Newton polytope correspond to divergences. \texttt{trDataGiven} (default \texttt{\{\}}) accepts pre-computed tropical data, skipping the \texttt{Polymake} call. When \texttt{STIntegrate} is used, tropical data for all gauge candidates is computed concurrently upfront and passed through this argument. \texttt{forceUs} (default \texttt{\{\}}) pins specific $u$-variables for both the NP continuation and the subtraction formula. \hfill \decosix

\subsubsection{\label{app:STFactor}Scaling analysis: \texttt{STFactor}}

\texttt{STFactor} factorizes products inside powers that depend on $\varepsilon$: it splits $(a\cdot b)^{c\varepsilon+d}$ into $a^{c\varepsilon+d}\, b^{c\varepsilon+d}$ whenever one of the factors is manifestly positive for positive integration variables:
\begin{lstlisting}[extendedchars=true,mathescape=true,language=Mathematica]
	STFactor[expression]
\end{lstlisting}
A typical usage is to rescale the integration variables along a ray $\rho$ of the Newton polytope (e.g., $x_i \to x_i/\lambda$ for each component $\rho_i > 0$) and then apply \texttt{STFactor} to the result. The factorized form makes the $\lambda$-dependence explicit, revealing whether the integrand diverges or converges along that ray. \hfill \decosix

\subsubsection{\label{app:STMapSeries}Series mapping: \texttt{STMapSeries}}

\texttt{STMapSeries} applies a given function to each coefficient of a Mathematica \texttt{SeriesData} object and returns the result as a new \texttt{SeriesData}. For example, if the locally finite integrand has been expanded in $\varepsilon$ via \texttt{Series}, one can integrate each order separately:
\begin{lstlisting}[extendedchars=true,mathescape=true,language=Mathematica]
	STMapSeries[function, series]
\end{lstlisting}
A typical usage is \texttt{STMapSeries[HyperIntica[\#, \{x1, x2\}]\&, series]}, which calls \texttt{HyperIntica} on the integrand at each order in $\varepsilon$ and reassembles the results into a series.
\hfill \decosix

\subsection{\label{app:STFasterFubini}Linear reducibility: \texttt{STFasterFubini}}

\texttt{STFasterFubini} finds a linearly reducible integration order for a given set of polynomials, using an optimized implementation of the Fubini reduction algorithm discussed in Sec.~\ref{sec:linred}. It is called internally by \texttt{STIntegrate} to determine valid integration orders before dispatching to \texttt{HyperIntica}.\footnote{Users running
\textsc{Mathematica} $<$ 14.x may encounter a kernel crash during the linear-order finding step, caused by a bug in the \textsc{Flint} library (\href{https://github.com/flintlib/flint/issues/1998}{\texttt{fmpz\_mpoly\_pfrac\_init}: internal error}) shipped with that version.
The bug affects \texttt{FactorList} on certain multivariate polynomials arising in complicated examples
inside the internal function \texttt{STFubini}, and is
fixed in \textsc{Mathematica} 14.x.} The syntax is:
\begin{lstlisting}[extendedchars=true,mathescape=true,language=Mathematica]
	STFasterFubini[groupPolynomials, variables,
	"SolverBound" -> $10^9$,
	"FindRoots"   -> False,
	"Heuristic"   -> "LeafCountLinear"]
\end{lstlisting}

\paragraph{Arguments.}
The argument \texttt{groupPolynomials} is a list of polynomial lists (see, e.g., Sec.~\ref{sec:hyperInticaEx}), one per counter-term integrand; the algorithm finds an order that is linearly reducible simultaneously for all groups. The \texttt{variables} is the list of integration variables. The function returns a permutation of \texttt{variables} giving a valid integration order, or \texttt{NOLR} if none is found. \hfill \decosix

\paragraph{Options.}
\texttt{SolverBound} (default $10^9$) caps the number of terms in polynomials during discriminant and resultant computations; polynomials above this bound are skipped, which can speed up the search at the cost of potentially missing some valid orders. \texttt{FindRoots -> True} attempts to factor univariate polynomials of degree greater than one by introducing auxiliary root variables, extending the class of linearly reducible integrals at the cost of increased complexity (default \texttt{False}). \texttt{"Heuristic"} (default \texttt{"LeafCountLinear"}) controls how candidate orders are scored. See the description under \texttt{STIntegrate} (see Par.~\ref{par:B}) for the full list of available heuristics.
\hfill \decosix
\newcommand{\sgn}{\mathrm{sgn}}

\section{\label{app:HyperIntica}Notation used in \texttt{HyperIntica}}

This appendix collects the conventions used by \texttt{HyperIntica}, following those of Panzer's \texttt{HyperInt} \cite{Panzer_2015}. The material overlaps with the review in Sec.~\ref{sec:integration} but is presented here in self-contained form for reference.

\paragraph{Hyperlogarithms and words.}

As introduced in Sec.~\ref{sec:integration}, the hyperlogarithm $\Hlog(z; w_1, \ldots, w_n)$ is the iterated integral
\begin{equation}
\Hlog(z; w_1, \ldots, w_n) = \int_{0 < t_n < \cdots < t_1 < z} \frac{\d t_1}{t_1 - w_1}  \frac{\d t_2}{t_2 - w_2}  \cdots  \frac{\d t_n}{t_n - w_n}\,,
\end{equation}
or equivalently $\Hlog(z; w_1, \ldots, w_n) = \int_0^z \frac{\d t}{t - w_1}\, \Hlog(t; w_2, \ldots, w_n)$ with $\Hlog(z; \emptyset) = 1$. The sequence $w = [w_1, \ldots, w_n]$ is called a \emph{word}, its entries are the \emph{letters}, and the integer $n = |w|$ is the \emph{weight}. For example, the relevant alphabets are $\Sigma = \{0, 1\}$ for classical multi-zeta values and $\Sigma = \{-1, 0, 1\}$ for alternating Euler sums.

Internally, \texttt{HyperIntica} stores linear combinations of hyperlogarithms as \emph{wordlists}: lists of (coefficient, word) pairs
\begin{equation}
\mathtt{wordlist} = \bigl\{ \{c_1, w_1\}, \{c_2, w_2\}, \ldots \bigr\} \quad \longleftrightarrow \quad \sum_i c_i \cdot \Hlog(z; w_i)\,.
\end{equation}

\paragraph{Shuffle product.}

The shuffle product $u \shuffle v$ of two words $u = [u_1, \ldots, u_m]$ and $v = [v_1, \ldots, v_n]$ is the sum over all interleavings that preserve the relative order within each word (cf.\ \eqref{eq:shuffle2}):
\begin{equation}
u \shuffle v = \sum_{\sigma \in \Sigma(m,n)} \sigma(u, v)\,,
\end{equation}
where $\Sigma(m,n)$ is the set of $(m,n)$-shuffles. It satisfies the recursion
\begin{equation}\label{eq:recSh}
[a, u'] \shuffle [b, v'] = \bigl[a, (u' \shuffle [b, v'])\bigr] + \bigl[b, ([a, u'] \shuffle v')\bigr]\,,
\end{equation}
with $u \shuffle \emptyset = \emptyset \shuffle u = u$, and the key algebraic property is that hyperlogarithms satisfy a shuffle algebra:
\begin{equation}\label{eq:shuffle}
\Hlog(z; u) \cdot \Hlog(z; v) = \sum_{w \in u \shuffle v} \Hlog(z; w)\,.
\end{equation}

\paragraph{Shuffle regularization.}

Iterated integrals diverge when letters coincide with integration boundaries. Shuffle regularization exploits the algebra \eqref{eq:shuffle} to systematically extract finite values. The basic idea is that the shuffle product lets us isolate divergent contributions; for instance, for a word $[a, 0]$ with one trailing zero,
\begin{equation}
\Hlog(z; a, 0) = \Hlog(z; a) \cdot \log(z) - \Hlog(z; 0, a)\,,
\end{equation}
where $\Hlog(z; 0, a)$ is convergent as $z \to 0$ and the divergence is carried entirely by $\log(z)$. Regularization then discards the pure-logarithm divergences while preserving all algebraic relations.

As discussed in the main text, \texttt{HyperIntica} uses three regularizations, all following this pattern. They are summarized in the table below:

\begin{center}
\renewcommand{\arraystretch}{1.6}
\begin{tabular}{@{} c c c c @{}}
\toprule
\textbf{Reg.} & \textbf{Boundary} & \textbf{Divergent word structure} & \textbf{Divergence} \\
\midrule
$\Reg_0$ & $z \to 0$ & $[w_1, \ldots, w_{n-k}, \underbracket[0.4pt]{0, \ldots, 0}_{k}]$ & $\dfrac{\log(z)^k}{k!}$ \\[6pt]
$\Reg_\sigma$ & $z \to \sigma$ & $[\underbracket[0.4pt]{\sigma, \ldots, \sigma}_{k}, w_{k+1}, \ldots, w_n]$ & $\dfrac{(-\log(\sigma-z))^k}{k!}$ \\[6pt]
$\Reg_\infty$ & $z \to \infty$ & $[w_1, \ldots, w_{n-k}, \underbracket[0.4pt]{-1, \ldots, -1}_{k}]$ & $\dfrac{\log(z)^k}{k!}$ \\
\bottomrule
\end{tabular}
\end{center}

Explicitly, for a word with $k$ divergent letters at the relevant boundary, the regularization shuffles with the pure-divergent word and applies an alternating sign:
\begin{align*}
\Reg_0 \Hlog(z; [w_1, \ldots, w_{n-k}, 0^k]) &\equiv \sum_{u \in [w_1, \ldots, w_{n-k-1}] \shuffle [0]^k} (-1)^k  \Hlog(z; [u, w_{n-k}])\,, \\
\Reg_\sigma \Hlog(\sigma; [\sigma^k, w_{k+1}, \ldots, w_n]) &\equiv \sum_{u \in [\sigma]^k \shuffle [w_{k+2}, \ldots, w_n]} (-1)^k  \Hlog(\sigma; [w_{k+1}, u])\,, \\
\Reg_\infty \Hlog(\infty; [w_1, \ldots, w_{n-k}, (-1)^k]) &\equiv \sum_{u \in [w_1, \ldots, w_{n-k-1}] \shuffle [-1]^k} (-1)^k  \Hlog(\infty; [u, w_{n-k}])\,.
\end{align*}
Words consisting entirely of the divergent letter regularize to zero. Note the asymmetry: $\Reg_\sigma$ acts on \emph{leading} entries (closest to the upper bound $z$) while $\Reg_0$ acts on \emph{trailing} entries (closest to the lower bound $0$), reflecting the ordering $0 < t_n < \cdots < t_1 < z$ in the iterated integral. For illustration,
\begin{align*}
\Reg_0 \Hlog(z; [a, 0]) &= -\Hlog(z; [0, a])\,, \\
\Reg_1 \Hlog(1; [1, a]) &= -\Hlog(1; [a, 1])\,, \\
\Reg_1 \Hlog(1; [1, a, b]) &= -\Hlog(1; [a, 1, b]) - \Hlog(1; [a, b, 1])\,, \\
\Reg_\infty \Hlog(\infty; [-1]) &= 0\,.
\end{align*}

\paragraph{Periods.} The fully regularized periods combine both boundary regularizations:
\begin{subequations}
    \begin{align}
\mathtt{zeroOnePeriod}(w) &= \Reg_1 \Reg_0 \Hlog(1; w)\,, \\
\mathtt{zeroInfPeriod}(w) &= \Reg_\infty \Reg_0 \Hlog(\infty; w)\,.
\end{align}
\end{subequations}
After regularization, the results are finite linear combinations of convergent hyperlogarithms, expressible in terms of multi-zeta values. A crucial property is that shuffle regularization respects the shuffle product \cite{Panzer:2015ida},
\begin{equation}
\Reg\bigl(\Hlog(z; u) \cdot \Hlog(z; v)\bigr) = \Reg\left( \sum_{w \in u \shuffle v} \Hlog(z; w) \right)\,,
\end{equation}
ensuring consistency of the integration algorithm.

\paragraph{Multi-zeta values and classical functions as hyperlogarithms.}

The multiple polylogarithm (MPL) with positive integer indices $n_1, \ldots, n_k$ and arguments $z_1, \ldots, z_k \in \mathbb{C}$ is
\begin{equation}
\Li_{n_1, \ldots, n_k}(z_1, \ldots, z_k) = \sum_{m_1 > m_2 > \cdots > m_k \geq 1} \frac{z_1^{m_1} z_2^{m_2} \cdots z_k^{m_k}}{m_1^{n_1} m_2^{n_2} \cdots m_k^{n_k}}\,,
\end{equation}
converging for $|z_1| \leq 1$ (with $|z_1| < 1$ or $n_1 \geq 2$ at the boundary). For nonzero integers $n_1, \ldots, n_k$ with $|n_1| \geq 2$, the multiple zeta value (MZV), also called alternating Euler sum when negative indices appear, is the specialization
\begin{equation}
\begin{split}
    \zeta(n_1, n_2, \ldots, n_k) &\equiv \Li_{|n_1|, \ldots, |n_k|}\!\left(\frac{n_1}{|n_1|}, \ldots, \frac{n_k}{|n_k|}\right) 
    \\&
    = \sum_{m_1 > \cdots > m_k \geq 1} \frac{(\sgn n_1)^{m_1} \cdots (\sgn n_k)^{m_k}}{m_1^{|n_1|} \cdots m_k^{|n_k|}}\,,
\end{split}
\end{equation}
where $\sgn n_i = n_i/|n_i| \in \{-1,+1\}$. When all $n_i > 0$, this is a classical MZV. When at least one index is negative, the alternating signs $(-1)^{m_i}$ give an alternating Euler sum. The weight is $|n_1|+\cdots+|n_k|$ and the depth is $k$. Some familiar special cases are
\begin{align*}
\zeta(2) = \pi^2/6, \quad \zeta(-2) = -\pi^2/12, \quad \zeta(-1) = -\log 2\,,
\\
    \zeta(1, 2) = \zeta(3), \quad \zeta(-1, 2) = -\tfrac{3}{2} \log(2)\, \zeta(2) + \zeta(3)\,.
\end{align*}
In \texttt{HyperIntica}, these can be generated using the rule
\begin{lstlisting}[extendedchars=true,mathescape=true,language=Mathematica]
mzv[1, 2, 1, 2, 1] //. mzvAllReductions
(* Out:  -(1/30) Pi^4 Zeta[3] + 3 Zeta[7] *)
\end{lstlisting}
up to weight seven.

The connection to hyperlogarithms is that, for positive indices $n_1, \ldots, n_k$ with $n_1 \geq 2$  \cite{Panzer:2015ida},
\begin{equation}
\zeta(n_1, \ldots, n_k) = (-1)^k  \Hlog\Bigl(1; \underbracket[0.4pt]{0, \ldots, 0}_{n_k - 1}, 1, \ldots, \underbracket[0.4pt]{0, \ldots, 0}_{n_1 - 1}, 1\Bigr)\,.
\end{equation} More generally, for $n_i \in \mathbb{Z} \setminus \{0\}$ the word uses letters in $\{-1, 0, 1\}$, with each index $n_i$ contributing $|n_i| - 1$ zeros followed by a letter encoding the sign transitions.

For reference, we collect the standard dictionary. The natural logarithm is $\log(z) = \Reg_0 \Hlog(z; 0)$, while the classical polylogarithm $\Li_n(z) = \sum_{k=1}^\infty z^k/k^n$ reads
\begin{equation}
\Li_n(z) = -\Hlog\bigl(z; \underbracket[0.4pt]{0, \ldots, 0}_{n-1}, 1\bigr)\,,
\end{equation}
giving in particular $\Li_1(z) = -\log(1-z) = -\Hlog(z; 1)$, $\Li_2(z) = -\Hlog(z; 0, 1)$, and $\Li_3(z) = -\Hlog(z; 0, 0, 1)$. More generally, the MPL is expressed as
\begin{equation*}
\Li_{n_1, \ldots, n_k}(z_1, \ldots, z_k) = (-1)^k  \Hlog\Bigl(1; \underbracket[0.4pt]{0, \ldots, 0}_{n_k - 1}, \frac{1}{\prod_{j=k}^{k} z_j}, \ldots, \underbracket[0.4pt]{0, \ldots, 0}_{n_1 - 1}, \frac{1}{\prod_{j=1}^{k} z_j}\Bigr)\,.
\end{equation*}

\bibliographystyle{JHEP}
\bibliography{refs}

\end{document}